\documentclass[%
reprint,
superscriptaddress,
amsmath,amssymb,
aps,
floatfix,
nofootinbib,
longbibliography
]{revtex4-1}

\usepackage{graphicx}
\usepackage{subcaption}
\usepackage{rotating}
\usepackage{dcolumn}
\usepackage{bm}
\usepackage{braket}
\usepackage{amsmath}
\usepackage{mathtools}
\usepackage[colorlinks,allcolors=blue]{hyperref} 
\usepackage{hyperref}

\newcommand{\expect}[1]{\left\langle #1 \right\rangle}

\begin{document}
	
	\raggedbottom
	
	\preprint{APS/123-QED}
	
	\title{Quantum-Limited Acoustoelectric Amplification in a Piezoelectric-2DEG Heterostructure}
	
	\author{Eric Chatterjee}
	\affiliation{Wyant College of Optical Sciences, The University of Arizona, Tucson, AZ, USA}
    \author{Daniel Soh}
    \affiliation{Wyant College of Optical Sciences, The University of Arizona, Tucson, AZ, USA}
	\author{Matt Eichenfield}
    \email{matt.eichenfield@colorado.edu}
    \affiliation{Wyant College of Optical Sciences, The University of Arizona, Tucson, AZ, USA}
    \affiliation{Electrical, Computer \& Energy Engineering, University of Colorado, Boulder, CO, USA}
    \affiliation{Sandia National Laboratories, Albuquerque, NM, USA}

	\date{\today}
	
	\begin{abstract}

    We provide a quantum mechanical description of phonon amplification in a heterostructure consisting of a two-dimensional electron gas (2DEG) stacked on top of a piezoelectric material. An applied drift voltage effectively creates a population inversion in the momentum states of the 2DEG electrons, giving rise to spontaneous emission of phonons. Once an acoustic wave is launched, the pumped electrons release phonons via stimulated emission, returning to depleted ground states before being pumped back to the excited states. We show that whereas efficient amplification using a 1D electron gas requires the acoustic wavelength to roughly equal the mean distance between electrons, a 2DEG enables efficient amplification for any wavelength greater than this distance. We derive the imaginary and real parts of the 2DEG’s first-order acoustic susceptibility as functions of electronic drift velocity in specific limits and derive the gain per unit length for the signal and the quantum noise, with the gain matching the classical result in the short-electronic-lifetime (low-mobility) regime. Moreover, we analyze the gain clamping due to pump depletion and calculate the maximum achievable intensity. Our results provide a framework for designing novel acoustic devices including a quantum phononic laser and phase-insensitive quantum phononic amplifiers.
		
	\end{abstract}
	
	\pacs{Valid PACS appear here}
	\maketitle

\section{Introduction}

Quantum sources of microwave frequency acoustic waves can be achieved in many systems such as cavity and traveling wave optomechanics with optical and microwave frequency electromagnetic fields \cite{fiore2011storingoptical, galland2014heraldedsingle, sollner2016deterministicsingle, diamandi2025optomechanicalcontrol}, Raman processes \cite{anderson2018twocolor}, as well as piezoelectric transduction of superconducting qubits and the microwave photons they can be made to produce \cite{mirhosseini2020superconductingqubit, urmey2024stableoptomechanical}. These quantum acoustic sources  promise to lead to linear mechanical quantum computing \cite{cleland2004superconductingqubit, ruskov2012coherentphonons, satzinger2018quantumcontrol, pistolesi2021proposalnanomechanical, qiao2023splittingphonons}, quantum interferometry of phonons \cite{cohen2015phononcounting, hong2017hanburybrown, weaver2018phononinterferometry, bienfait2020quantumerasure, zivari2022onchip}, generating squeezed phononic states \cite{ma2021nonclassical} using phonons to mediate microwave-to-optical frequency conversion \cite{zhao2025quantumenabled}, and using phonons as quantum memories \cite{oconnell2010macroscopicmechanical, bozkurt2025mechanicalquantum}.

In previous work, we have shown experimentally that the integration of semiconductors on the surface of piezoelectric materials at room temperature can produce phonon lasers featuring large amplitudes but also significant thermal noise, and we have shown that the same systems can be used to induce giant nonlinearities between phonons \cite{hackett2024giantelectron}. As in the case of quantum photonics, harnessing these phononic nonlinearities to produce such phenomena as parametric down-conversion, squeezing, 3- and 4-wave-mixing processes for frequency conversion, and even means for producing single phonons requires a large-amplitude, ultra-coherent, stable phonon source acting as a phonon laser. The first step toward producing such a source is understanding the dynamics of a phonon amplifier in the quantum regime, where the noise is limited by quantum effects instead of classical effects such as thermal noise and carrier diffusion. 

Beyond the production of low-noise sources of coherent phonons, the ability to perform low-noise and phase-preserving amplification is also crucial to the development of quantum phononic systems. As an analogy, high-electron-mobility transistor (HEMT) based amplifiers provide large, phase-preserving gain and low noise (although not as low an additive noise as Josephson parametric amplifiers), which have made them a workhorse of quantum science and technology for microwave systems for decades. An analogous high-gain, low-noise, phase-preserving phonon amplifier could play an equally crucial role for quantum phononics, and here we study such a system.

Recently, we have harnessed surface acoustic waves \cite{lange2008surfaceacoustic, ding2013surfaceacoustic, schuetz2015universalquantum, mandal2022surfaceacoustic} to demonstrate the first solid-state and electrically injected surface-acoustic-wave phonon lasers \cite{wendt2026electricallyinjected} by heterogeneous integration of a bulk compound semiconductor with a piezoelectric medium in a cavity, as well as traveling-wave phonon amplifiers with high gain and low noise \cite{hackett2023nonreciprocal}. However, these devices are far from the fundamental limits of acoustoelectric amplification and phonon lasing, which would generally require reducing thermal noise, increasing mobility, and decreasing electric field screening. Specifically, the existing devices cannot operate in these regimes due to carrier freeze-out at cryogenic temperatures.

Here, we study the quantum limits of amplification for an acoustoelectric phonon amplifier by analyzing the fundamental interactions between the quantum states of a 2D electron gas and piezoelectric acoustic waves. Due to the piezoelectric nature of the acoustic medium, each phonon carries an electric field, which penetrates into the 2DEG and interacts with the electrons there via electric dipole interaction. The electric field thus mediates the electron-phonon coupling \cite{chatterjee2024abinitio}. A dc voltage is applied to the 2DEG, effectively pumping the electron distribution and giving rise to an inverted population that amplifies a propagating surface acoustic wave through stimulated emission. The key advantages provided by the 2DEG are an intrinisic conduction-band population, vastly increased mobility, and vastly reduced electric field screening compared to a bulk semiconductor (thus even further enhancing the electron-phonon coupling strength). To the last point, it is worth noting that strong acoustoelectric effects have been observed in other low-dimensional materials such as graphene \cite{bandhu2013macroscopicacoustoelectric, bandhu2016controllingproperties, hernandezminguez2018interactionsurface, mou2024gatetunable} or transition metal dichalcogenides \cite{preciado2015scalablefabrication, kalameitsev2019valleyacoustoelectric}.

Our principal focus is to theoretically determine the gain and saturation effects as functions of drift velocity, carrier concentration, and mobility for these acoustoelectric amplifiers. To do so, we analyze the electron-phonon interaction in the quantum regime, deriving the phonon emission rate per unit electric field intensity, as well as the electric field dielectric screening (which governs the electric field intensity per phonon). Next, we examine the clamping dynamics, demonstrating a tradeoff between gain and maximum acoustic field intensity. Crucially, we show that our findings for the gain in the short-electronic-lifetime (low-mobility) limit match the established classical results.

The manuscript is organized as follows: In Sec.~\ref{sec: Connecting the Classical and Quantum Pictures}, we connect the classical treatment of the acoustic wave interacting with the current distribution to the quantum picture consisting of individual electron-phonon interactions. In Sec.~\ref{sec: Deriving the Electric Susceptibility}, we derive the real and imaginary parts of the 2DEG's first-order acoustic susceptibility (which respectively govern the phonon emission rate per unit electric field intensity and the electric field intensity per phonon) in 3 regimes: high-mobility/low-drift-velocity, low-mobility, and high-drift-velocity. In Sec.~\ref{sec: Calculating the Amplifier Gain}, we synthesize the susceptibilities to calculate the amplifier gain, with the result in the low-mobility and high-mobility/low-drift-velocity regimes matching past literature. Finally, in Sec.~\ref{sec: Clamping}, we expand the Hilbert space to include both the phonon and electron states and perform a full-quantum analysis to derive the maximum phonon intensity where the pump depletion rate overcomes the excited-state repopulation rate for the electrons, causing the gain to become clamped.

\section{Connecting the Classical and Quantum Pictures}
\label{sec: Connecting the Classical and Quantum Pictures}

\begin{figure}[!tb]
	\centering
	\includegraphics[width=\linewidth]{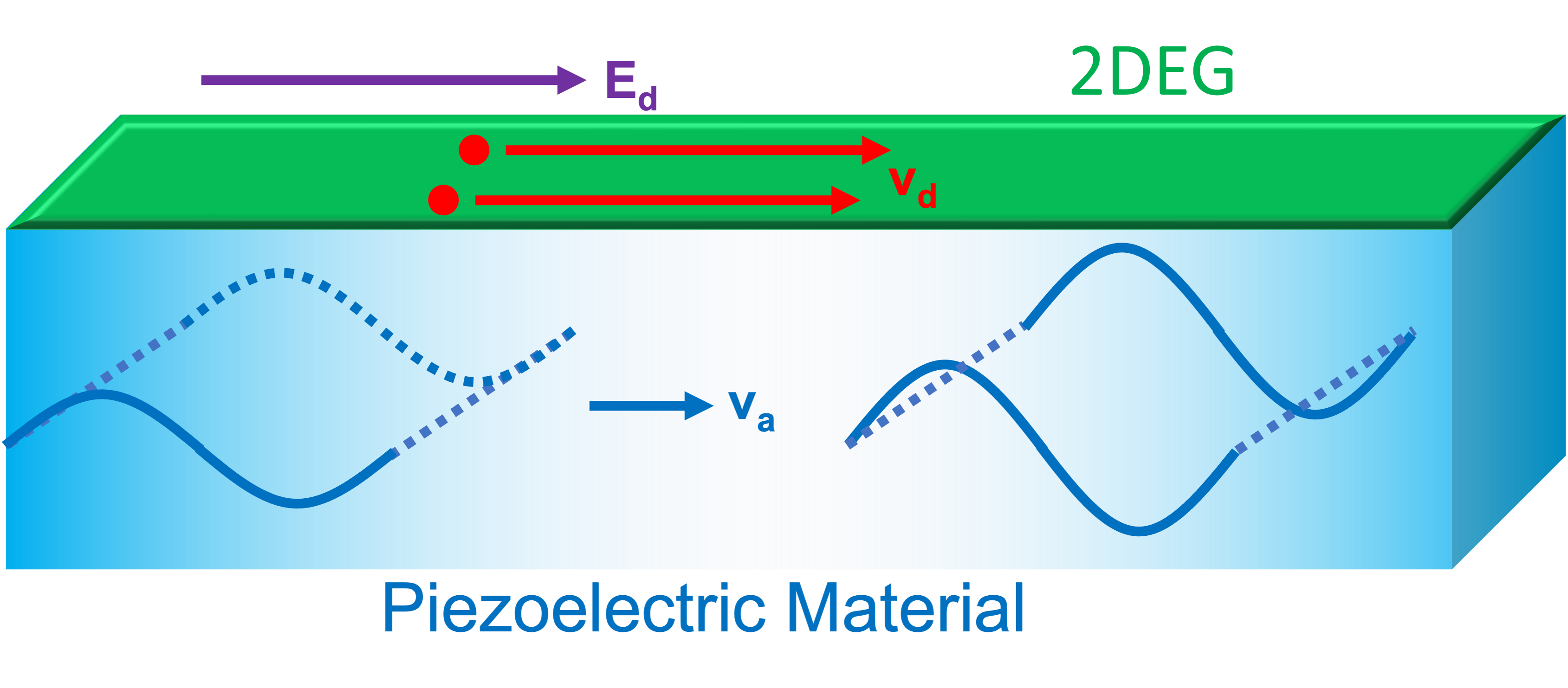}
	\caption{Diagram of acoustoelectric amplifier. Note that the external dc electric field $\bm{E_d}$ provides the energy for amplifying the traveling acoustic wave.}
	\label{fig:amplifierdiagram}
\end{figure}

An acoustoelectric amplifier consists of a semiconductor layer on top of a piezoelectric material that carries an acoustic pulse. The diagram of the setup is shown in Fig.~\ref{fig:amplifierdiagram}.
A dc electric field $\bm{E_d}$ is applied to the semiconductor layer, imparting a drift velocity $v_d$ onto the electrons. Next, an acoustic pulse is launched into the piezoelectric material, where the energy of the pulse is split between lattice vibrations and an associated longitudinal electric field. This electric field penetrates into the semiconductor layer, where it interacts with the traveling electrons. Classically, if the electron drift velocity exceeds the propagation velocity $v_a$ of the acoustic wave, then some of the electrons' kinetic energy is transferred to the acoustic field, causing the electrons to slow down and the acoustic field intensity to be amplified \cite{eichenfield2015acoustoelectric}. As a result, the overall process consists of the dc electric field providing the energy needed to amplify the traveling acoustic wave.

\begin{figure}[!tb]
	\centering
	\includegraphics[width=\linewidth]{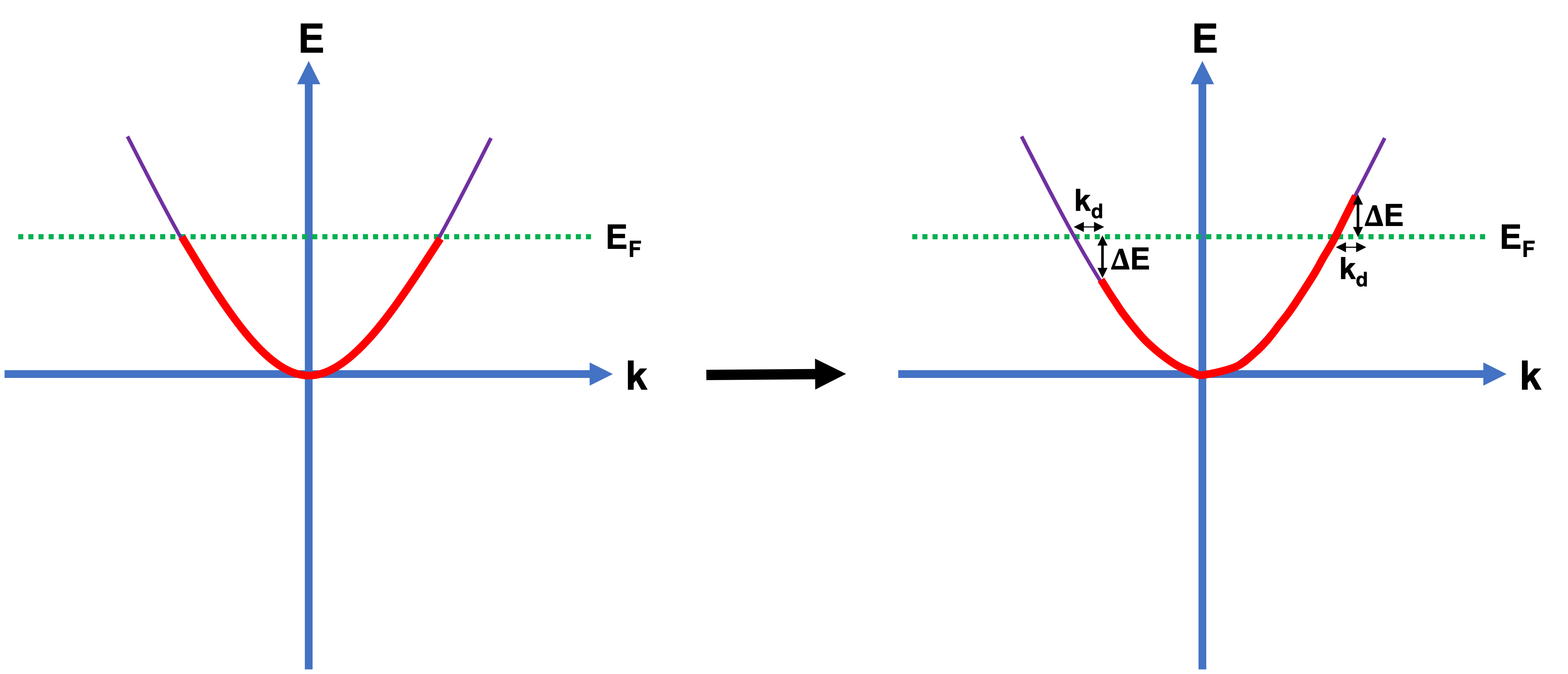}
	\caption{Shift in semiconductor electronic spectrum due to drift electric field for the 1D case. Note that the range of occupied states shifts by the drift wavevector $k_d$, raising (lowering) the highest occupied energy level for carriers traveling along (against) the dc electric field by $\Delta E$ in the limit $k_d \ll k_F$.}
	\label{fig:bandevolution}
\end{figure}

In order to understand the detailed noise dynamics of a low-intensity amplified signal, it is essential to map the aforementioned classical picture onto a quantum picture. Here, the key requirement for amplification is the presence of an inverted population of semiconductor electrons, analogous to a laser. Intuitively, when a dc electric field is applied to a semiconductor, its electron distribution should shift by a wavevector $\bm{k_d} = m \bm{v_d}/\hbar$, where $\bm{v_d}$ is the electron drift velocity induced by the field. If we consider the 1D case, for a sufficiently small drift velocity ($v_d \ll v_F$, where $v_F$ is the Fermi velocity), the positive-momentum states immediately above the Fermi level become newly populated, while the negative-momentum states immediately below the Fermi level are depopulated (where we define the positive direction as the direction of the electric field). The evolution of the electronic spectrum is depicted in Fig.~\ref{fig:bandevolution}.
This provides an inverted population, in that an acoustic pulse can stimulate the transition of electrons in the newly occupied states above the Fermi level to the newly empty states below the Fermi level through emission of phonons into the acoustic field, thereby amplifying the field.

Using the band picture, we can derive the condition for the minimum electron drift velocity required to achieve amplification of a spatially delocalized acoustic signal. In order to achieve net phonon emission, the following condition must be met based on energy and momentum conservation:
\begin{equation} \label{eq: condition on maximum slope}
\frac{2 \Delta E}{2 k_F} > \hbar v_a,
\end{equation}
where $v_a$ is the propagation velocity of the acoustic wave. In the aforementioned low-drift-velocity regime ($v_d << v_F$), $\Delta E$ becomes proportional to the Fermi wavevector $k_F$:
\begin{equation} \label{eq: energy shift}
\Delta E \approx \frac{\hbar^2 k_F}{m} k_d = \hbar v_d k_F.
\end{equation}
Note that the maximum slope is therefore invariant in $k_F$, as desired. Consequently, Eq.~\eqref{eq: condition on maximum slope} reduces to the following condition for net amplification:
\begin{equation}
v_a < \frac{2 \Delta E}{2 \hbar k_F} \approx v_d.
\end{equation}
This matches the classically predicted intuitive result. Specifically, if the carrier drift velocity exceeds the acoustic wave's propagation velocity, then a sufficiently inverted carrier population will exist such that the relaxation rate of the highest energy carriers through phonon emission exceeds the excitation rate of lower energy carriers through phonon absorption.

We note that in this 1D system, stimulated emission can only occur if $q \approx 2k_F$, where $q$ is the acoustic wavevector and $k_F$ is the Fermi momentum of the electron gas. This corresponds to the regime where the acoustic wavelength is on the same scale as the average nearest-neighbor spacing between the electrons. However, in a realistic system, the carrier density yields a nearest-neighbor electron spacing in the range of tens of nanometers, whereas for a microwave-frequency acoustic signal in the low-end of the GHz range, the wavelength is in the micron range (assuming a speed of sound of about 4000 m/s in the piezoelectric material). Consequently, a more realistic assumption is that $q \ll 2 k_F$. For a 1D electron gas, this process is suppressed by the fact that the pair of electronic states (initial and final) resonant and phase-matching with the phonon lie far below the Fermi level, corresponding to an occupied final electronic state (to which a transition is prohibited due to Pauli exclusion). This necessitates an electron gas with higher dimensionality.

To achieve amplification in the long-wavelength regime $q \ll k_F$, we thus turn to a 2D electron gas (2DEG). Here, the presence of the second degree of freedom ensures that if we project the electronic states onto the $k_x$-axis, there will be multiple bands crossing the Fermi level, each featuring a different value of the transverse momentum $k_y$. For a sufficiently high $k_y$, the longitudinal component of the Fermi momentum $k_F$ will be low enough such that we can find an electronic state pair resonant and phase-matching with the emitted phonon where the initial (final) state is occupied (unoccupied). 

In the 2D regime, near zero temperature and in the absence of the applied dc voltage, the states within the Fermi circle ($k < k_F$) will be occupied, while those outside will be unoccupied. If we apply a dc voltage corresponding to an average electron drift momentum $\bm{k_{d,\mathrm{full}}} = \hat{x}k_d + \hat{y}k_\perp$ (where $\hat{x}$ is the propagation axis of the acoustic wave and $\hat{y}$ is the transverse in-plane axis), then the circle shifts in phase space such that it will be centered at $(k_x,k_y) = (k_d,k_\perp)$. We thus use a new coordinate system in phase space $(k_x',k_y')$, defined as $k_x' = k_x - k_d$ and $k_y' = k_y - k_\perp$, such that the shifted Fermi circle is centered at $(k_x',k_y') = (0,0)$.

Note that since the phonon travels along the $x$-axis, an electron-phonon interaction conserves the electron's $y$-momentum while shifting its $x$-momentum. This accords with the intuition that an observer in the frame co-moving with the drifting electrons will not see the transverse component of the wave's propagation, since the wave's amplitude is uniform along the transverse axis. Moreover, in a free electron gas, the energy varies separately with the momenta along the two axes (i.e., the variation of the energy with momentum along a given axis is independent of the momentum along the perpendicular axis). Therefore, all functions stemming from the interaction vary only with the $x$-component of the electron momentum (i.e., with $k_d$). We thus ignore the $y$-component $k_\perp$ in our analysis. 

Then, we conceptually deduce that in order for net emission to occur from initial electronic states $k_{x,i}$ to final states $k_{x,f}$, the former must feature an excess of electrons relative to the latter, which is satisfied under the following condition:
\begin{equation} \label{eq: conceptual condition 2D amplification}
|k_{x,i}'| < |k_{x,f}'|,
\end{equation}
which is equivalent to the condition $k_d > k_c$, where $k_c = (k_{x,i} + k_{x,f})/2$. Quantitatively, for a given initial-final state pair $k_{x,i}$ and $k_{x,f} = k_{x,i} - q$, the range of $k_y$ values for which initial occupied states pair with final unoccupied states is calculated as follows:
\begin{equation} \label{eq: ky span generic}
k_{y,\textrm{span}} = 2 \Big(\sqrt{k_F^2 - k_{x,i}'^2} - \sqrt{k_F^2 - k_{x,f}'^2}\Big).
\end{equation}
It is worth noting that $k_{y,\mathrm{span}}$ is positive only if $(k_{x,i}')^2 < (k_{x,f}')^2$. This confirms the conceptual condition laid out in Eq.~\eqref{eq: conceptual condition 2D amplification}. 

We seek to map the required relationship between $k_{x,i}'$ and $k_{x,f}'$ onto a corresponding condition for drift velocity. As discussed above, we have defined the midpoint between $k_{x,i}$ and $k_{x,f}$ as $k_c$, such that $k_{x,i} = k_c + q/2$ and $k_{x,f} = k_c - q/2$. We solve for $k_c$ by determining the resonance condition where the energy difference between a state at $k_{x,i}$ and a corresponding state at $k_{x,f} = k_{x,i} - q$ equals the phonon energy $\hbar v_s q$:
\begin{align}
\begin{split}
\hbar v_s q &= \frac{\hbar^2}{2 m} \Big(k_{x,i}^2 - k_{x,f}^2\Big) = \frac{\hbar^2 q k_c}{m}, \\
k_c &= \frac{m v_s}{\hbar}.
\end{split}
\end{align}
The midpoint thus represents the electronic momentum for which the electronic velocity equals the phonon velocity. Intuitively, this corresponds to the fact that a resonant interaction occurs when the dipole velocity (i.e., the average velocity between the initial and final electronic states) equals the phonon velocity. Since $k_d$ must be greater than $k_c$ in order for net amplification to occur, the condition $v_d > v_s$ must be met, matching the classical picture.

\section{Deriving the Electric Susceptibility}
\label{sec: Deriving the Electric Susceptibility}

The key step in determining the amplification gain is deriving the first-order electric susceptibility $\chi^{(1)}(-\omega_0)$, where $\omega_0$ is the frequency of the traveling acoustic field. The imaginary part of $\chi^{(1)}$ governs the net phonon emission rate per unit volume per unit intensity of the electric field, while the real and imaginary parts combine with external screening mechanisms to govern the material's overall dielectric screening, thus inversely controlling the electric field intensity per phonon. In the following subsections, we will analytically derive $\textrm{Im}[\chi^{(1)}(-\omega_0)]$ in the high-mobility/low-drift-velocity, low-mobility, and high-drift-velocity limits. In addition, we will perform a numerical calculation for the overall range of mobilities and drift velocities. In all regimes, we will start with a model based on the Lindhard theory \cite{lindhard1953propertiesgas}, building on our previous analysis of the susceptibilities in the zero-drift-velocity case \cite{chatterjee2024abinitio}. It is worth noting, however, that the Lindhard theory treats scattering-induced electronic state collapse by adding an anti-Hermitian term to the state's Hamiltonian (corresponding to the state broadening). Although this accounts for the population decay in the initial state, it does not take the impact on the potential final states for the scattered electron into consideration. In the high-mobility limit, the low scattering rate ensures that this impact is negligible. However, for a generic mobility, we will need to apply the Mermin correction \cite{mermin1970lindharddielectric}, which ensures electron number conservation by modeling the scattering process through a quantum master equation treatment rather than through an anti-Hermitian effective Hamiltonian:
\begin{equation} \label{eq: lindhard to mermin conversion}
\chi^{(1)}(q,-\Omega,\gamma) = \frac{-\Omega + i\gamma}{-\Omega + i\gamma \frac{\chi^{(1)}_L(q,-\Omega,\gamma)}{\chi^{(1)}_L(q,0,0)}} \chi^{(1)}_L(q,-\Omega,\gamma),
\end{equation}
where $\chi^{(1)}_L(q,\Omega,\gamma)$ represents the Lindhard susceptibility for a given phonon wavevector $q$, effective phonon frequency $\Omega$, and electronic decay rate $\gamma$ (defined as $\gamma = q_e/(m\mu)$, where $q_e$, $m$, and $\mu$ are the electron charge, effective electron mass, and mobility, respectively). The effective phonon frequency is defined as the phonon frequency in the frame of the drifting electrons, thus yielding $\Omega = (v_s - v_d)q$. Note that in the coming subsections, we will label $\chi^{(1)}(q,-\Omega,\gamma)$ and $\chi^{(1)}_L(q,-\Omega,\gamma)$ as $\chi^{(1)}(-\omega_0)$ and $\chi^{(1)}_L(-\omega_0)$, respectively, for purposes of brevity.

\subsection{General Derivation for Imaginary Lindhard Susceptibility}
\label{sec: General Derivation for Imaginary Lindhard Susceptibility}

\begin{figure*}[!tb]
	\centering
	\begin{subfigure}{\columnwidth}
		\centering
	    \includegraphics[width=0.95\linewidth]{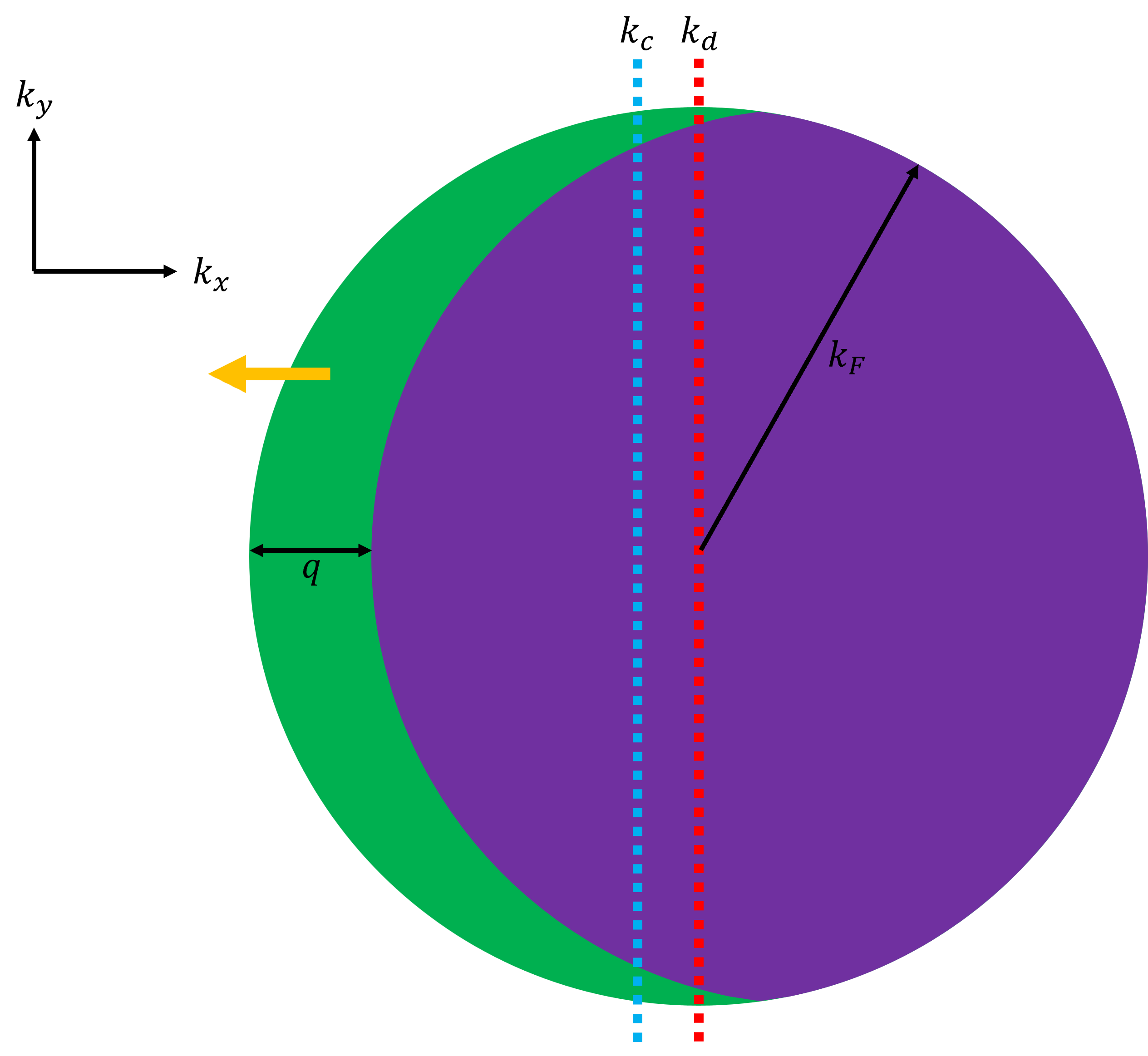}
	    \caption{}
	    \label{}
	\end{subfigure}
	\begin{subfigure}{\columnwidth}
		\centering
	    \includegraphics[width=\linewidth]{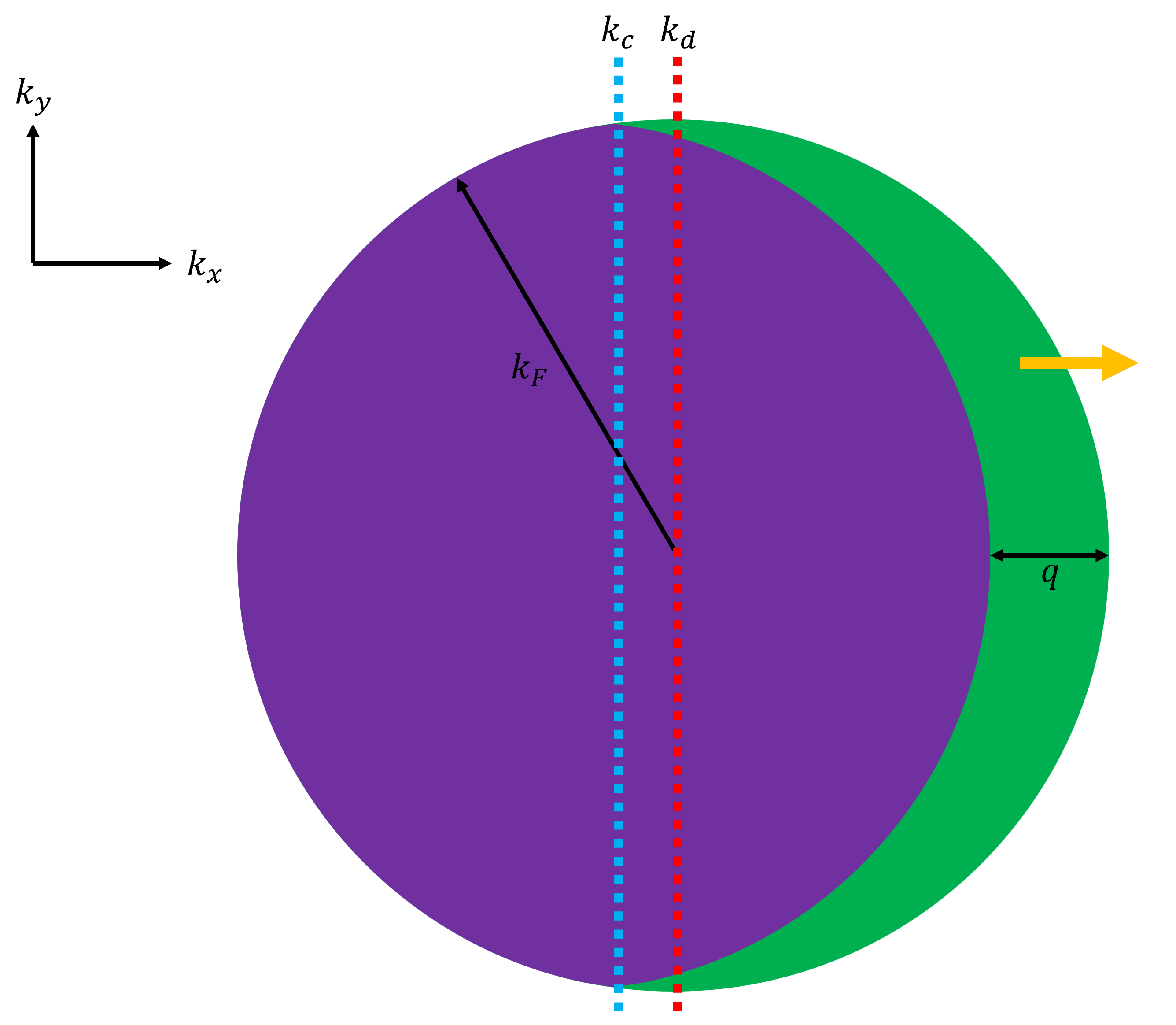}
		\caption{}
		\label{}
	\end{subfigure}
    \caption{Phase-space diagram of the regions $S$ (a), and $S'$ (b), depicted as the green-shaded crescent-shaped areas, representing the electrons that can emit (absorb) phonons of wavevector $\bm{q} = q\hat{x}$ due to the corresponding final states being unoccupied.}
	\label{fig:phasespacelowmobility}
\end{figure*}

We start with the general derivation for the imaginary part of the Lindhard susceptibility. Figure~\ref{fig:phasespacelowmobility}(a) depicts (in green) the range of initial states $(k_x,k_y)$ from which an electron can transition to an unoccupied final state $(k_x-q,k_y)$ by emitting a phonon of wavevector $q$, while Fig.~\ref{fig:phasespacelowmobility}(b) depicts the range of states in which an electron can absorb a phonon of wavevector $q$ and transition to an unoccupied state $(k_x+q,k_y)$.
The regions are labeled as $S$ and $S'$, respectively. Note that $S$ and $S'$ are reflections of each other in phase space about $k_x' = 0$ (i.e., $k_x = k_d$). The effect of the drift velocity manifests itself in setting the balance between the absorption and emission rates, thus governing the net emission (or absorption) rate. To that end, the absorptive detuning for an initial state at $k_x' \in S'$ and the emissive detuning for the corresponding initial state at $-k_x' \in S$, respectively, take the following form:
\begin{align}
\label{eq: absorptive detuning}
\omega_{k_x'+k_d+q,k_x'+k_d} - \omega_0 &= (\Delta \omega)_{k_x'} + (v_d - v_s) q, \\
\label{eq: emissive detuning}
\omega_{-k_x'+k_d-q,-k_x'+k_d} + \omega_0 &= \omega_{k_x'-k_d+q,k_x'-k_d} + \omega_0 \\
&= (\Delta \omega)_{k_x'} - (v_d - v_s) q,
\end{align}
where $v_d = \hbar k_d/m$, and the average of the two detunings is defined as $(\Delta \omega)_{k_x'}$:
\begin{equation} \label{eq: average detuning}
(\Delta \omega)_{k_x'} = \frac{\hbar q (q + 2k_x')}{2m}.
\end{equation}
Note that this value is always positive, since $k_x' > -q/2$ for all $k_x' \in S'$. It is immediately apparent that if $v_d < v_s$, the detuning is greater for the emissive case, leading the absorptive process to be favored. An example of this is the zero-bias case, where net phonon absorption takes place at any mobility \cite{chatterjee2024abinitio}. On the other hand, if $v_d > v_s$, the detuning is greater for the absorptive case, and the emissive process is favored. Moreover, the discrepancy between the absorptive and emissive detuning values increases with drift velocity, ensuring a similar increase for the net emission rate with the drift velocity. 

It is worth noting, however, that the average detuning is invariant in the drift velocity. This average value in turn equals the electronic transition frequency (i.e., the difference between the final and initial electronic state frequencies) in the case of an unbiased electronic distribution. As such, the effect of the drift velocity is equivalent to the scenario where the effective speed of sound is shifted from $v_s \rightarrow v_s - v_d$. This matches the intuitive deduction that in the frame of reference of the drifting electrons, the acoustic wave undergoes a Doppler shift to an effective frequency $(v_s - v_d)q$, labeled as $\Omega$ above. Consequently, $\chi^{(1)}_L(-\omega_0)$ will indeed serve as the $\chi^{(1)}_L(q,-\Omega,\gamma)$ input in the Lindhard-to-Mermin conversion in Eq.~\eqref{eq: lindhard to mermin conversion}.

The specific methods for deriving the imaginary parts of the Lindhard susceptibility are shown in Appendices~\ref{sec: Imaginary Part of Chi(1) High-Mobility/Low-Drift-Velocity Calculation} and~\ref{sec: Imaginary Part of Chi(1) Low-Mobility Limit} (for the high-mobility/low-drift-velocity and low-mobility limits, respectively). The general method is described by the following integral \cite{chatterjee2024abinitio}:
\begin{equation}
\textrm{Im}[\chi^{(1)}_L(-\omega_0)] = \beta \int_{S'} dk_x' k_{y,\mathrm{span}}(k_x') \textrm{Im}[f_{\mathrm{net},k_x'}^{(1)}(-\omega_0)],
\end{equation}
where $\beta$ is a constant function of the 2DEG and piezoelectric material parameters \cite{chatterjee2024abinitio}, proportional to the square of the dipole moment for an electron that has undergone interaction with the phonon field and inversely proportional to the 2DEG thickness:
\begin{equation}
\beta = \frac{q_e^2 v_s^2}{2 \pi^2 \hbar \epsilon_0 t_\mathrm{2DEG} \omega_0^2},
\end{equation}
and $f_{\mathrm{net},k_x'}^{(1)}$ is defined as the net of the inverse complex detuning values for the counter-resonant (absorptive) process for an electron at $k_x' \in S'$ and the corresponding resonant (emissive) process for an electron at $-k_x \in S$ (see Appendix~\ref{sec: Imaginary Part of Chi(1) Low-Mobility Limit}). For each electron-phonon interaction process, the inverse complex detuning governs the interaction probability, since it serves as measurement of the closeness of the interaction to resonance. Specifically, the real part corresponds to the probability of a coherent interaction, while the imaginary part corresponds to the probability of a quantum jump (i.e., a phonon absorption or emission event). The imaginary part takes the following form in terms of the average detuning from Eq.~\eqref{eq: average detuning} and the absorptive and emissive offsets shown in Eqs~\eqref{eq: absorptive detuning} and~\eqref{eq: emissive detuning}:
\begin{align}
\begin{split} \label{eq: imaginary f net}
\textrm{Im}[f_{\mathrm{net},k_x'}^{(1)}(-\omega_0)] &= \frac{\gamma}{((\Delta \omega)_{k_x'} - (v_d - v_s)q)^2 + \gamma^2} \\
&\quad - \frac{\gamma}{((\Delta \omega)_{k_x'} + (v_d - v_s)q)^2 + \gamma^2}.
\end{split}
\end{align}
As mentioned earlier, $(\Delta \omega)_{k_x'}$ is positive for all $k_x' \in S'$, since the lower bound of $k_x'$ in $S'$ is $-q/2$ (see Fig.~\ref{fig:phasespacelowmobility}(b)). The electron-phonon detuning for the absorption and emission processes is thus the same positive value when $v_d = v_s$. As the drift velocity increases, the emissive detuning decreases, while the absorptive detuning increases, and vice versa. Therefore, as expected, a higher drift velocity favors emission, while a lower drift velocity favors absorption. The relative values of $v_d$ and $v_s$ should thus determine the sign of $\textrm{Im}[\chi^{(1)}_L(-\omega_0)]$, which becomes apparent from expanding Eq.~\eqref{eq: imaginary f net} and taking the numerator:
\begin{align}
\begin{split} \label{eq: imaginary f net numerator}
&\gamma \Big((\Delta \omega)_{k_x'} + (v_d - v_s)q)^2 + \gamma^2\Big) \\
&\quad - \gamma \Big((\Delta \omega)_{k_x'} - (v_d - v_s)q)^2 + \gamma^2\Big) \\
&= 4 \gamma (\Delta \omega)_{k_x'} (v_d - v_s) q \\
&= 2 \gamma (v_d - v_s) \frac{\hbar q^2 (q + 2k_x')}{m}.
\end{split}
\end{align}
Since the denominator of each term consists of a sum of squares, the common denominator is always positive. As a result, the sign of $\textrm{Im}[\chi^{(1)}(-\omega_0)]$ equals that of the numerator. We thus analyze the above expression for the numerator. As previously discussed, $q + 2k_x' > 0$ for all values of $k_x' \in S'$. Therefore, if $v_d > v_s$ ($v_d < v_s$), $\textrm{Im}[\chi^{(1)}(-\omega_0)]$ is positive (negative), indicating net emission and amplification (net absorption and attenuation).

\subsection{General Derivation for Real Lindhard Susceptibility}
\label{sec: General Derivation for Real Lindhard Susceptibility}

We now turn to the real part. In general, this is solved by using the same methodology as in Eq.~\eqref{eq: low-mobility chi(1) integral} but replacing the imaginary parts of the inverse detuning with the corresponding real parts, yielding the following integral:
\begin{equation}
\textrm{Re}[\chi^{(1)}(-\omega_0)] = \beta \int_{S'} dk_x' k_{y,\mathrm{span}}(k_x') \textrm{Re}[f_{\mathrm{net},k_x'}^{(1)}(-\omega_0)],
\end{equation}
where $\textrm{Re}[f_{\mathrm{net},k_x'}^{(1)}]$ is defined in a manner analogous to Eq.~\eqref{eq: imaginary f net}:
\begin{align}
\begin{split} \label{eq: real f net}
\textrm{Re}[f_{\mathrm{net},k_x'}^{(1)}(-\omega_0)] &= \frac{(\Delta \omega)_{k_x'} - (v_d - v_s)q}{((\Delta \omega)_{k_x'} - (v_d - v_s)q)^2 + \gamma^2} \\
&\quad + \frac{(\Delta \omega)_{k_x'} + (v_d - v_s)q}{((\Delta \omega)_{k_x'} + (v_d - v_s)q)^2 + \gamma^2},
\end{split}
\end{align}
where the first and second terms represent negative-velocity and positive-velocity dipole formation, respectively (where the dipole velocity will be defined relative to the drift velocity throughout this section). We will make approximations for the relative size of $\gamma$ for the analytical calculations, while solving rigorously in the numerical calculations.

\begin{figure*}[!tb]
	\centering
	\begin{subfigure}{\columnwidth}
		\centering
	    \includegraphics[width=\linewidth]{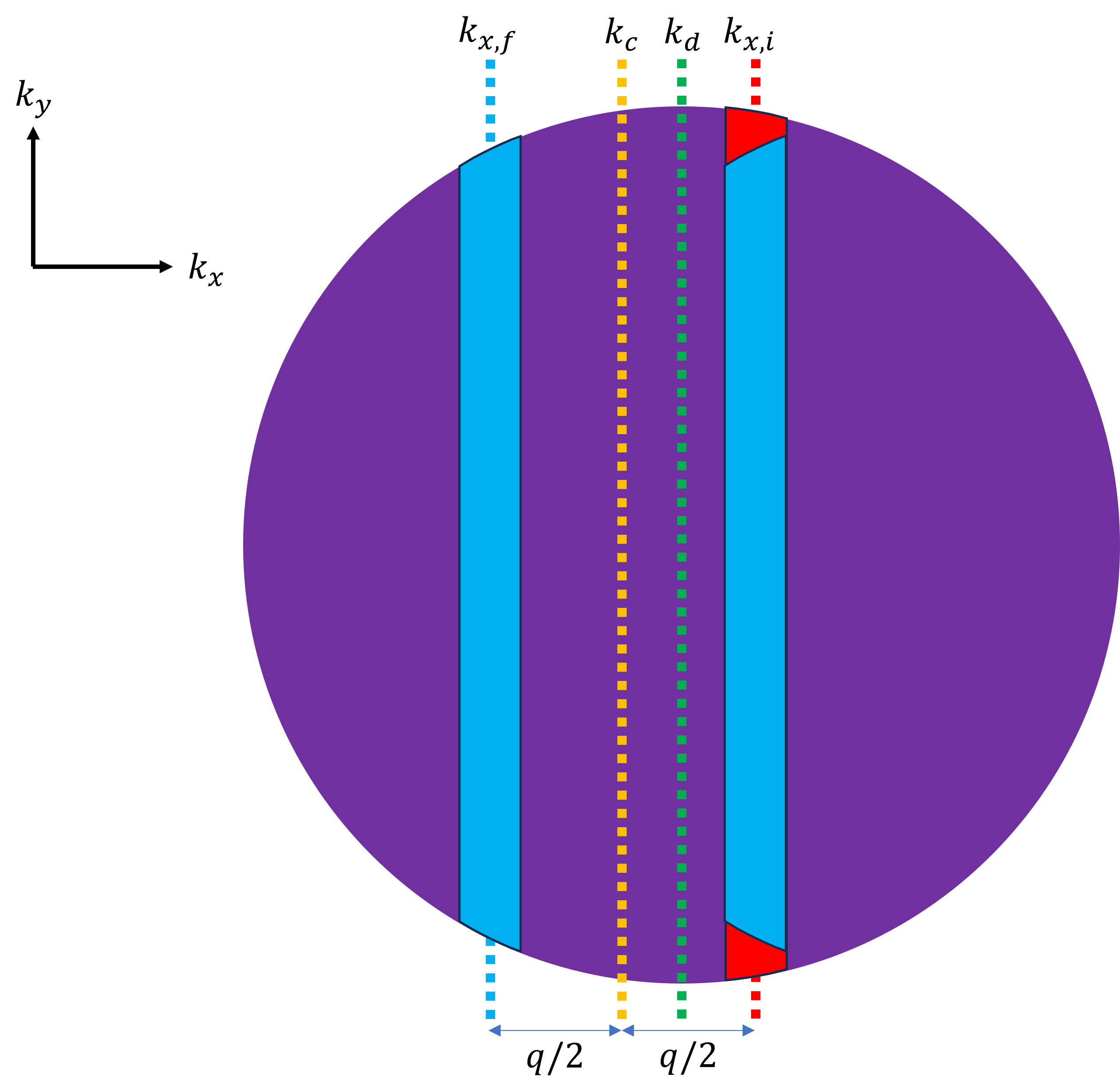}
	    \caption{}
	    \label{}
	\end{subfigure}
	\begin{subfigure}{\columnwidth}
		\centering
	    \includegraphics[width=\linewidth]{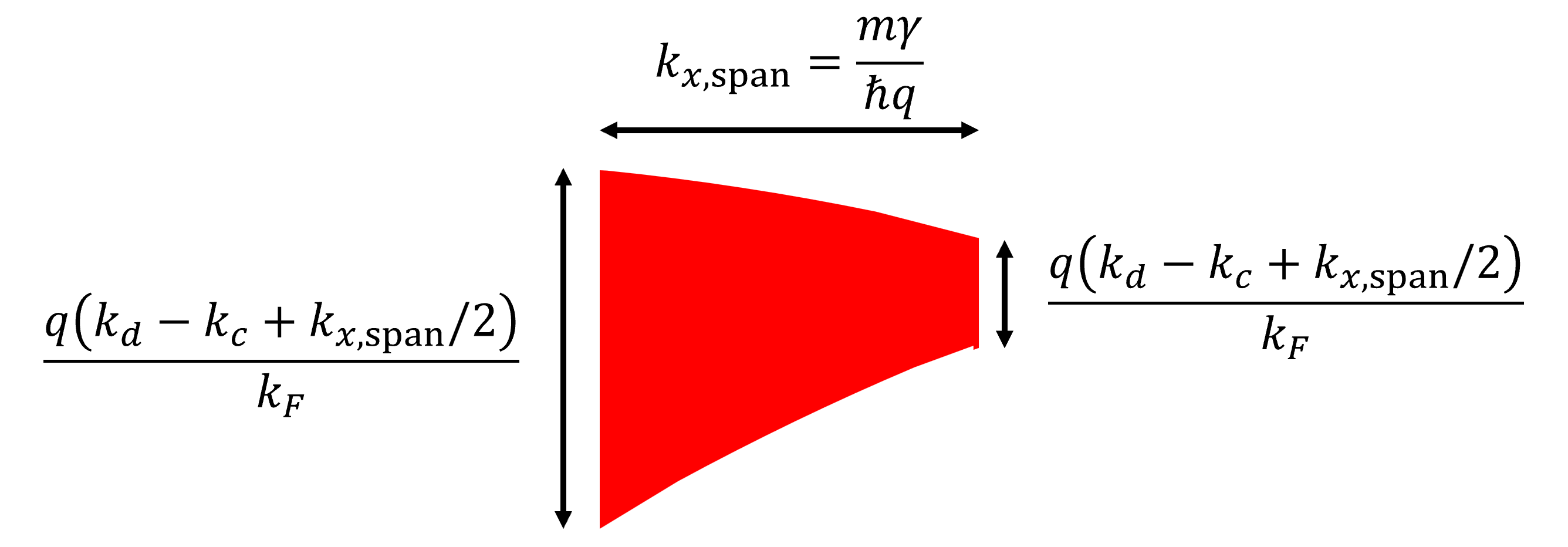}
		\caption{}
		\label{}
	\end{subfigure}
    \caption{Phase-space diagram (a) of the acoustoelectric amplification process in the high-mobility/low-drift-velocity limit given an acoustic field propagating in the $+\hat{x}$-direction with wavevector $q$, along with dimensions of each emissive region (b). Note that in (a), the red-shaded regions represent near-resonance electrons capable of emitting phonons and transitioning to unoccupied states, while the right-hand-side light-blue-shaded region (in between the 2 red regions) represents near-resonance electrons forbidden from emitting phonons since the corresponding final states (represented by the left-hand-side light-blue-shaded region) are occupied. This comports with the fact that the red-shaded regions are subsets of the broader emission-allowed region $S$.}
	\label{fig:phasespace}
\end{figure*}

\subsection{High-Mobility, Low-Drift-Velocity Regime}
\label{sec: High-Mobility Regime}

Here, we analyze the high-mobility/low-drift-velocity regime (which applies when $\gamma/q, v_d \ll v_F$), starting with deriving the imaginary part of the Lindhard susceptibility. Since $k_d \ll k_F$ in this limit, and since $q \ll k_F$ in general in the systems we consider, the resonant initial and final wavevectors will be close to the middle of the Fermi circle (i.e., $|k_{x,i}'|,|k_{x,f}'| \ll k_F$). As we have established in our previous work \cite{chatterjee2024abinitio}, if we combine this low-drift-velocity condition with the high-mobility condition, then the overall 2DEG-phonon interaction will be dominated by electrons in the near-resonance regime, which features a phase-space span of $k_{x,\mathrm{span}} = m\gamma/(\hbar q)$. Here, the $k_y$-span of initial occupied states that couple to final unoccupied states is approximated by applying Eq.~\eqref{eq: ky span generic} and taking the limit $|k_{x,i}'|,|k_{x,f}'| \ll k_F$:
\begin{align}
\begin{split} \label{eq: ky span high mobility low drift}
k_{y,\textrm{span}} &\approx 2 k_F \bigg(-\frac{k_{x,i}'^2}{2 k_F^2} + \frac{k_{x,f}'^2}{2 k_F^2}\bigg) \\
&= \frac{-q (k_{x,f}' + k_{x,i}')}{k_F} \\
&= \frac{2q (k_d - k_c - \Delta k_x)}{k_F},
\end{split}
\end{align}
where $\Delta k_x = k_x - k_{x,0}$ represents the offset of the wavevector from the resonant value. As this expression shows, if the initial momentum is more positive than the resonant value, then the $k_y$ span is reduced, and vice versa. The phase-space dynamics of the interaction in the high-mobility/low-drift-velocity regime are shown in Fig.~\ref{fig:phasespace}. Note that the circle is centered at $k_x = k_d$, while the midpoint $k_c$ between the initial and final electron wavevectors ($k_{x,i}$ and $k_{x,f}$, respectively) in the emission process is to the left of the center (i.e., $k_c < k_d$). The blue band represents the range of $k_y$ values for which the final states are already occupied, thus preventing phonon emission by electrons in that range. On the other hand, the red zones represent the initial occupied states that couple to unoccupied final states, enabling emission. As such, the red regions meet 2 conditions: first, the phonon emission process for these electrons is near-resonance, and second, phonon emission couples these electrons to unoccupied final states. The $k_y$-span of the red regions reduces as $k_x$ increases, confirming the result in Eq.~\eqref{eq: ky span high mobility low drift}.

The imaginary part of $\chi^{(1)}_L(-\omega_0)$ in this regime is solved by integrating over the valid phase-space region as shown in Appendix~\ref{sec: Imaginary Part of Chi(1) High-Mobility/Low-Drift-Velocity Calculation}, yielding the following result:
\begin{equation} \label{eq: imaginary chi(1) Lindhard high mobility low drift main}
\textrm{Im}[\chi^{(1)}_L(-\omega_0)] \approx \frac{q_e^2 v_s^2 (v_d - v_s) m^2}{\pi \hbar^3 \epsilon_0 t_\mathrm{2DEG} \omega_0^2 k_F}.
\end{equation}
Note that the result is identical to the imaginary part of $\chi^{(1)}$ calculated in Ref.~\cite{chatterjee2024abinitio}, except for the replacement $v_s^3 \rightarrow v_s^2 (v_d - v_s)$ in the numerator. Regarding this replacement, it is worth noting that if $v_d > v_s$ ($v_d < v_s$), this expression is positive (negative), corresponding to net phonon emission (absorption), i.e., amplification (attenuation) of the phonon field, as expected. In the case of zero bias voltage (i.e., $v_d = 0$), the result reduces to the opposite of the imaginary part of $\chi^{(1)}$ from Ref.~\cite{chatterjee2024abinitio}. This is because the previous result shows the net absorption rate, whereas this result shows the net emission rate.

\begin{figure*}[!tb]
	\centering
	\begin{subfigure}{\columnwidth}
		\centering
	    \includegraphics[width=\linewidth]{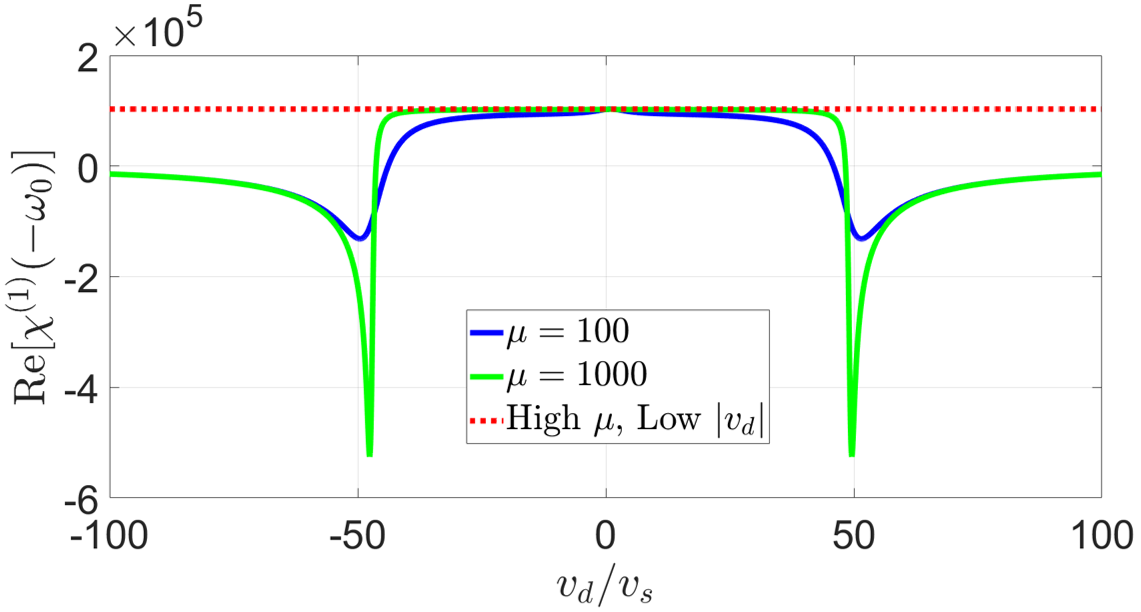}
	    \caption{}
	    \label{}
	\end{subfigure}
	\begin{subfigure}{\columnwidth}
		\centering
	    \includegraphics[width=\linewidth]{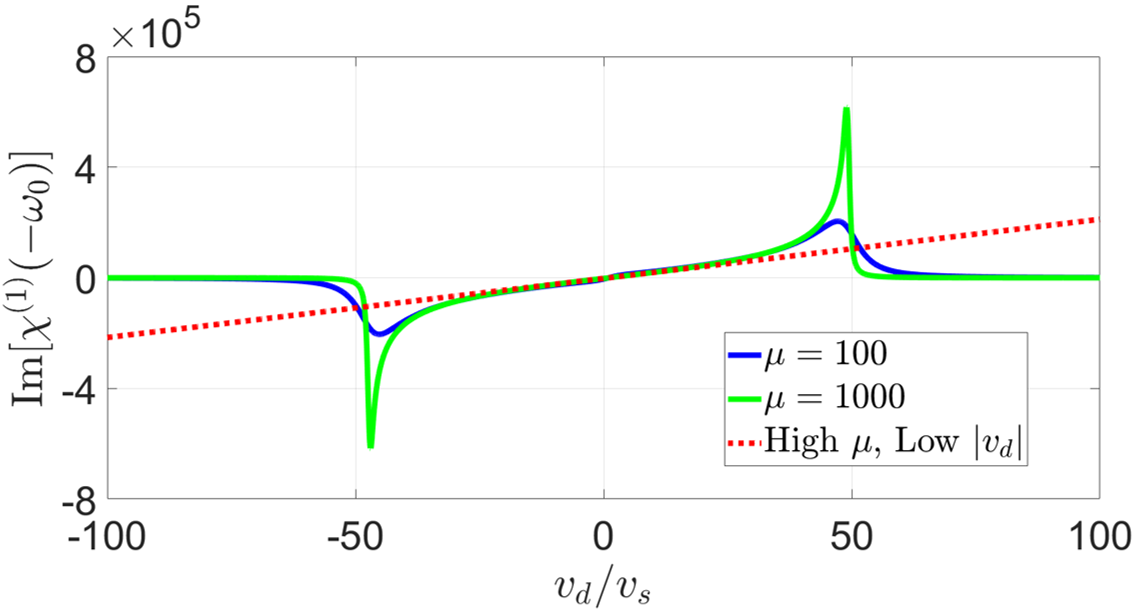}
		\caption{}
		\label{}
	\end{subfigure}
    \caption{Numerical results for the real (a) imaginary (b) parts of $\chi^{(1)}(-\omega_0)$ as functions of the drift velocity $v_d$, given a mobility (in units of $\textrm{m}^2/(\textrm{V} \cdot \textrm{s})$) of 100 (solid, blue), or 1000 (solid, green), along with the analytical results (dotted, red) for the high-mobility/low-drift-velocity limit $\gamma/q,v_d \ll v_F$. We assume a phonon angular frequency of $\omega_0 = 2\pi \times 10^9 \textrm{ s}^{-1}$, speed of sound $v_s = 4 \times 10^3$ m/s, a carrier density $n = 2 \times 10^{15} \textrm{ m}^{-2}$, a carrier effective mass $m = 0.067 m_0$, and a 2DEG thickness $t_\mathrm{2DEG} = 2 \times 10^{-8} \textrm{ m}$.}
	\label{fig:chi1highmulowvd}
\end{figure*}

The linear relationship between the imaginary part of $\chi^{(1)}_L$ and the drift-velocity-phonon-velocity mismatch $v_d - v_s$ fundamentally relates to the fact that increasing the drift velocity shifts the carrier distribution without affecting the initial and final wavevectors $k_{x,i}$ and $k_{x,f}$ that meet the resonance condition for electron-phonon interaction. If we use the previously discussed $k_x'$ coordinate system (where we maintain $k_x' = 0$ as the center line vertically bisecting the Fermi circle regardless of how shifted the circle is from $k_x = 0$), then an increase in the drift velocity leads to a leftward (negative) shift in the $k_x'$ values of the initial and final wavevectors (i.e., $k_{x,i}'$ and $k_{x,f}'$ become more negative as $v_d$ increases). Recall that, for our analytical calculation in the high-mobility regime, we used the approximation that both of the initial and final wavevectors are close to the center line of the Fermi circle (i.e., $|k_{x,i}'|,|k_{x,f}'| \ll k_F$, which corresponds to the low-drift-velocity limit $v_d \ll v_F$), as is the midpoint $k_c = mv_s/\hbar$ (i.e., $|k_d - k_c| \ll k_F$). The $k_y$-vs-$k_x'$ curvature of the Fermi circle in this region is approximately parabolic. Since $k_{y,\mathrm{span}}$ (corresponding to the excess number of electrons at $k_{x,i}$ relative to $k_{x,f}$) is proportional to the average slope of the circle between the initial and final wavevectors (which equals the slope of the circle at $k_x = k_c$ given $q \ll k_F$), the approximate parabolic curvature of the circle in this regime ensures that $k_{y,\mathrm{span}}$ varies linearly with $-(k_{x,i}' + k_{x,f}') = k_d - k_c = v_d - v_s$, as implied in Eq.~\eqref{eq: ky span high mobility low drift}. Consequently, the high-mobility Lindhard susceptibility also varies linearly with $v_d - v_s$ in the regime $|v_d| \ll v_F$.

Next, we derive the real part of the Lindhard susceptibility in the high-mobility/low-drift-velocity limit. Here, the fact that the spectral broadening is much smaller than the detuning for any transition (i.e., $\gamma \ll |(\Delta \omega)_{k_x'} \pm (v_d - v_s)q|$) causes each term in Eq.~\eqref{eq: real f net} to approximately reduce to the real inverse detuning:
\begin{align}
\begin{split}
&\textrm{Re}[f_{\mathrm{net},k_x'}^{(1)}(-\omega_0)] \approx \\
&\quad \frac{1}{(\Delta \omega)_{k_x'} - (v_d - v_s)q} + \frac{1}{(\Delta \omega)_{k_x'} + (v_d - v_s)q}.
\end{split}
\end{align}
We integrate this over $S'$ using the long-acoustic-wavelength approximation $q \ll k_F$, corresponding to the same approximations for $k_{y,\mathrm{span}}(k_x')$ and $(\Delta \omega)_{k_x'}$ as in Eq.~\eqref{eq: low mobility imaginary chi(1)} and yielding the following: 
\begin{align}
\begin{split} \label{eq: real chi(1) Lindhard high mobility low drift}
&\textrm{Re}[\chi^{(1)}_L(-\omega_0)] \\
&= \beta \int_{S'} dk_x' k_{y,\mathrm{span}}(k_x') \\
&\quad \times \bigg(\frac{1}{(\Delta \omega)_{k_x'} - (v_d - v_s)q} + \frac{1}{(\Delta \omega)_{k_x'} + (v_d - v_s)q}\bigg) \\
&\approx \beta \int_0^{k_F} dk_x' \frac{2 q k_x'}{\sqrt{k_F^2 - k_x'^2}} \\
&\quad\quad \times \Bigg(\frac{1}{\frac{\hbar q}{m} k_x' - (v_d - v_s)q} + \frac{1}{\frac{\hbar q}{m} k_x' + (v_d - v_s)q}\Bigg) \\
&= \beta \frac{2 \pi m}{\hbar} \\
&= \frac{q_e^2 v_s^2 m}{\pi \hbar^2 \epsilon_0 t_\mathrm{2DEG} \omega_0^2}.
\end{split}
\end{align}
Therefore, as long as the drift velocity is well below the Fermi velocity, the real part of $\chi^{(1)}_L$ is uniform in the high-mobility limit, reducing to the previously calculated zero-drift result \cite{chatterjee2024abinitio}. This is due to the fact that as the drift velocity is increased, the increase in the negative-velocity dipole formation probability due to decreased emissive detuning is cancelled out by the decrease in the positive-velocity dipole formation probability due to increased absorptive detuning. 

Finally, we convert these Lindhard results to the Mermin susceptibilities. To that end, we note that both the real and imaginary parts of the Lindhard susceptibility are invariant in mobility in this regime, thus equaling the same result at any electronic decay rate $\gamma$ as they would in the limit of undamped electronic states where $\gamma = 0$. Moreover, since the ratio between the imaginary and real parts is approximately $(v_d - v_s)/v_F$ (where $v_F = \hbar k_F/m$ is the Fermi velocity), and since $|v_d - v_F| \ll v_F$ in the low-drift-velocity limit, the overall Lindhard screening is dominated by the real part. In turn, the fact that $\textrm{Re}[\chi^{(1)}_L]$ is invariant in drift velocity implies that it is the same for any effective phonon frequency $\Omega$ as in the limit of zero effective speed of sound where $\Omega = 0$. Consequenly, for all $\Omega$ and $\gamma$ satisfying the high-drift-velocity/low-mobility condition, we can make the following approximation:
\begin{equation} \label{eq: chi(1)(q,0,0) Lindhard}
\chi^{(1)}_L(q,-\Omega,\gamma) \approx \frac{q_e^2 v_s^2 m}{\pi \hbar^2 \epsilon_0 t_\mathrm{2DEG} \omega_0^2} = \chi^{(1)}_L(q,0,0).
\end{equation}
This implies that the Mermin screening approximately equals the Lindhard screening, per Eq.~\eqref{eq: lindhard to mermin conversion}, yielding the following Mermin real and imaginary susceptibilities:
\begin{align}
\label{eq: imag chi(1) high mobility low drift}
\textrm{Im}[\chi^{(1)}(-\omega_0)] &\approx \frac{q_e^2 v_s^2 (v_d - v_s) m^2}{\pi \hbar^3 \epsilon_0 t_\mathrm{2DEG} \omega_0^2 k_F}, \\
\label{eq: real chi(1) high mobility low drift}
\textrm{Re}[\chi^{(1)}(-\omega_0)] &\approx \frac{q_e^2 v_s^2 m}{\pi \hbar^2 \epsilon_0 t_\mathrm{2DEG} \omega_0^2}.
\end{align}
In the following subsections, we will use the second equality in Eq.~\eqref{eq: chi(1)(q,0,0) Lindhard} to define $\chi^{(1)}_L(q,0,0)$ in the Lindhard-to-Mermin conversions.

Figure~\ref{fig:chi1highmulowvd} depicts the real and imaginary parts of the susceptibility as functions of drift velocity in the high-mobility/low-drift-velocity regime, given a phonon frequency of 1 GHz (corresponding to $\omega_0 = 2\pi \times 10^9 \textrm{ s}^{-1}$), speed of sound $v_s = 4 \times 10^3 \textrm{ m}/\textrm{s}$, a carrier density a carrier density $n = 2 \times 10^{15} \textrm{ m}^{-2}$, a carrier effective mass $m = 0.067 m_0$, and a 2DEG thickness $t_\mathrm{2DEG} = 2 \times 10^{-8} \textrm{ m}$. The Fermi velocity for the given carrier density and electron effective mass relates to the speed of sound as $v_F \approx 48v_s$. For $|v_d| \lesssim v_F$, the numerical results for the specific mobility values $\mu = 100$ or $1000 \textrm{ m}^2/(\textrm{V} \cdot \textrm{s})$ match the analytical predictions, featuring a constant (linear) relationship between the real (imaginary) susceptibility and the drift velocity.

\subsection{Low-Mobility Regime}
\label{sec: Low-Mobility Regime}

We now turn to the low-mobility limit (corresponding to $v_d, v_F \ll \gamma/q$), starting again with the imaginary Lindhard susceptibility. Here, the spectral broadening of the electronic states causes the energy conservation requirement to be relaxed. Therefore, the probability that any electron in the green-shaded crescent-shaped region in Fig.~\ref{fig:phasespacelowmobility} emits a phonon, or that any electron in the corresponding region in Fig.~\ref{fig:phasespacelowmobility} absorbs a phonon, is roughly uniform for all electrons in the respective regions. The $k_y$-span of each region at a given $k_x$ is given as follows in the long-wavelength limit $q \ll k_F$ \cite{chatterjee2024abinitio}:
\begin{equation}
k_{y,\mathrm{span}}(k_x) \approx \frac{2 q |k_x'|}{\sqrt{k_F^2 - k_x'^2}}.
\end{equation}
The approximate uniformity of the transition probability for all interactions also ensures that we need to account for both the resonant (emission) and counter-resonant (absorption) processes, which counteract each other. 

Appendix~\ref{sec: Imaginary Part of Chi(1) Low-Mobility Limit} shows the detailed derivation for the imaginary part of the Lindhard susceptibility in the low-mobility regime, yielding the following result:
\begin{align}
\begin{split} \label{eq: imaginary chi(1) Lindhard low mobility}
\textrm{Im}[\chi^{(1)}_L(-\omega_0)] &\approx \bigg(\frac{v_d - v_s}{v_s}\bigg) \frac{q_e^2 \omega_0 k_F^2}{\pi \epsilon_0 m t_\mathrm{2DEG} \gamma^3} \\
&= \bigg(\frac{v_d - v_s}{v_s}\bigg) \frac{2 m^2 \omega_0 n \mu^3}{\epsilon_0 q_e t_\mathrm{2DEG}},
\end{split}
\end{align}
where in the second line, we make the replacement $k_F^2 \rightarrow 2\pi n$ in order to express the imaginary part of $\chi^{(1)}$ as a function of carrier density, in addition to applying the substitution $\gamma = q_e/(m\mu)$. The cubic rather than linear dependence on mobility is attributable to the fact that the absorptive and emissive contributions destructively interfere with each other, with the rates of the 2 processes becoming increasingly balanced as the mobility is lowered. The result in terms of carrier density accounts for the high-temperature case $k_B T \gg \hbar v_F k_F$, where the carrier density exceeds the value predicted from the Fermi wavevector due to Fermi-Dirac broadening, demonstrating that $\textrm{Im}[\chi^{(1)}_L]$ for a given carrier density and mobility is invariant in temperature in both the medium-temperature and high-temperature regimes. This can be attributed to different reasons for the 2 regimes. In the medium-temperature regime, as the thermal energy is increased from $k_B T \sim \hbar v_F q$ to $k_B T \sim \hbar v_F k_F$, the fraction of electrons participating in electron-phonon interaction increases linearly as $k_B T/(\hbar v_F q)$ (corresponding to an effective widening of the crescent in Fig.~\ref{fig:phasespacelowmobility}), while the average interaction probability for each eligible electron is attenuated by a factor $\hbar v_F q/(k_B T)$ due to the corresponding attenuation in the Fermi-Dirac occupation number difference between the initial and final electronic states, with the changes cancelling each other out. In the high-temperature regime, increasing the spectral broadening beyond $k_B T \sim \hbar v_F k_F$ causes the average electron-phonon detuning $\expect{(\Delta \omega)_{k_x'}}$ to increase as approximately $\sqrt{k_B T/(\hbar v_F k_F)}$, causing the raw interaction probability per electron (i.e., the probability assuming occupied initial and unoccupied final electronic states) to increase by the same factor. However, the average occupation number difference between initial and final electronic states is reduced by the same factor, thus cancelling out the increase in the raw interaction probability. 

It is worth noting that this result is equivalent to that calculated for the low-mobility case in Ref.~\cite{chatterjee2024abinitio}, except for the factor $(v_d - v_s)/v_s$. As with the high-mobility case, if $v_d > v_s$ ($v_d < v_s$), the expression is positive (negative), corresponding to phonon-field amplification (attenuation). Also, as in the high-mobility limit, the case of $v_d = 0$ yields a result equal and opposite to that for $\textrm{Im}[\chi^{(1)}(\omega_0)]$ in Ref.~\cite{chatterjee2024abinitio}, demonstrating that the emission rate in this case is opposite to the previously calculated absorption rate, as desired.

Next, we analyze the real part of the Lindhard susceptibility in the low-mobility limit. Unlike the high-mobility case, we approximate that the spectral broadening is much \textit{larger} than the detuning for any transition (i.e., $\gamma \gg |(\Delta \omega)_{k_x'} \pm (v_d - v_s)q|$), causing each term in Eq.~\eqref{eq: real f net} to approximately become linear in the real detuning. As a result, the drift-velocity-dependent offsets for the positive-velocity and negative-velocity dipole formation processes cancel out, reducing Eq.~\eqref{eq: real f net} to a linear function in $(\Delta \omega)_{k_x'}$:
\begin{align}
\begin{split}
&\textrm{Re}[f_{\mathrm{net},k_x'}^{(1)}(-\omega_0)] \\
&\approx \frac{(\Delta \omega)_{k_x'} - (v_d - v_s)q}{\gamma^2} + \frac{(\Delta \omega)_{k_x'} + (v_d - v_s)q}{\gamma^2} \\
&= \frac{2 (\Delta \omega)_{k_x'}}{\gamma^2}.
\end{split}
\end{align}
This is identical to the corresponding term for the real part of the Lindhard susceptibility in the low-mobility regime from Ref.~\cite{chatterjee2024abinitio}. Therefore, integrating over $S'$ leads to the same result, independent of the drift velocity:
\begin{equation} \label{eq: real chi(1) Lindhard low mobility}
\textrm{Re}[\chi^{(1)}_L(-\omega_0)] \approx \frac{q_e^2 k_F^2}{2\pi \epsilon_0 m t_\mathrm{2DEG} \gamma^2} = \frac{m n \mu^2}{\epsilon_0 t_\mathrm{2DEG}},
\end{equation}
where we made the replacement $k_F^2 \rightarrow 2\pi n$ in the second step. For the same reason as with the imaginary part of $\chi^{(1)}_L$, the variation of the real part of $\chi^{(1)}_L$ with carrier density and mobility is identical in the medium-temperature ($v_F k_F \gtrsim k_B T/\hbar \gtrsim v_F q$) and the high-temperature ($k_B T/\hbar \gtrsim v_F k_F$) regimes, particularly because the raw electron-phonon interaction rates for both the real and imaginary cases feature the same variation (i.e., a linear variation) with the average electron-phonon detuning $\expect{(\Delta \omega)_{k_x'}}$ (see the discussion in the derivation for the low-mobility result for $\textrm{Im}[\chi^{(1)}]$ for an explanation of how this ensures a temperature-invariant interaction rate per electron per mobility-cubed in the high-temperature limit). 

\begin{figure*}[!tb]
	\centering
	\begin{subfigure}{\columnwidth}
		\centering
	    \includegraphics[width=\linewidth]{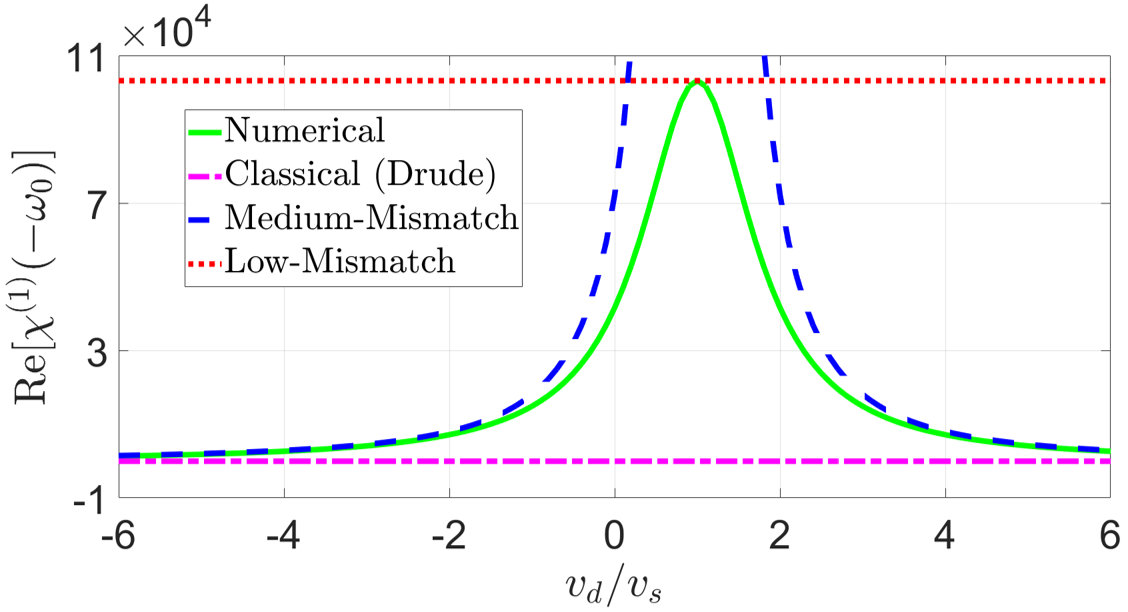}
	    \caption{}
	    \label{}
	\end{subfigure}
	\begin{subfigure}{\columnwidth}
		\centering
	    \includegraphics[width=\linewidth]{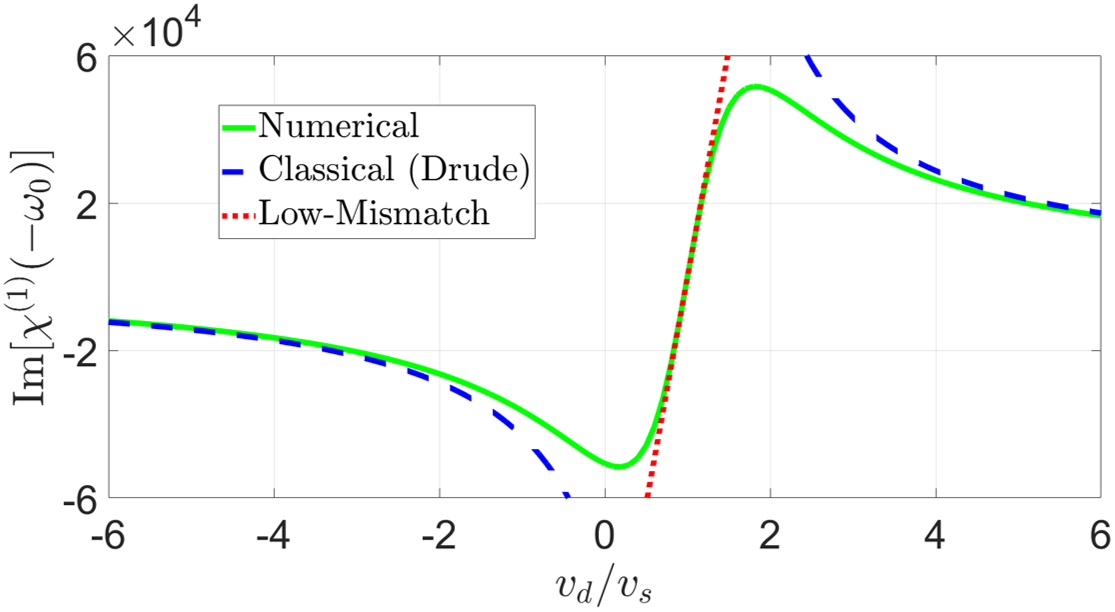}
		\caption{}
		\label{}
	\end{subfigure}
    \caption{Numerical results (solid, green) for the real (a) imaginary (b) parts of $\chi^{(1)}(-\omega_0)$ as functions of the drift velocity $v_d$, given a mobility of $0.3 \textrm{ m}^2/(\textrm{V} \cdot \textrm{s})$), along with the analytical results for the low-mobility limit $v_d,v_F \ll \gamma/q$ in the low-velocity-mismatch (dotted, red), medium-velocity-mismatch (dashed, blue) and high-velocity-mismatch (dash-dotted, magenta) regimes. Note that the high-mismatch results for both the real and imaginary parts, as well as the medium-mismatch result for the imaginary part, match the corresponding classical results. We assume a phonon angular frequency of $\omega_0 = 2\pi \times 10^9 \textrm{ s}^{-1}$, speed of sound $v_s = 4 \times 10^3$ m/s, a carrier density $n = 2 \times 10^{15} \textrm{ m}^{-2}$, a carrier effective mass $m = 0.067 m_0$, and a 2DEG thickness $t_\mathrm{2DEG} = 2 \times 10^{-8} \textrm{ m}$.}
	\label{fig:chi1lowmu}
\end{figure*}

As with the high-mobility/low-drift-velocity limit, the invariance of the real part of $\chi^{(1)}_L$ in drift velocity is explained by the changes in negative-velocity and positive-velocity dipole formation probabilities counteracting each other as the drift velocity is changed. This is especially clear in the low-mobility case since the changes in emissive detuning and absorptive detuning with respect to drift velocity cancel each other out. Here, the ratio between $\textrm{Im}[\chi^{(1)}_L]$ and $\textrm{Re}[\chi^{(1)}_L]$ (from Eq.~\eqref{eq: low mobility imaginary chi(1)} becomes the following:
\begin{equation} \label{eq: low-mobility imaginary vs real Lindhard chi(1) ratio}
\frac{\textrm{Im}[\chi^{(1)}_L(-\omega_0)]}{\textrm{Re}[\chi^{(1)}_L(-\omega_0)]} \approx \bigg(\frac{v_d - v_s}{v_s}\bigg) \frac{2 m \omega_0 \mu}{q_e} = \frac{2 (v_d - v_s) q}{\gamma}.
\end{equation}
Since $v_d, v_s \ll \gamma$ in the low-mobility regime, the real part is dominant in the overall Lindhard screening. However, this real part is much lower than the real Lindhard screening in the high-mobility/low-drift-velocity limit, as evidenced by taking the ratio between Eq.~\eqref{eq: real chi(1) Lindhard low mobility} and~\eqref{eq: real chi(1) Lindhard high mobility low drift}:
\begin{equation}
f = \frac{\hbar^2 k_\mathrm{max}^2 \omega_0^2 \mu^2}{2 q_e^2 v_s^2} = \frac{1}{2}\bigg(\frac{v_\mathrm{max} q}{\gamma}\bigg)^2,
\end{equation}
where we define $k_\mathrm{max} = (2\pi n)^{1/2}$ in order to account for the case of $\gamma \gtrsim v_F k_F$. Note that $f \ll 1$ in the low-mobility limit, since $v_\mathrm{max} \ll \gamma/q$. Since the real part of the Lindhard susceptibility is dominant in both the low-mobility and high-mobility cases, we can make the approximation $f \approx \chi^{(1)}_L(q,-\Omega,\gamma)/\chi^{(1)}_L(q,0,0)$, yielding the following Lindhard-to-Mermin conversion:
\begin{equation}
\chi^{(1)}(q,-\Omega,\gamma) \approx \frac{-\Omega + i\gamma}{-\Omega + if\gamma} \chi^{(1)}_L(q,-\Omega,\gamma).
\end{equation}
The real and imaginary parts of the Lindhard susceptibility thus map to the corresponding parts of the Mermin susceptibility in the following manner:
\begin{align}
\begin{split}
\textrm{Re}[\chi^{(1)}] &\approx \frac{(\Omega^2 + f\gamma^2) \textrm{Re}[\chi^{(1)}_L] + \Omega \gamma \textrm{Im}[\chi^{(1)}_L]}{\Omega^2 + f^2 \gamma^2} \\
&\approx \frac{-\Omega^2 + f\gamma^2}{\Omega^2 + f^2\gamma^2} \textrm{Re}[\chi^{(1)}_L],
\end{split}
\\
\begin{split}
\textrm{Im}[\chi^{(1)}] &\approx \frac{(\Omega^2 + f\gamma^2) \textrm{Im}[\chi^{(1)}_L] - \Omega \gamma \textrm{Re}[\chi^{(1)}_L]}{\Omega^2 + f^2 \gamma^2} \\
&\approx -\frac{\Omega \gamma}{\Omega^2 + f^2\gamma^2} \textrm{Re}[\chi^{(1)}_L],
\end{split}
\end{align}
where the first approximation in both expressions is based on the condition $f \ll 1$, while the second approximation for both lines derives from substituting $\textrm{Im}[\chi^{(1)}_L] \approx -(2\Omega/\gamma) \textrm{Re}[\chi^{(1)}_L]$ per Eq.~\eqref{eq: low-mobility imaginary vs real Lindhard chi(1) ratio}. Specifically, for the second expression, this substitution renders the first term of the numerator much smaller than the second term, since $\gamma \gg \Omega$ and $f \ll 1$ in the low-mobility limit. Note that both the imaginary part of the Mermin susceptibility is dominated by the real rather than the imaginary Lindhard susceptibility.

The susceptibility results can be divided into 3 regimes for the velocity mismatch (i.e., the difference between the drift velocity and the speed of sound): high-mismatch ($v_\mathrm{max}/\sqrt{2} \lesssim |v_d - v_s| \ll \gamma/q$), intermediate-mismatch ($v_\mathrm{max}^2 q/(2\gamma) \lesssim |v_d - v_s| \lesssim v_\mathrm{max}/\sqrt{2}$), and low-mismatch ($|v_d - v_s| \lesssim v_\mathrm{max}^2 q/(2\gamma)$). In the high-mismatch regime, the real Lindhard susceptibity simply undergoes a sign flip to produce the real Mermin susceptibility, while being amplified by a factor of $-\gamma/\Omega = \gamma/((v_d - v_s)q)$ to produce the imaginary Mermin susceptibility:
\begin{align}
\label{eq: low-mobility high-mismatch real chi(1)}
\textrm{Re}[\chi^{(1)}(-\omega_0)] &\approx -\textrm{Re}[\chi^{(1)}_L(-\omega_0)] \approx -\frac{m n \mu^2}{\epsilon_0 t_\mathrm{2DEG}}, \\
\begin{split}
\label{eq: low-mobility high-mismatch imag chi(1)}
\textrm{Im}[\chi^{(1)}(-\omega_0)] &\approx -\frac{\gamma}{\Omega} \textrm{Re}[\chi^{(1)}_L(-\omega_0)] \\
&\approx \frac{n q_e \mu}{\epsilon_0 t_\mathrm{2DEG} (v_d - v_s) q}.
\end{split}
\end{align}
As such, in the high-mismatch regime, the imaginary susceptibility is greater than the real susceptibility. Note that both the real and imaginary parts here match the classical results from the Drude model. On the other hand, in the intermediate-mismatch regime, while the imaginary Mermin susceptibility relates to the real Lindhard susceptibility in the same manner as in the high-mismatch regime (thus also matching the classical Drude result), the real Mermin susceptibility is now produced by amplifying the real Lindhard susceptibility by a factor of $f\gamma^2/\Omega^2 = v_\mathrm{max}^2/(2(v_d - v_s)^2)$:
\begin{align}
\begin{split}
\label{eq: low-mobility mid-mismatch real chi(1)}
\textrm{Re}[\chi^{(1)}(-\omega_0)] &\approx \frac{1}{2} \bigg(\frac{v_\mathrm{max}}{v_d - v_s}\bigg)^2 \textrm{Re}[\chi^{(1)}_L(-\omega_0)] \\
&\approx \frac{\pi \hbar^2 n^2 \mu^2}{\epsilon_0 t_\mathrm{2DEG} m (v_d - v_s)^2},
\end{split}
\\
\label{eq: low-mobility mid-mismatch imag chi(1)}
\textrm{Im}[\chi^{(1)}(-\omega_0)] &\approx \frac{n q_e \mu}{\epsilon_0 t_\mathrm{2DEG} (v_d - v_s) q}.
\end{align}
Here, the imaginary part is also greater than the real part, but the ratio changes to $\Omega/(f\gamma) = 2(v_d - v_s)/(v_\mathrm{max}^2 q)$ (which by definition is greater than 1 in the intermediate-mismatch regime). Finally, in the low-mismatch regime, the real part reaches a peak at the high-mobility result for the Lindhard real part, while the imaginary part becomes linear in the velocity mismatch:
\begin{align}
\label{eq: low-mobility low-mismatch real chi(1)}
\textrm{Re}[\chi^{(1)}(-\omega_0)] &\approx \frac{1}{f} \textrm{Re}[\chi^{(1)}_L(-\omega_0)] = \frac{q_e^2 v_s^2 m}{\pi \hbar^2 \epsilon_0 t_\mathrm{2DEG} \omega_0^2}, \\
\begin{split}
\label{eq: low-mobility low-mismatch imag chi(1)}
\textrm{Im}[\chi^{(1)}(-\omega_0)] &\approx -\frac{\Omega}{f^2 \gamma} \textrm{Re}[\chi^{(1)}_L(-\omega_0)] \\
&\approx \frac{q_e^3 m^2 (v_d - v_s)}{\pi^2 \epsilon_0 \hbar^4 n t_\mathrm{2DEG} q^3 \mu}.
\end{split}
\end{align}
In this limit, the real part becomes greater than the imaginary part by a factor $f\gamma/\Omega = v_\mathrm{max}^2 q/(2(v_d - v_s))$. In the case where $v_d = v_s$, the imaginary part goes to 0, while the real part converges onto the high-mobility result as discussed above.

Figure~\ref{fig:chi1lowmu} depicts the real and imaginary parts of the susceptibility in terms of drift velocity in the low-mobility regime. We use the same material parameters as in Fig.~\ref{fig:chi1highmulowvd}, except for changing the mobility to $0.3 \textrm{ m}^2/(\textrm{V} \cdot \textrm{s})$ for the numerical example. It is interesting to note that our quantum model broadens the Drude model's divergence for the imaginary part of the susceptibility around $v_d = v_s$. This provides a more physically complete picture, since a finite carrier density and a finite dipole length (limited by the half-wavelength of the applied ac electric field) ensure that even a maximally polarized electron gas cannot fully cancel out the applied field, thus rendering the divergence of the imaginary part at $v_d = v_s$ physically untenable.

\subsection{High-Drift-Velocity Regime}
\label{sec: High-Drift-Velocity Regime}

Finally, we discuss the case of a high drift velocity, where the drift momentum significantly exceeds both the Fermi momentum as well as the spectral broadening in phase space due to electronic decay (i.e., $v_d \gg v_F, \gamma/q$, which also yields $v_d \gg v_s$, since $v_F \gg v_s$). We first solve for the imaginary Lindhard susceptibility by re-calculating the integral from Eq.~\eqref{eq: low mobility imaginary chi(1)} using these assumptions:
\begin{widetext}
\begin{align} 
\begin{split} \label{eq: imag chi(1) Lindhard high vd}
&\textrm{Im}[\chi^{(1)}_L(-\omega_0)] \\
&= \beta \gamma \int_{S'} dk_x' k_{y,\mathrm{span}}(k_x') \bigg(\frac{1}{((\Delta \omega)_{k_x'} - (v_d - v_s)q)^2 + \gamma^2} - \frac{1}{((\Delta \omega)_{k_x'} + (v_d - v_s)q)^2 + \gamma^2}\bigg) \\
&\approx \beta \gamma \int_{S'} dk_x' k_{y,\mathrm{span}}(k_x') \frac{4 q v_d (\Delta \omega)_{k_x'}}{q^4 v_d^4} \\
&\approx \frac{4 \beta \gamma}{q^3 v_d^3} \int_0^{k_F} dk_x' \frac{2 q k_x'}{\sqrt{k_F^2 - k_x'^2}} \frac{\hbar q k_x'}{m} \\
&= \bigg(\frac{4 \beta \gamma v_s^3}{\omega_0^3 v_d^3}\bigg) \bigg(\frac{\pi \hbar \omega_0^2 k_F^2}{2 m v_s^2}\bigg) \\
&= \frac{2 n q_e^3}{\epsilon_0 m^2 t_\mathrm{2DEG} \omega_0^3} \frac{v_s^3}{\mu v_d^3}.
\end{split}
\end{align}
\end{widetext}
In the second line, we apply the aforementioned approximations $v_d \gg v_F \gg v_s$ and $v_d \gg \gamma/q$. This corresponds to the fact that the detuning is approximately the same for all transitions, equaling $v_d q$ for absorptive processes and $-v_d q$ for emissive processes. In the third line, we apply the long-wavelength approximation $k_F \gg q$, as in the low-mobility/low-drift-velocity case. In the last line, we apply the replacements $k_F^2 \rightarrow 2\pi n$ and $\gamma \rightarrow q_e/(m\mu)$.

As this result shows, the imaginary Lindhard susceptibility drops as $1/v_d^3$ in the high-drift-velocity regime. This is because in this regime, the carrier distribution has already been pushed far enough in momentum space such that no electron-phonon interaction in the near-resonance range (i.e., an interaction where the detuning is less than the electronic spectral broadening) is available. Therefore, increasing the drift velocity pushes the carrier distribution even farther away from resonance. It is also worth noting the inverse variation with the mobility $\mu$. This is because a lower mobility relaxes the resonance requirement, expanding the near-resonance window by increasing the spectral broadening. Since all electron-phonon interactions in the high-drift-velocity regime are off-resonance, this has the effect of increasing the interaction strength.

Turning to the real part of the Lindhard susceptibility, given the approximations $v_d \gg v_F \gg v_s$ and $v_d \gg \gamma/q$, the result of the sum over negative-velocity and positive-velocity dipole formation processes as encapsulated by Eq.~\eqref{eq: real f net} changes to the following:
\begin{align}
\begin{split} 
&\textrm{Re}[f_{\mathrm{net},k_x'}^{(1)}(-\omega_0)] \\
&\approx \frac{(\Delta \omega)_{k_x'} - (v_d - v_s)q}{(qv_d)(-qv_d)} + \frac{(\Delta \omega)_{k_x'} + (v_d - v_s)q}{(qv_d)(-qv_d)} \\
&= -\frac{2 (\Delta \omega)_{k_x'}}{q^2 v_d^2}.
\end{split}
\end{align}
As in the low-mobility limit, this result is also linear in $(\Delta \omega)_{k_x'}$. Integrating over $S'$ thus yields the following result:
\begin{equation} \label{eq: real chi(1) Lindhard high vd}
\textrm{Re}[\chi^{(1)}_L(-\omega_0)] \approx -\frac{n q_e^2}{\epsilon_0 m t_\mathrm{2DEG} \omega_0^2} \frac{v_s^2}{v_d^2}.
\end{equation}
Similarly to the low-drift-velocity/high-mobility case, the result is invariant in mobility. This is because in both cases, the average dipole-phonon detuning is vastly greater than the electronic spectral broadening. However, unlike the low-drift-velocity cases (where the real part of $\chi^{(1)}_L$ is invariant in the drift velocity), $\textrm{Re}[\chi^{(1)}_L]$ in the high-drift-velocity regime declines with increasing drift velocity as $1/v_d^2$. This is due to the fact that whereas changes in the absorptive detuning and emissive detuning counteract each other as the drift velocity is increased within the low-drift-velocity regime, both of the detuning values increase in magnitude if the drift velocity is increased within the high-drift-velocity regime. As a result, shifting the drift velocity higher suppresses both the negative-velocity and the positive-velocity dipole formation processes. 

It is also worth considering the medium-temperature and high-temperature cases ($v_Fk_F \gtrsim k_B T/\hbar \gtrsim v_F q$ and $k_B T/\hbar \gtrsim v_Fk_F$, respectively). Although these thermally-broadened regimes are naturally associated with the low-mobility limit (since the thermal broadening sets the baseline for the electronic spectral broadening), the fact that $v_d \gg v_F$ in the high-drift-velocity limit implies that there may exist a temperature window in the medium-temperature range associated with a high-mobility regime (where the spectral broadening is less than the electron-phonon detuning). It is important to note that as the temperature is varied within this window, the change in the number of transition-allowed electrons is roughly cancelled out by the change in the average Fermi-Dirac occupation number difference between initial and final electronic states, as explained in the previous subsection. On the other hand, the high-temperature range can generally be assumed to fall within the low-mobility regime, since $q \ll k_F$, implying that the drift velocities featuring significant gain are close enough to the Fermi velocity such that $v_d q \ll v_F k_F$.

\begin{figure*}[!tb]
	\centering
	\begin{subfigure}{\columnwidth}
		\centering
	    \includegraphics[width=\linewidth]{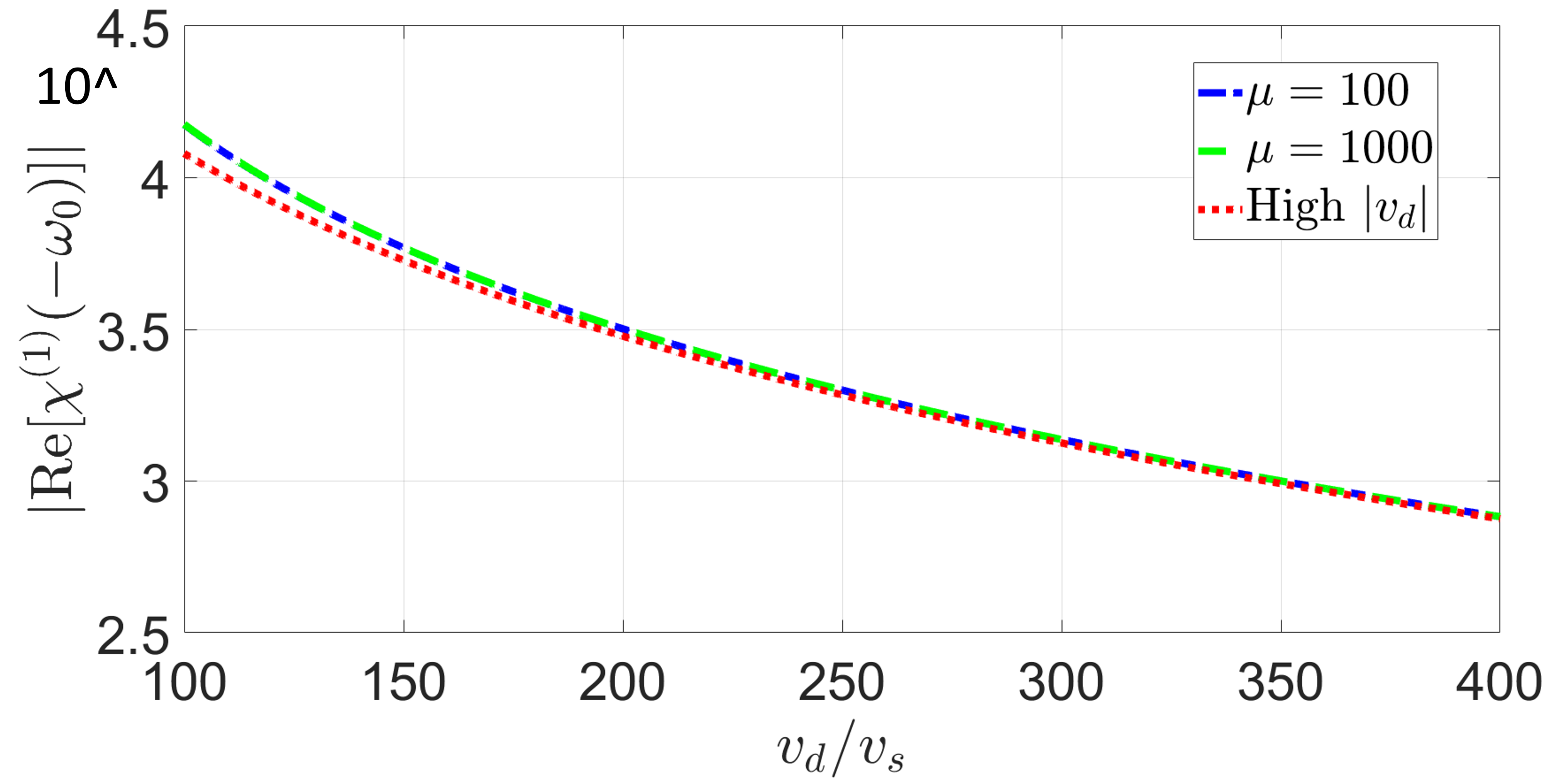}
	    \caption{}
	    \label{}
	\end{subfigure}
	\begin{subfigure}{\columnwidth}
		\centering
	    \includegraphics[width=\linewidth]{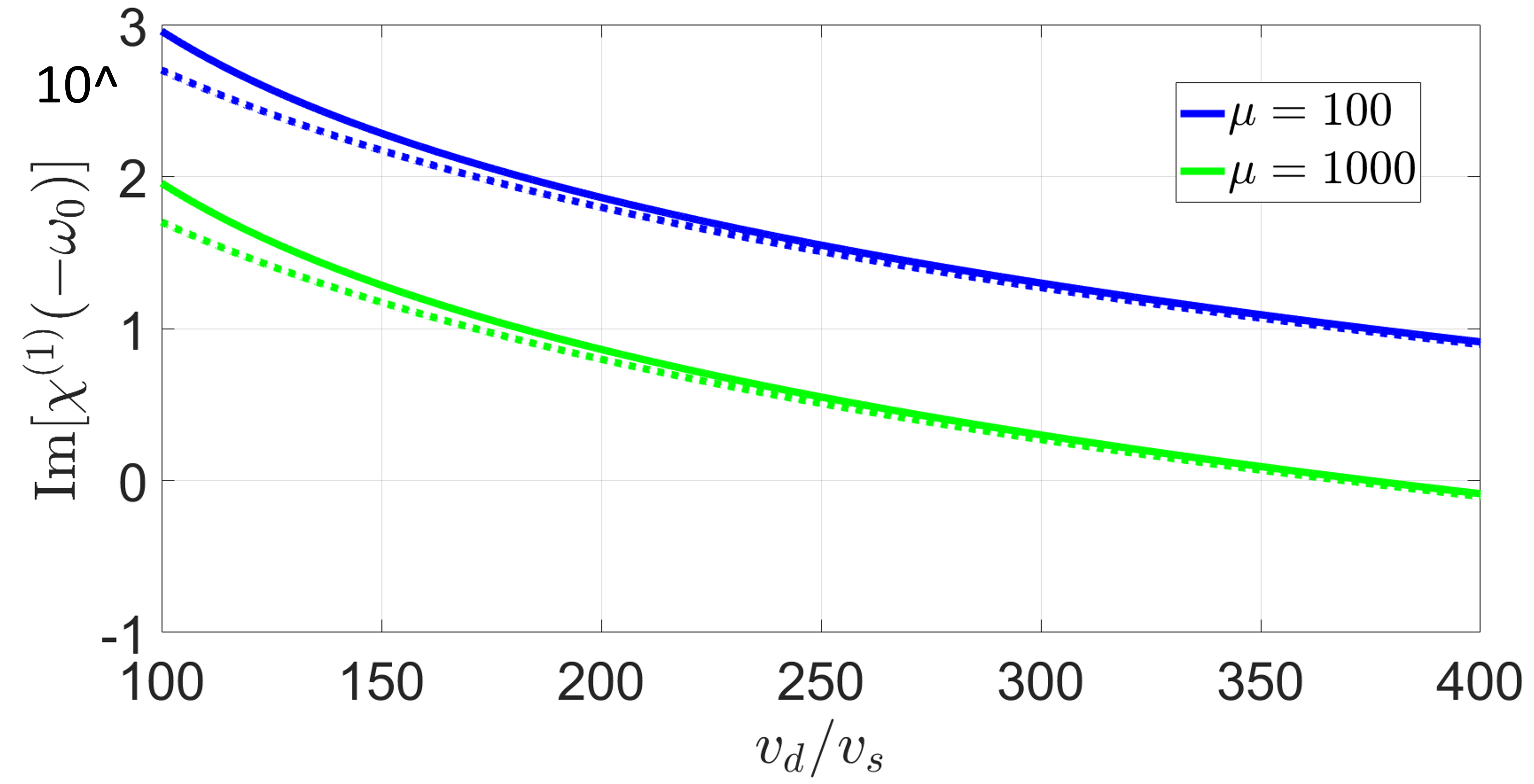}
		\caption{}
		\label{}
	\end{subfigure}
    \caption{(a) Numerical results for the real part of $\chi^{(1)}(-\omega_0)$ given mobility values of 100 (dash-dotted, blue) and 1000 (dashed, green) $\textrm{m}^2/(\textrm{V} \cdot \textrm{s})$, along with analytical results (dotted, red) for the high-drift-velocity limit $\gamma/q,v_F \ll v_d$; (b) Analogous results for the imaginary part, with solid and dotted lines representing numerical and analytical results, respectively, for mobilities of 100 (blue, upper) and 1000 (green, lower) $\textrm{m}^2/(\textrm{V} \cdot \textrm{s})$. We assume a phonon angular frequency of $\omega_0 = 2\pi \times 10^9 \textrm{ s}^{-1}$, speed of sound $v_s = 4 \times 10^3$ m/s, a carrier density $n = 2 \times 10^{15} \textrm{ m}^{-2}$, a carrier effective mass $m = 0.067 m_0$, and a 2DEG thickness $t_\mathrm{2DEG} = 2 \times 10^{-8} \textrm{ m}$.}
	\label{fig:chi1highvd}
\end{figure*}

The ratio between the imaginary and real parts of $\chi^{(1)}_L$ is roughly inverted from the low-mobility case:
\begin{equation} \label{eq: imaginary to real chi(1) Lindhard ratio high vd}
\frac{\textrm{Im}[\chi^{(1)}_L(-\omega_0)]}{\textrm{Re}[\chi^{(1)}_L(-\omega_0)]} \approx -\frac{2 q_e v_s}{m \omega_0 \mu v_d} = -\frac{2 \gamma}{v_d q},
\end{equation}
where we used the result for the imaginary part from Eq.~\eqref{eq: imag chi(1) Lindhard high vd}. Since $v_d \gg \gamma/q$ in the high-drift-velocity regime, the imaginary part is negligible compared to the real part. Consequently, the Lindhard susceptibility can be approximated as the real part shown in Eq.~\eqref{eq: real chi(1) Lindhard high vd}, leading to the following ratio between the Lindhard susceptibility in this limit and that given $\gamma = \Omega = 0$ (see Eq.~\eqref{eq: real chi(1) Lindhard high mobility low drift}):
\begin{equation}
f' = \frac{\chi^{(1)}_L(q,-\Omega,\gamma)}{\chi^{(1)}_L(q,0,0)} \approx -\frac{1}{2} \bigg(\frac{v_F}{v_d}\bigg)^2.
\end{equation}
In the high-drift-velocity limit, $|f'| \ll 1$ since $v_F \ll |v_d|$. Combined with the fact that $\gamma \ll |\Omega|$ in this limit, the second term in the Lindhard-to-Mermin conversion factor from Eq.~\eqref{eq: lindhard to mermin conversion} can be neglected, causing the expression to approximately simplify to the following:
\begin{equation}
\chi^{(1)}(q,-\Omega,\gamma) \approx \bigg(1 - i\frac{\gamma}{\Omega}\bigg) \chi^{(1)}_L(q,-\Omega,\gamma).
\end{equation}
Given the fact that the imaginary Lindhard susceptibility is much smaller than the real Lindhard susceptibility (see Eq.~\eqref{eq: imaginary to real chi(1) Lindhard ratio high vd}), along with the fact that $|\gamma/\Omega| \ll 1$, the real Mermin susceptibility approximately reduces to the real Lindhard susceptibility:
\begin{equation} \label{eq: real chi(1) Mermin high vd}
\textrm{Re}[\chi^{(1)}(-\omega_0)] \approx \textrm{Re}[\chi^{(1)}_L(-\omega_0)] \approx -\frac{n q_e^2}{\epsilon_0 m t_\mathrm{2DEG} \omega_0^2} \frac{v_s^2}{v_d^2}.
\end{equation}
On the other hand, the imaginary Mermin susceptibility receives inputs of similar magnitudes from both the real and imaginary parts of the Lindhard susceptibility, as evidenced by substituting $\textrm{Im}[\chi^{(1)}_L] \approx (2\gamma/\Omega) \textrm{Re}[\chi^{(1)}_L]$ into the Lindhard-to-Mermin conversion:
\begin{align}
\begin{split} \label{eq: imag chi(1) Mermin high vd}
\textrm{Im}[\chi^{(1)}(-\omega_0)] &\approx \textrm{Im}[\chi^{(1)}_L(-\omega_0)] - \frac{\gamma}{\Omega} \textrm{Re}[\chi^{(1)}_L(-\omega_0)] \\
&\approx \frac{\gamma}{\Omega} \textrm{Re}[\chi^{(1)}_L(-\omega_0)] \\
&\approx \frac{n q_e^3}{\epsilon_0 m^2 t_\mathrm{2DEG} \omega_0^3} \frac{v_s^3}{\mu v_d^3}.
\end{split}
\end{align}
Note that this is almost identical to the imaginary part of the Lindhard susceptibility, except for a factor-of-2 reduction. 

The real and imaginary parts of the susceptibility in the high-drift-velocity limit are shown in Fig.~\ref{fig:chi1highvd}. Note that these are in log-log scale, with the first plot actually depicting the \textit{opposite} of the real susceptibility (since the real susceptibility in this regime is negative). As expected, both the real and imaginary parts decline with increasing drift velocity in this regime. However, while the real part for any given drift velocity is invariant in mobility, the imaginary part is about 10 times lower for a mobility of $1000$ compared to $100 \textrm{ m}^2/(\textrm{V} \cdot \textrm{s})$, reflecting the aforementioned inverse variation between the imaginary susceptibility and the mobility.

\subsection{Analysis}
\label{sec: Analysis}

Having solved for the susceptibility in various limits, we now specifically analyze the behavior in the high-mobility limit across drift velocities. We start with the imaginary part. As shown in Fig.~\ref{fig:chi1highvd}(b), the numerical result spikes well above the analytically predicted values in the high-mobility regime as $|v_d|$ approaches $v_F$. On the other hand, once $|v_d|$ surpasses $v_F$, the susceptibility rapidly drops off below the analytical results extrapolated from the low-drift-velocity limit. These shifts are particularly pronounced at higher mobility values. 

The linear rise of the imaginary part of $\chi^{(1)}$ with the drift velocity in the low-drift-velocity limit $|v_d| \ll v_F$ is explained by the approximately parabolic curvature of the Fermi circle around the resonant states in that limit, as discussed previously in this section. However, if the Fermi circle is shifted enough such that $|k_{x,i}'|$ and $|k_{x,f}'|$ are no longer far smaller than $k_F$ (corresponding to $v_d$ no longer being well below $v_F$), then the $k_y$-vs-$k_x'$ curvature of the Fermi circle becomes significantly sharper than a parabolic curvature. Moreover, the curvature keeps increasing as $|k_{x,i}'|$ and $|k_{x,f}'|$ approach $k_F$. Consequently, $k_{y,\mathrm{span}}$ and thus the high-mobility susceptibility increase nonlinearly as $v_d$ approaches $v_F$, as seen in Fig.~\ref{fig:chi1highvd}(b). The excess number of initial-wavevector electrons (relative to final-wavevector electrons) is maximized once $|k_{x,f}'|$ reaches $k_F$, since the number of occupied states at the final wavevector reaches zero. As the Fermi circle is shifted beyond this (i.e., as $|k_{x,i}'|$ approaches $k_F$), the number of occupied states at the initial wavevector also rapidly drops toward zero, explaining why the high-mobility susceptibility sharply drops after crossing the peak. 

Next, we analyze the real part. As discussed earlier, the real susceptibility is invariant in drift velocity in the low-drift-velocity limit, since shifting the drift velocity in this limit approximately conserves the overall dipole formation probability despite changing the balance between positive-velocity and negative-velocity dipoles. However, as shown in Fig.~\ref{fig:chi1highvd}(a), once the drift velocity crosses the Fermi velocity, the real part of $\chi^{(1)}$ sharply shifts to a negative value, before dropping rapidly toward zero. The sign flip and sharp peak around $|v_d| = v_F$ is explained by the fact that the average emissive detuning crosses over resonance and shifts from positive to negative. Subsequently, the monotonous drop in the magnitude is because both the emissive and absorptive detuning amplitudes rise as $|v_d|$ is increased beyond $v_F$, suppressing both negative- and positive-velocity dipole formation processes, thus steadily reducing the electric field screening.

\section{Calculating the Amplifier Gain}
\label{sec: Calculating the Amplifier Gain}

We now merge the imaginary and real parts of $\chi^{(1)}$ to calculate the gain per unit length. The gain is composed of a product of two quantities: the phonon stimulated emission rate per unit electric field intensity, and the field intensity per phonon. The former is governed by the imaginary part of $\chi^{(1)}$ in a straightforwardly proportional manner. The latter, however, is inversely proportional to the amplitude-squared of the 2DEG's dielectric screening. This screening, in turn, relates to $\chi^{(1)}$ as follows:
\begin{align}
\begin{split}
|\epsilon(-\omega_0)| &= 
\epsilon_0 a(|\omega_0|) \Bigg(\Big(\epsilon_p + \epsilon_0 a(|\omega_0|)\textrm{Re}[\chi^{(1)}(-\omega_0)]\Big)^2 \\
&\quad + \Big(\epsilon_0 a(|\omega_0|)\textrm{Im}[\chi^{(1)}(-\omega_0)]\Big)^2\Bigg)^{1/2},
\end{split}
\end{align}
where $\epsilon_p$ is the piezoelectric material's dielectric coefficient. Note that $a(|\omega|) \sim t_\mathrm{2DEG} q$ is the attenuation factor for the screening due to the thin-film nature of the 2DEG. The gain per unit length is solved by merging the results for the dielectric screening and the imaginary part of $\chi^{(1)}$. To this end, we double the amplitude attenuation per unit length $\alpha$ from Ref.~\cite{chatterjee2024abinitio} and convert to dB-based units by multiplying by 10/ln(10), yielding:
\begin{equation} \label{eq: acoustoelectric gain in dB}
G = \frac{10}{\ln{10}} \frac{\epsilon_0 C^2 t_\mathrm{2DEG} \omega_0}{v_s L_z} \frac{\textrm{Im}[\chi^{(1)}(-\omega_0)]}{|\epsilon(-\omega_0)|^2},
\end{equation}
where we have assumed that the 2DEG covers the surface of the piezoelectric material. It is worth noting that in the high-mobility/low-drift-velocity limit, this result matches the past findings by Mosekilde \cite{mosekilde1974quantum}, as calculated in Appendix~\ref{sec: Gain Comparison in Low-Drift-Velocity/High-Mobility Limit}. 

\begin{figure*}[!tb]
	\centering
	\begin{subfigure}{\columnwidth}
		\centering
	    \includegraphics[width=\linewidth]{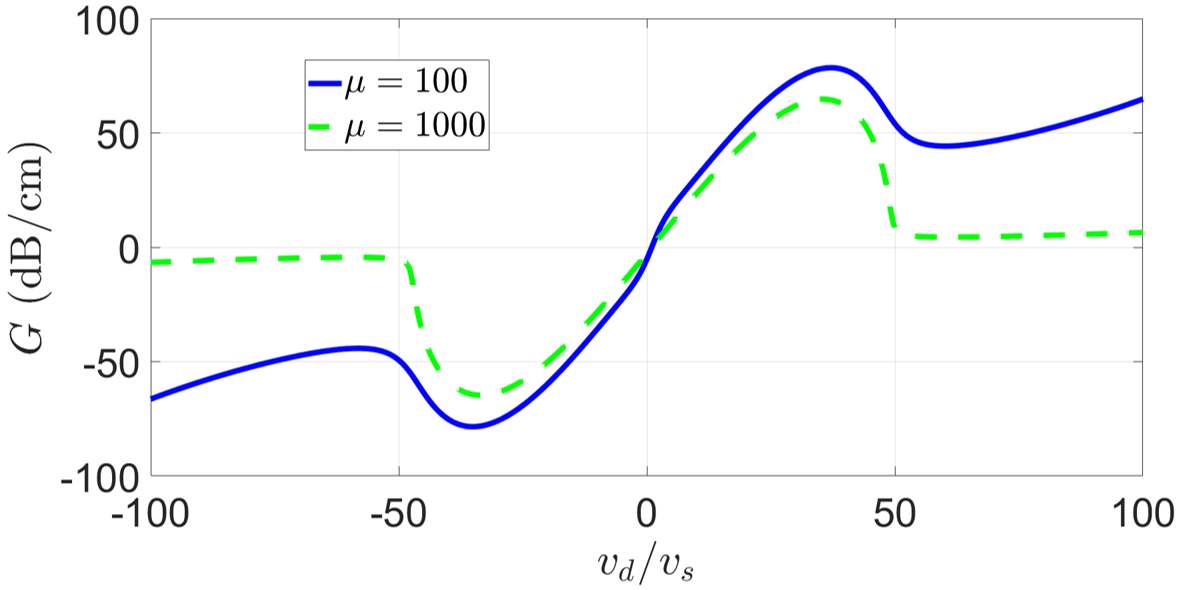}
	    \caption{}
	    \label{}
	\end{subfigure}
	\begin{subfigure}{\columnwidth}
		\centering
	    \includegraphics[width=\linewidth]{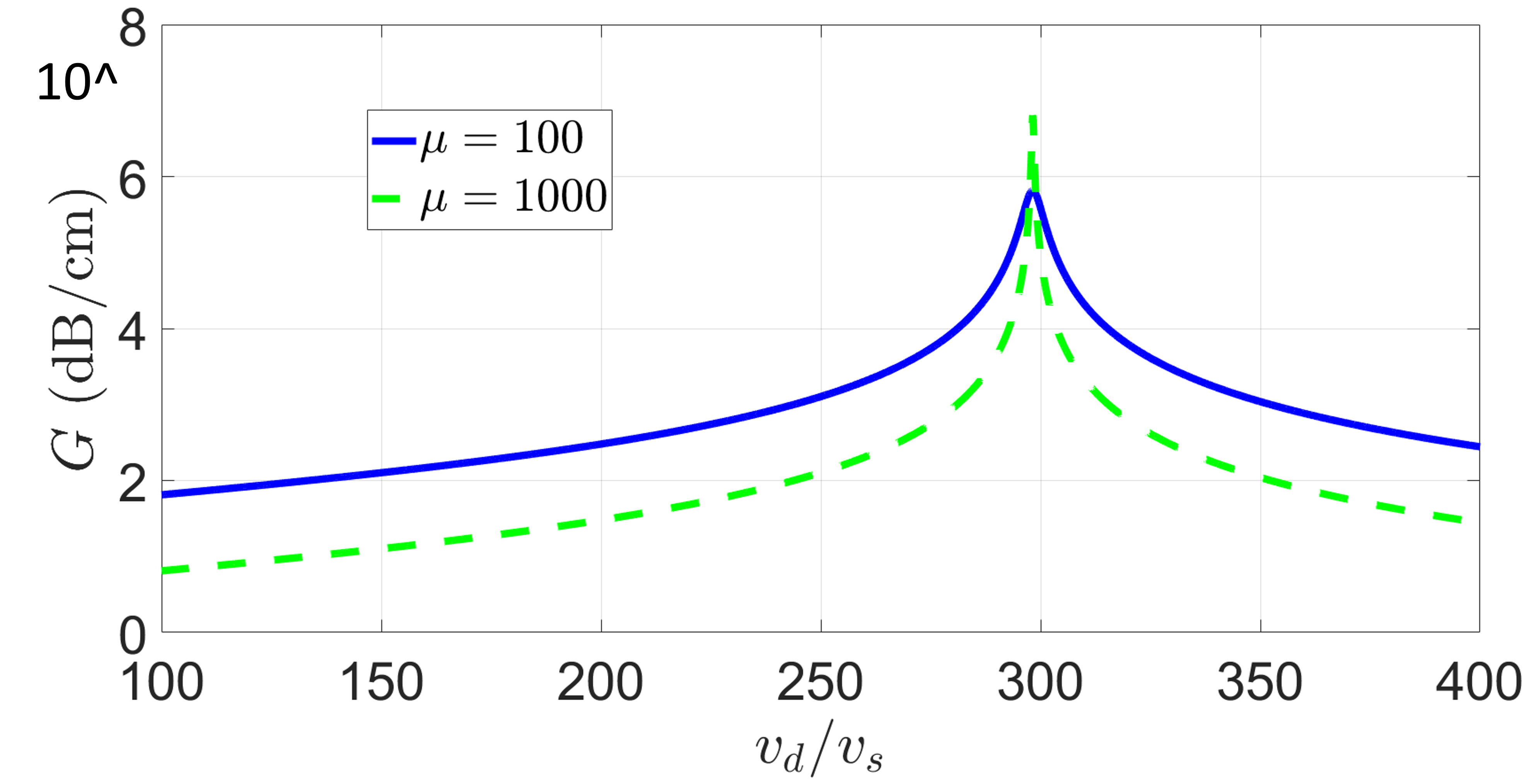}
		\caption{}
		\label{}
	\end{subfigure}
    \caption{Numerical results for the amplifier waveguide gain per unit length $G$ as a function of the drift velocity $v_d$ for lower (a) and higher (b) drift velocities, given mobilities (in units of $\textrm{m}^2/(\textrm{V} \cdot \textrm{s})$) of 100 (solid, blue), or 1000 (dashed, green). We assume a phonon angular frequency of $\omega_0 = 2\pi \times 10^9 \textrm{ s}^{-1}$, speed of sound $v_s = 4 \times 10^3$ m/s, a carrier density $n = 2 \times 10^{15} \textrm{ m}^{-2}$, a carrier effective mass $m = 0.067 m_0$, a 2DEG thickness $t_\mathrm{2DEG} = 2 \times 10^{-8} \textrm{ m}$, and a mode depth $L_z = v_s/\omega_0$.}
    \label{fig:gainwaveguide}
\end{figure*}

Figure~\ref{fig:gainwaveguide} shows the gain per unit length for the amplifier waveguide for given mobility values in the high-mobility regime, assuming an acoustic field mode depth $L_z \sim v_s/\omega_0$. Counter-intuitively, the peak gain is maximized in the high-drift-velocity regime (at about 5 times the Fermi velocity given our parameters). This is explained by the fact that for $|v_d| > v_F$, an increasing drift velocity yields a tradeoff: The drop in the imaginary part of $\chi^{(1)}$ reduces the phonon emission rate per unit electric field intensity, while the drop in the dielectric screening increases the field intensity per phonon. In the high-drift-velocity regime, the latter rises faster than the former drops (at rates of $v_d^4$ and $v_d^{-3}$, respectively). As a result, the net effect is for the gain to rise with drift velocity in the high-drift-velocity regime. However, the dielectric screening amplitude eventually reaches its floor once the 2DEG screening roughly cancels out the piezoelectric material's screening, at which point the electric field intensity per phonon and thus the gain sharply rise. Thereafter, the screening starts to rise again, asymptotically approaching $\epsilon_p$, thus causing the gain to decrease from the sharp peak. 

It is worth specifically analyzing the dependence of the peak gain on mobility (within the high-mobility range) and carrier density. In the high-drift-velocity regime, the gain is counterintuitively lower for higher mobilities (except in the vicinity of the peak). This is because the imaginary part of the susceptibility, and thus the phonon emission rate per unit electric field intensity, scales inversely with mobility, while the electric field intensity per phonon is roughly invariant in mobility since the dielectric screening is dominated by the mobility-independent real susceptibility. 
\begin{figure*}[!tb]
    \begin{subfigure}{\columnwidth}
		\centering
	    \includegraphics[width=\linewidth]{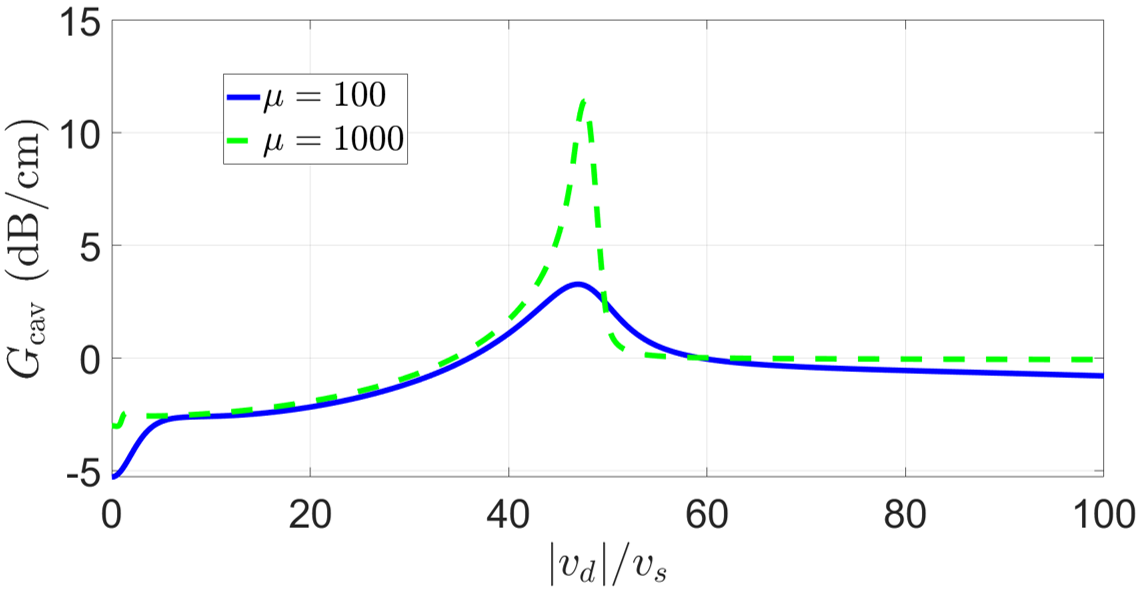}
	    \caption{}
	    \label{}
	\end{subfigure}
    \begin{subfigure}{\columnwidth}
		\centering
	    \includegraphics[width=\linewidth]{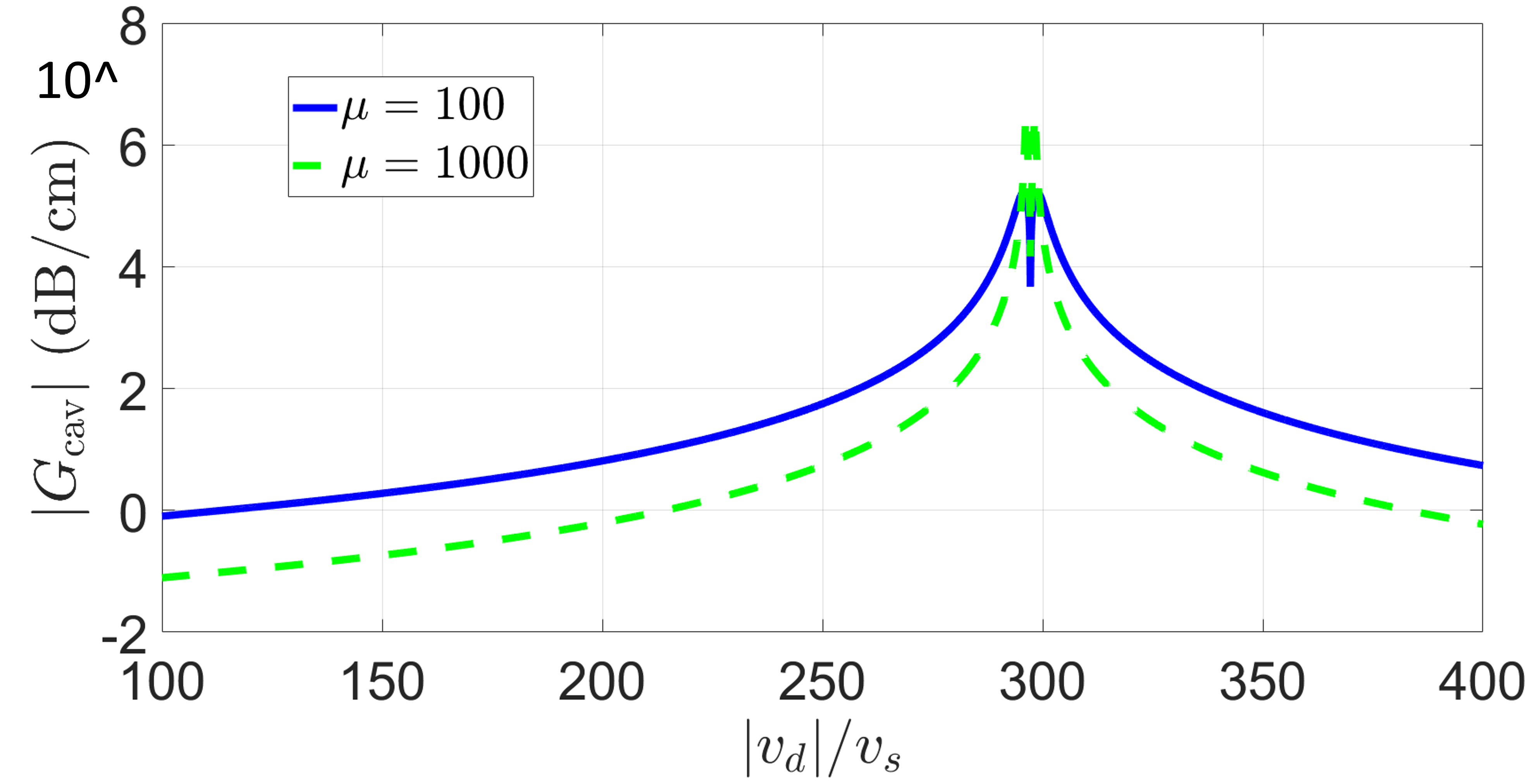}
		\caption{}
		\label{}
	\end{subfigure}
    \caption{Numerical results for the amplifier cavity gain per unit length $G_\mathrm{cav}$ as a function of the drift speed $|v_d|$ for lower (a) and higher (b) drift speeds, given mobilities (in units of $\textrm{m}^2/(\textrm{V} \cdot \textrm{s})$) of 100 (solid, blue), or 1000 (dashed, green). Note that in the high-drift-velocity range, the gain is negative (positive) for $|v_d| < 297 v_s$ ($|v_d| > 297 v_s$). We assume a phonon angular frequency of $\omega_0 = 2\pi \times 10^9 \textrm{ s}^{-1}$, speed of sound $v_s = 4 \times 10^3$ m/s, a carrier density $n = 2 \times 10^{15} \textrm{ m}^{-2}$, a carrier effective mass $m = 0.067 m_0$, a 2DEG thickness $t_\mathrm{2DEG} = 2 \times 10^{-8} \textrm{ m}$, and a mode depth $L_z = v_s/\omega_0$.}
	\label{fig:gaincavity}
\end{figure*}
However, the gain peaks at higher values for higher mobilities, due to the fact that the peak corresponds to the zeroing out of the real part of the susceptibility, making the imaginary part dominant in the screening and causing the electric field intensity to scale as the inverse-squared of the imaginary susceptibility. The peak gain thus varies inversely with the imaginary susceptibility, hence increasing with mobility. We can confirm this quantitatively by deriving the optimal drift velocity $v_{d,\mathrm{opt}}$ (i.e., the drift velocity where the real part of the 2DEG screening cancels out the piezoelectric material's screening):
\begin{align}
\begin{split}
\epsilon_p &= -\epsilon_0 a \textrm{Re}[\chi^{(1)}(v_d)], \\
v_{d,\mathrm{opt}} &\approx \bigg(\frac{a q_e^2 v_s^2}{m t_\mathrm{2DEG} \omega_0^2 \epsilon_p}\bigg)^{1/2} n^{1/2} \approx \bigg(\frac{q_e^2}{m q \epsilon_p}\bigg)^{1/2} n^{1/2},
\end{split}
\end{align}
where we substituted the high-drift-velocity result for the real part of $\chi^{(1)}$ from Eq.~\eqref{eq: real chi(1) Mermin high vd}, along with the relationship $a \sim t_\mathrm{2DEG} q$. Note that the optimal drift velocity is invariant in the mobility, matching the plot in Fig.~\ref{fig:gainwaveguide}(b). On the other hand, the optimal drift velocity increases with the carrier density as $n^{1/2}$, thus making it proportional to the Fermi velocity. The corresponding maximum gain is calculated by substituting this into the imaginary part of $\chi^{(1)}$ in the high-drift-velocity regime from Eq.~\eqref{eq: imag chi(1) Mermin high vd} and then into the gain equation from Eq.~\eqref{eq: acoustoelectric gain in dB}, yielding:
\begin{align}
\begin{split}
G_\mathrm{max} &= \frac{10}{\ln{10}} \frac{\epsilon_0 C^2 t_\mathrm{2DEG} \omega_0}{v_s L_z} \frac{1}{a^2 \epsilon_0^2 \textrm{Im}[\chi^{(1)}(-\omega_0)]} \\
&\approx \frac{10}{\ln{10}} \bigg(\frac{C^2 t_\mathrm{2DEG}^{1/2} m^{1/2} \omega_0}{a^{1/2} v_s L_z \epsilon_p^{3/2}}\bigg) \mu n^{1/2} \\
&\approx \frac{10}{\ln{10}} \bigg(\frac{C^2 m^{1/2} q^{3/2}}{\epsilon_p^{3/2}}\bigg) \mu n^{1/2},
\end{split}
\end{align}
where in the first line, we reduced the screening to only the imaginary part since the real part cancels out, and in the last line, we substituted $a \sim t_\mathrm{2DEG} q$ and $L_z \sim 1/q$. As we intuitively deduced above, the maximum gain increases with the mobility. It is also worth noting that it increases with carrier density. This is because the positive correlation between optimal drift velocity and carrier density, combined with the strong negative correlation between $\textrm{Im}[\chi^{(1)}]$ and the drift velocity, cause $\textrm{Im}[\chi^{(1)}]$ to decline with increasing carrier density. The peak gain's inverse variation with this imaginary screening causes it to rise with carrier density.

We now turn to the cavity gain per unit length in the high-mobility regime, depicted in Fig.~\ref{fig:gaincavity} as a function of the drift speed $|v_d|$. This is determined by averaging the forward-propagating and backward-propagating waveguide gains:
\begin{equation}
G_\mathrm{cav}(|v_d|) = \frac{1}{2} \Big(G(v_d) + G(-v_d)\Big).
\end{equation}
The fact that the waveguide gain is centered at $v_d = v_s$ rather than $v_d = 0$ breaks the velocity symmetry, yielding non-zero net cavity gain. In the low-drift-velocity regime, the linear variation between the waveguide gain and the drift velocity ensures that the cavity gain approximately features a stable negative value, roughly equaling the waveguide gain at zero drift velocity. As the drift velocity reaches the Fermi velocity, the waveguide gain's saturation causes the cavity gain to become positive and reach a peak, thereafter dropping back down into the negative range. This is because the negative waveguide gain in the negative drift velocity regime flattens and reverses at lower drift speeds than the corresponding positive gain in the positive drift velocity regime, causing the net gain to increase before the waveguide gain magnitude starts to increase again, thus making the cavity net gain negative again. A similar effect occurs around the waveguide gain peak ($v_d = 297v_s$ given our parameters), where the fact that this peak occurs at a lower drift speed for the negative drift velocity case than for the positive case causes the cavity gain to exhibit a biphasic peak and shift from negative to positive net gain.

It is also useful to consider the amplifier waveguide gain in the low-mobility regime and compare it to classical results. To that end, the fact that the real part of $\chi^{(1)}$ declines with decreasing mobility in the low-mobility regime (since the real susceptibility varies as $\mu^2$ in this regime, as derived in the previous section) implies that $\epsilon_0 a |\textrm{Re}[\chi^{(1)}| \ll \epsilon_p$ (i.e., the piezoelectric material's screening dominates the real part of the dielectric response). We express the imaginary part of $\chi^{(1)}$ for most of the drift velocity range except for the immediate vicinity of the speed of sound (i.e., the low-velocity-mismatch regime discussed in Sec.~\ref{sec: Low-Mobility Regime}) by re-writing Eq.~\eqref{eq: low-mobility high-mismatch imag chi(1)} (or equivalently, Eq.~\eqref{eq: low-mobility mid-mismatch imag chi(1)}) in terms of the 3D carrier density $n_\mathrm{3D} = n/t_\mathrm{2DEG}$:
\begin{equation} \label{eq: imaginary chi(1) high mid mismatch}
\textrm{Im}[\chi^{(1)}(-\omega_0)] \approx \frac{n_\mathrm{3D} q_e \mu}{\epsilon_0 (v_d - v_s) q} = -\frac{\sigma}{\epsilon_0 \omega_0 \Gamma},
\end{equation}
where $\sigma$ is the 2DEG conductivity and $\Gamma = 1 - v_d/v_s$ is the Doppler shift coefficient experienced by the drifting electrons. Substituting this, along with the relationship $a \sim t_\mathrm{2DEG}/L_z$ for the thin-film attenuation factor, into Eq.~\eqref{eq: acoustoelectric gain in dB}, we find that the gain per unit length reduces to the following in the medium- and high-mismatch cases:
\begin{align}
\begin{split} \label{eq: gain mid high mismatch}
G &\approx \frac{10}{\ln{10}} \epsilon_0 (K_0^2 \epsilon_p) a q \frac{-\sigma/(\epsilon_0 \omega_0 \Gamma)}{\epsilon_p^2 + (\epsilon_0 a\sigma/(\epsilon_0 \omega_0 \Gamma))^2} \\
&= \frac{10}{\ln{10}} K_0^2 q \frac{(v_d/v_s - 1) a \sigma/(\epsilon_p \omega_0)}{(v_d/v_s - 1)^2 + (a \sigma/(\epsilon_p \omega_0))^2},
\end{split}
\end{align}
where $K_0^2 = e^2/(\epsilon_p \kappa)$ represents the baseline electromechanical coupling for the piezoelectric material. The imaginary wavevector shift (determined by converting $G$ to the amplitude attenuation per unit length $\alpha$ using the ratio $\alpha/G = \ln{(10)}/20$) matches the classical result \cite{kino1971normalmode} if we make the replacement $a\sigma/(\epsilon_p \omega_0) \rightarrow R\omega_c/\omega_0$ (where $R$ and $\omega_c$ are the space-charge reduction factor and the dielectric relaxation frequency, respectively \cite{kino1971normalmode}). 

\begin{figure*}[!tb]
	\centering
	\begin{subfigure}{\columnwidth}
		\centering
	    \includegraphics[width=\linewidth]{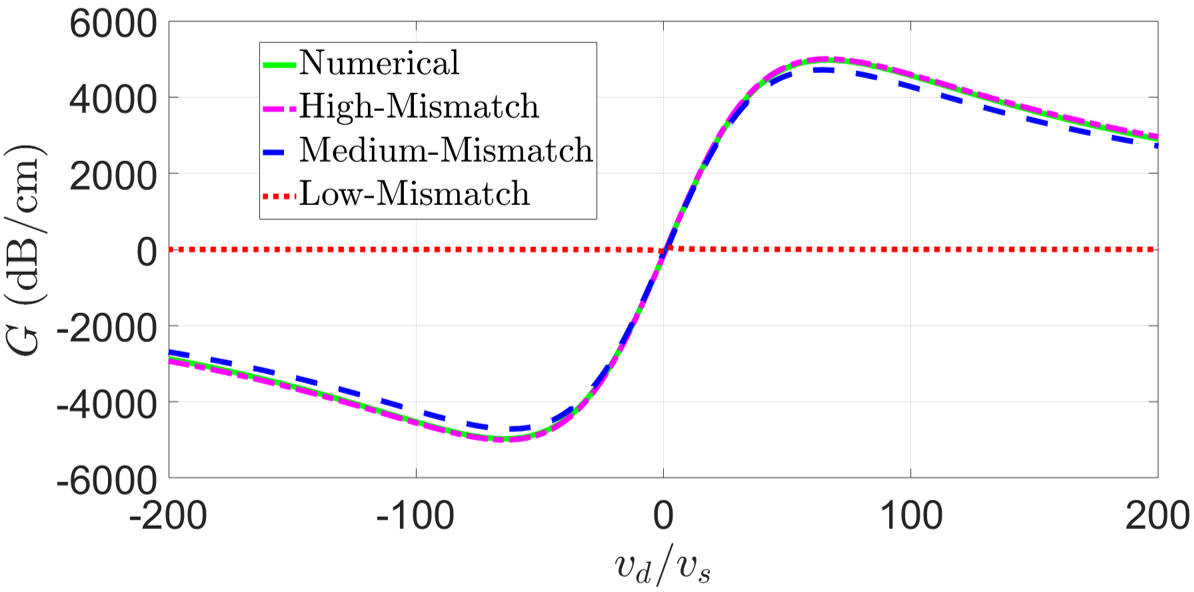}
	    \caption{}
	    \label{}
	\end{subfigure}
	\begin{subfigure}{\columnwidth}
		\centering
	    \includegraphics[width=\linewidth]{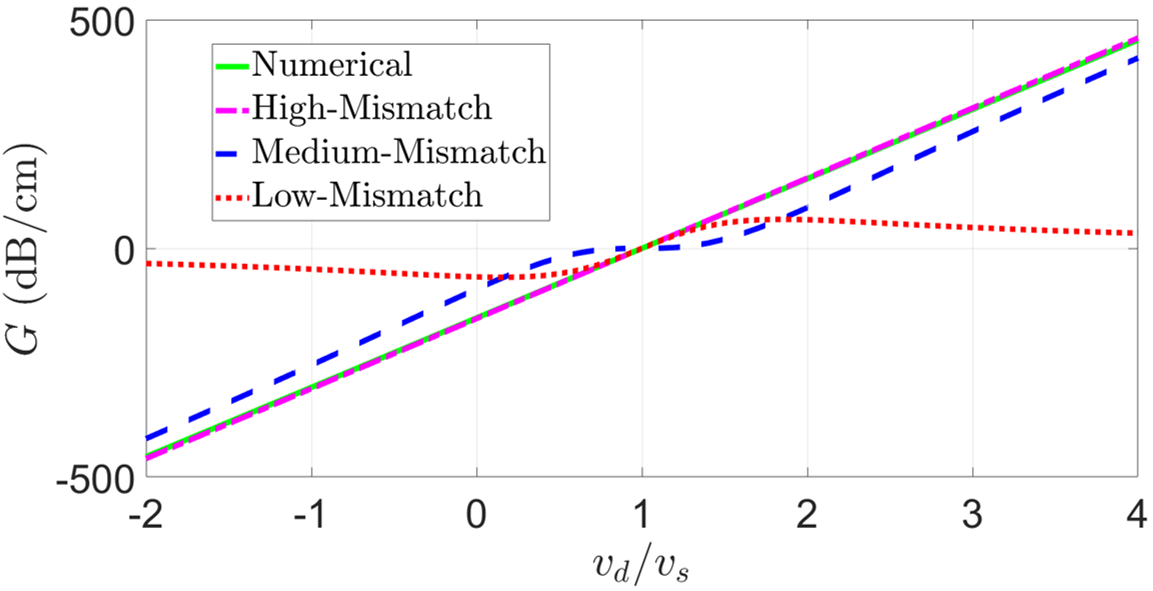}
		\caption{}
		\label{}
	\end{subfigure}
    \caption{Numerical results (solid, green) for the amplifier waveguide gain per unit length $G$ as a function of the drift velocity $v_d$, given a mobility of $0.3 \textrm{ m}^2/(\textrm{V} \cdot \textrm{s})$), along with the analytical results for the low-mobility limit $v_d,v_F \ll \gamma/q$ in the low-velocity-mismatch (dotted, red), medium-velocity-mismatch (dashed, blue) and high-velocity-mismatch (dash-dotted, magenta) regimes. Note that the high-mismatch results match the classical results. We assume a phonon angular frequency of $\omega_0 = 2\pi \times 10^9 \textrm{ s}^{-1}$, speed of sound $v_s = 4 \times 10^3$ m/s, a carrier density $n = 2 \times 10^{15} \textrm{ m}^{-2}$, a carrier effective mass $m = 0.067 m_0$, and a 2DEG thickness $t_\mathrm{2DEG} = 2 \times 10^{-8} \textrm{ m}$.}
	\label{fig:gainlowmu}
\end{figure*}

We can also demonstrate a match between the low-mobility quantum results for the gain and the classical gain in the low-velocity-mismatch regime. Here, the real part of the susceptibility sharply peaks, causing the 2DEG's own screening field to dominate the dielectric response (i.e., $\epsilon_0 a |\textrm{Re}[\chi^{(1)}| \gg \epsilon_p, \epsilon_0 a |\textrm{Im}[\chi^{(1)}|$). Substituting Eqs.~\eqref{eq: low-mobility low-mismatch real chi(1)} and~\eqref{eq: low-mobility low-mismatch imag chi(1)} into Eq.~\eqref{eq: acoustoelectric gain in dB} yields the following gain in the low-mismatch regime:
\begin{align}
\begin{split}
G &\approx \frac{10}{\ln{10}} \epsilon_0 (K_0^2 \epsilon_p) a q \frac{q_e^3 m^2 (v_d - v_s)/(\pi^2 \epsilon_0 \hbar^4 n t_\mathrm{2DEG} q^3 \mu)}{a^2 \epsilon_0^2 \Big(q_e^2 v_s^2 m/(\pi \hbar^2 \epsilon_0 t_\mathrm{2DEG} \omega_0^2)\Big)^2} \\
&= \frac{10}{\ln{10}} \epsilon_0 (K_0^2 \epsilon_p) a q \frac{(v_d - v_s) q}{\epsilon_0 a^2 n_\mathrm{3D} q_e \mu} \\
&= \frac{10}{\ln{10}} K_0^2 q \frac{(v_d/v_s - 1) \epsilon_p \omega_0}{a \sigma} 
\end{split}
\end{align}
Since $|\sigma/(\epsilon_0 \omega_0 \Gamma)| \gg \epsilon_p$, and thus $a\sigma/(\epsilon_p \omega_0) \gg |v_d/v_s - 1|$, the gain expression in the low-mismatch limit is approximately equivalent to Eq.~\eqref{eq: gain mid high mismatch}, thus matching the classical result as well. 

Figure~\ref{fig:gainlowmu} depicts the gain per unit length for a mobility of $1 \textrm{ m}^2/(\textrm{V} \cdot \textrm{s})$ for a broader (a) and zoomed-in (b) range of drift velocities. It is worth noting that the numerically determined gain consistently matches the classical Drude-model results (given by the high-mismatch results). Considering the overall range of drift velocities, the low-mobility gain is maximized when the 2 denominator terms in Eq.~\eqref{eq: gain mid high mismatch} expression match each other, yielding the following optimal drift velocity:
\begin{equation}
v_{d,\mathrm{opt}} \approx v_s \bigg(1 + \frac{a \sigma}{\epsilon_p \omega_0}\bigg) \approx v_s \bigg(1 + \frac{n q_e \mu}{\epsilon_p v_s}\bigg).
\end{equation}
The optimal drift velocity in the low-mobility regime thus increases with both carrier density and mobility, featuring a linear relationship with each parameter given $a \sigma \gg \epsilon_p \omega_0$. At this drift velocity, the fractional term in the gain equation simply reduces to $1/2$, yielding:
\begin{equation}
G_\mathrm{max} \approx \frac{5}{\ln{10}} K_0^2 q,
\end{equation}
which is independent in both carrier density and mobility (within the low-mobility range).

Finally, we briefly discuss the effect of temperature. As discussed in Sec.~\ref{sec: Deriving the Electric Susceptibility}, the variation of the Lindhard susceptibilities with temperature is mediated approximately fully through carrier density and mobility. This implies that the Mermin susceptibilities, and therefore the overall gain, are also similarly invariant in temperature if carrier density and mobility are given quantities, a conclusion borne out by the fact that the low-mobility gain equation calculated above matches room-temperature experimental results \cite{wendt2026electricallyinjected}. It is also worth noting that since the initial phonon population in any given mode is set by the Bose-Einstein occupation number, the overall thermal energy gain for the system can be determined by performing a weighted sum of the single-mode gain derived above over all possible mode using the Bose-Einstein occupation numbers as the weighting factors.

\section{Clamping}
\label{sec: Clamping}

Having determined the amplifier gain, we now shift our focus to deriving the clamping condition. Here, it is essential to account for the general time-evolution of the phonon mode through a beam-splitter Hamiltonian covering both the phonon and the 2DEG electron modes. Assuming a beam-splitter Hamiltonian consisting of $N$ pumped electrons each coupling with the phonon field through a time-varying coupling coefficient $g_j(t)$ (corresponding to the $j^\textrm{th}$ electron), the noise and the signal intensity feature the same gain per unit length (as shown in Appendix~\ref{sec: Amplified Phonon Wave and Quantum Noise}), causing the signal-to-noise ratio to remain constant throughout the amplification process (analogously to a laser or maser). This common gain per unit length is given in terms of the coupling coefficient between a single electron and a phonon as follows:
\begin{equation} \label{eq: gain generic}
G = \frac{(g\tau)^2 N M}{L} = \frac{g^2 \tau N}{v_s},
\end{equation}
where $\tau$ is the time elapsed in a single amplifier segment (with a corresponding length of $v_s \tau$), $g = \expect{\Big|\expect{g_j(t)}_t\Big|^2}_j^{1/2}$ is root-mean-squared (over the $N$ electrons) of the time-averaged amplitude (averaged over the period $\tau$) of the coupling coefficient, and $M = L/(v_s \tau)$ is the number of amplifier segments. We can conceptualize $\tau$ as the effective reset time for an electron after it emits a phonon (i.e., the time the electron takes to return to the original state). Given an electronic decay rate $\gamma/2$ for the initial and final states each, the overall reset time is given approximately as $\tau \approx 1/\gamma$. 

Although the gain per unit length at low phonon intensity is proportional to $g^2 N$, the clamping condition will be solely dependent on $g$ itself, as we will show later in this section. As such, we aim to dis-aggregate $g$ and $N$ by separately deriving $g$. We start by solving the interaction coefficient $g_\mathrm{emit}$ corresponding to phonon emission by pumped electrons. This coefficient acquires a time-dependence due to electronic decay and electron-phonon detuning, as shown in the following composite interaction Hamiltonian, where we sum over the interactions with all electrons in the region $S$ (see Fig.~\ref{fig:phasespacelowmobility}(a)):
\begin{align}
\begin{split} \label{eq: waveguide Hamiltonian over S}
&H_\mathrm{emit}(t) \\
&= \hbar g_0 \sum_{j \in S} \Big(b_q^{\dag} e^{-i \omega_0 t} c_{k_j,k_j-q} e^{i \omega_{k_j,k_j-q} t - \gamma t} \\
&\quad + b_q  e^{i \omega_0 t} c_{k_j,k_j-q}^{\dag} e^{-i \omega_{k_j,k_j-q} t - \gamma t}\Big) \\
&= \hbar \sum_{j \in S} \Big(g_{\mathrm{emit},j}(t) b_q^{\dag} c_{k_j,k_j-q} + g^*_{\mathrm{emit},j}(t) b_q c_{k_j,k_j-q}^{\dag}\Big),
\end{split}
\end{align}
where $b_q^{(\dag)}$ is the annihilation (creation) operator for a phonon of wavevector $q$, $c_{k_j,k_j-q}^{(\dag)}$ takes an electron from a wavevector $k_j$ to $k_j-q$ ($k_j-q$ to $k_j$), and $g_{\mathrm{emit},j}(t)$ is defined as:
\begin{equation} \label{eq: g_emit(t)}
g_{\mathrm{emit},j}(t) = g_0 e^{(i(\omega_{k_j,k_j-q} - \omega_0) - \gamma)t},
\end{equation}
where $g_0$ is the following \cite{chatterjee2024abinitio}:
\begin{equation} \label{eq: baseline coupling}
g_0 = \frac{q_e E_\mathrm{zpf}(\omega_0)}{\hbar q} = \frac{q_e v_s E_\mathrm{zpf}(\omega_0)}{\hbar \omega_0}.
\end{equation}
Intuitively, $g_0$ represents fully resonant coupling between a single electronic dipole moment and the electric field associated with a single phonon, equaling $q_e/q$ and $E_\mathrm{zpf}$ respectively. The electron-phonon detuning manifests itself in oscillation of this coupling over time, while the spectral broadening (i.e., the electronic decay rate) provides an exponentially decaying envelope. Using the reset time $1/\gamma$ as the effective timespan of this decay, we average $g_\mathrm{emit}$ over this period:
\begin{equation} \label{eq: emission coupling time-averaged}
\expect{g_{\mathrm{emit},j}}_t = \gamma \int_0^{1/\gamma} dt g_{\mathrm{emit},j}(t) \approx \frac{-i g_0 \gamma}{(\omega_{k_j-q,k_j} + \omega_0) - i\gamma},
\end{equation}
where in the second step, we made the approximation $g_{\mathrm{emit},j}(0) - g_{\mathrm{emit},j}(1/\gamma) \approx 1$.

Next, we examine the phonon absorption process by electrons in the region $S'$ (see Fig.~\ref{fig:phasespacelowmobility}(b)). As with the emissive case, we incorporate the effects of detuning and spectral broadening through time-dependent terms, leading to the following Hamiltonian:
\begin{align}
\begin{split} \label{eq: waveguide Hamiltonian over S'}
&H_\mathrm{abs}(t) \\
&= \hbar g_0 \sum_{j \in S'} \Big(b_q e^{i \omega_0 t} c_{k_j+q,k_j}^{\dag} e^{-i \omega_{k_j+q,k_j} t - \gamma t} \\
&\quad + b_q^{\dag}  e^{-i \omega_0 t} c_{k_j+q,k_j} e^{i \omega_{k_j+q,k_j} t - \gamma t}\Big) \\
&= \hbar \sum_{j \in S'} \Big(g_{\mathrm{abs},j}^*(t) b_q c_{k_j+q,k_j}^{\dag} + g_{\mathrm{abs},j}(t) b_q^{\dag} c_{k_j+q,k_j}\Big),
\end{split}
\end{align}
where $g_{\mathrm{abs},j}(t)$ is defined as:
\begin{equation} \label{eq: g_abs(t)}
g_{\mathrm{abs},j}(t) = g_0 e^{(i(\omega_{k_j+q,k_j} - \omega_0) - \gamma)t}.
\end{equation}
We average this over the reset time $1/\gamma$, yielding:
\begin{equation} \label{eq: absorption coupling time-averaged}
\expect{g_{\mathrm{abs},j}}_t = \gamma \int_0^{1/\gamma} dt g_{\mathrm{abs},j}(t) \approx \frac{i g_0 \gamma}{(\omega_{k_j+q,k_j} - \omega_0) + i\gamma},
\end{equation}
We are now ready to determine the composite coupling coefficient $g$ by calculating the root-mean-squared of the interaction coefficient amplitudes. Since the electronic states undergoing absorption and emission are largely separate from each other (with emission and absorption undergone by the electrons in regions $S$ and $S'$ respectively), we separately consider the rms coefficients of the two types of processes and take the difference to determine the net emissive coefficient:
\begin{widetext}
\begin{align}
\begin{split} \label{eq: g^2}
g^2 &= \frac{1}{N} \Bigg(N_S \expect{\Big|\expect{g_{\mathrm{emit},j}(t)}_t\Big|^2}_{j \in S} - N_{S'} \expect{\Big|\expect{g_{\mathrm{abs},j}(t)}_t\Big|^2}_{j \in S'}\Bigg) \\
&\approx \frac{1}{N} \bigg(\sum_{j \in S} \frac{|g_0|^2 \gamma^2}{(\omega_{k_j-q,k_j} + \omega_0)^2 + \gamma^2} - \sum_{j \in S'} \frac{|g_0|^2 \gamma^2}{(\omega_{k_j+q,k_j} - \omega_0)^2 + \gamma^2}\bigg) \\
&= \frac{\epsilon_0 V_\mathrm{2DEG}}{\hbar} \frac{\gamma}{N} \textrm{Im}[\chi^{(1)}_L(-\omega_0)] \Big|E_\mathrm{zpf}(\omega_0)\Big|^2,
\end{split}
\end{align}
\end{widetext}
where $N = N_S + N_{S'}$, and the final line is determined by substituting in Eq.~\eqref{eq: baseline coupling} as well as the definition of $\chi^{(1)}$ \cite{chatterjee2024abinitio}. Substituting this into the generic gain expression from Eq.~\eqref{eq: gain generic} and setting the reset time to $\tau = 1/\gamma$ as previously discussed, we find that the gain per unit length takes the following form:
\begin{equation} \label{eq: gain per unit length}
G = \frac{g^2 \tau N}{v_s} \approx \frac{\epsilon_0 V_\mathrm{2DEG}}{\hbar v_s} \textrm{Im}[\chi^{(1)}_L(-\omega_0)] \Big|E_\mathrm{zpf}(\omega_0)\Big|^2.
\end{equation}
Substituting $|E_\mathrm{zpf}(\omega_0)|^2 = C^2\hbar\omega_0/(|\epsilon(\omega_0)|^2 V)$ for the zero-point electric field \cite{chatterjee2024abinitio} and multiplying by $10/\ln{(10)}$ to convert to dB-based units, the gain expression matches the amplification coefficient from Eq.~\eqref{eq: acoustoelectric gain in dB} as long as the imaginary Lindhard susceptibility equals the imaginary Mermin susceptibility, which holds fully in the high-mobility/low-drift-velocity limit and up to a factor of 2 in the high-drift-velocity limit. Intuitively, the time-evolution of the phonon population occurs through a series of parabolic increases, each spanning the reset time $\tau$ and corresponding to a coherent evolution in which the increase in the phonon population is balanced out by an equal and opposite decrease in the pumped electron population. As the reset time passes, however, the electron scattering processes cause the 2DEG to revert to a fully pumped set of states, causing a new coherent evolution process to begin. Therefore, the overall time-evolution consists of a steady exponential increase in the phonon population and a constant pumped electron population. 

Having derived the single electron-phonon coupling coefficient $g$, we now consider the clamping condition, which is a result of pumped-state depletion, specifically in the high-mobility regimes where the imaginary Lindhard and Mermin susceptibilities roughly match each other. The overall Hamiltonian is expressed in the following time-dependent form by adding Eqs.~\eqref{eq: waveguide Hamiltonian over S} and~\eqref{eq: waveguide Hamiltonian over S'} corresponding to phonon emission and absorption, respectively:
\begin{align}
\begin{split}
H &= \hbar \bigg(\sum_{j \in S} \Big(g_{\mathrm{emit},j}(t) b^{\dag} c_j + g_{\mathrm{emit},j}^*(t) b c_j^{\dag}\Big) \\
&\quad + \sum_{j \in S'} \Big(g_{\mathrm{abs},j}(t) b c_j^{\dag} + g_{\mathrm{abs},j}^*(t) b^{\dag} c_j\Big)\bigg),
\end{split}
\end{align}
where for $j \in S$, $c_j$ is defined as $c_{k_j,k_j-q}$ (corresponding to a carrier shifting from $k_j$ in the region $S$ \textit{to} $k_j-q$), while for $j \in S'$, $c_j$ is defined as $c_{k_j+q,k_j}$ (corresponding to a carrier shifting to $k_j$ in the region $S'$ \textit{from} $k_j+q$). The time-varying coupling coefficients $g_{\mathrm{emit},j}(t)$ and $g_{\mathrm{abs},j}(t)$ are defined as in Eqs.~\eqref{eq: g_emit(t)} and~\eqref{eq: g_abs(t)}, respectively. 

Per Appendix~\ref{sec: Detailed Derivation of Electron-Phonon Interaction}, the resulting change in the phonon population from $t = 0$ to the reset time $\tau$ takes the form of a superposition of Rabi oscillations with varying frequencies proportional to each coupling coefficient. Given the coupling coefficients in our 2-level carrier systems, this takes the following form:
\begin{align}
\begin{split}
&\Delta(b^{\dag}b)(\tau) \approx \\
&\quad\sum_{j \in S} \sin^2{\Big(|g_{j,\mathrm{emit}}| \sqrt{n_\mathrm{ph}} \tau\Big)} - \sum_{j \in S'} \sin^2{\Big(|g_{j,\mathrm{abs}}| \sqrt{n_\mathrm{ph}} \tau\Big)},
\end{split}
\end{align}
where $n_\mathrm{ph}$ is the phonon number, and $|g_{\mathrm{emit},j}|$ and $|g_{\mathrm{abs},j}|$ represent the amplitude of the time-averaged coefficients, which we solve based on Eqs.~\eqref{eq: emission coupling time-averaged} and~\eqref{eq: absorption coupling time-averaged} as:
\begin{align}
\label{eq: g(emit,j)}
|g_{\mathrm{emit},j}| &= \frac{|g_0| \gamma}{\sqrt{(\omega_{k_j-q,k_j} + \omega_0)^2 + \gamma^2}}, \\
\label{eq: g(abs,j)}
|g_{\mathrm{abs},j}| &= \frac{|g_0| \gamma}{\sqrt{(\omega_{k_j+q,k_j} - \omega_0)^2 + \gamma^2}}.
\end{align}
It is worth noting that the coupling coefficients become approximately uniform in the high-drift-velocity. Here, as discussed earlier, the spectral broadening $\gamma$ is negligible compared to the detuning, which in turn features a uniform amplitude of approximately $v_d q$. The coupling coefficients thus all approximately converge to the same value:
\begin{equation}
|g_{\mathrm{emit},j}| \approx \frac{|g_0| \gamma}{v_d q} \approx |g_{\mathrm{abs},j}|.
\end{equation}
The time-averaged coefficients are significantly attenuated from the maximum value $|g_0|$ due to the rapidly oscillatory behavior of the time-varying coefficients in the interval $t = 0$ to $\tau$. Therefore, the phonon population roughly oscillates at the single frequency $\omega_\mathrm{Rabi} = |g_0| \sqrt{n_\mathrm{ph}} \gamma/(v_d q)$. This causes the phonon density at clamping to be enhanced from the low-mobility result by a factor of $(v_d q/\gamma)^2 = (v_d \omega_0/(v_s \gamma))^2$, yielding:
\begin{equation}
\frac{n_\mathrm{ph}}{V} \approx \frac{\hbar \omega_0^3 v_d^2 |\epsilon(\omega_0)|^2}{q_e^2 v_s^4 C^2}.
\end{equation}
The phonon density here is independent of the reset time $\tau = 1/\gamma$, since the effective coupling coefficient varies inversely with the reset time.

Next, we examine the low-drift-velocity/high-mobility limit. This regime is complicated by the fact that the effective detuning (defined as the denominator in Eq.~\eqref{eq: g(emit,j)} or~\eqref{eq: g(abs,j)}) varies over a wide range (specifically, from $\gamma$ for resonant interactions to about $v_F q$ for maximally detuned interactions). The phonon population is thus constituted from a superposition of Rabi oscillations with vastly differing frequencies, resulting in a lack of a well-defined clamping value. We can, however, make 2 heuristic observations: first, that the carrier population is concentrated near the maximum detuning $v_F q$ (see Fig.~\ref{fig:phasespacelowmobility}), and second, that the maximum Rabi oscillation period is set by those maximally detuned interactions (since the effective coupling rate varies inversely with the effective detuning). As such, we can roughly estimate that the interactions with effective detuning amplitude $v_F q$ set the clamping dynamics, yielding the following set of uniform effective coupling coefficients:
\begin{equation}
|g_{\mathrm{emit},j}| \sim \frac{|g_0| \gamma}{v_F q} \sim |g_{\mathrm{abs},j}|.
\end{equation}
This yields an analogous result for the maximum phonon density as the high-drift-velocity result, except with the replacement $v_d \rightarrow v_F$:
\begin{equation}
\frac{n_\mathrm{ph}}{V} \sim \frac{\hbar \omega_0^3 v_F^2 |\epsilon(\omega_0)|^2}{q_e^2 v_s^4 C^2}.
\end{equation}
The phonon density here is also independent of the reset time due to the inverse variation between the effective coupling coefficient and the reset time.

It is interesting to note that for both regimes, the maximum phonon density increases with the screening as $|\epsilon(\omega_0)|^2$. This is due to the fact that a higher screening value suppresses the electron-phonon coupling by reducing the 2DEG electric field per phonon, requiring a higher phonon density in order for the overall interaction rate between each electron and the acoustic field to reach the carrier reset rate and start to deplete the pump. On the other hand, the gain at low phonon numbers varies inversely with $|\epsilon(\omega_0)|^2$, since suppressing the electron-phonon coupling results in a reduced gain. Therefore, there is a tradeoff in setting the screening value: a lower screening causes a faster gain but a reduced amplitude ceiling, and vice versa for a higher screening.

\section{Conclusion}

We have theoretically demonstrated a high-gain, low-noise, phase-preserving acoustic amplifier in a 2DEG-piezoelectric heterostructure. The 2DEG is pumped by a dc electric field, and we determine the gain as a function of drift velocity in various mobility regimes, with the analytical results matching the numerical simulations in the low-mobility, high-drift-velocity, and low-drift-velocity/high-mobility limits. Moreover, the results in the low-mobility and low-drift-velocity/high-mobility limits match established literature, further validating our findings.

Our method consisted of separately deriving the phonon emission rate per unit electric field intensity and the electric field intensity per phonon as functions of drift velocity and mobility. Furthermore, we derived the time-evolution of the quantum noise by expanding the Hilbert space to incorporate the electron modes along with the phonon modes, with the results showing that the phonon field's noise increases linearly with the amplitude in the constant-gain regime. Finally, using the same expanded Hilbert space, we also derived the gain clamping condition for the various drift velocity and mobility regimes. 

The high gain per unit length at microwave frequencies offered by this acoustoelectric amplifier design potentially unlocks a vast array of quantum and semi-classical phononic applications. For example, it can be used to amplify phonon sources for quantum optomechanical systems. Moreover, it promises to lead to ultra-coherent sources of phonons on a chip and ultra-narrow-linewidth microwave sources: phononic lasers. We have recently demonstrated such a phonon laser operating in a classical regime using a bulk semiconductor-piezoelectric heterostructure \cite{wendt2026electricallyinjected}, and we believe that we can obtain substantially narrower linewidths for such phonon lasers using this 2DEG system. In addition, we have shown both theoretically in the quantum regime \cite{chatterjee2024abinitio} and experimentally in the classical regime \cite{hackett2024giantelectron} that these heterostructures provide giant second- and third-order acoustic nonlinearities. Combining these giant phononic nonlinearities on the same chip as a phonon laser can yield more sophisticated quantum circuits. For example, the phonon laser could be used to pump a degenerate parametric oscillator to create a single-mode squeezed state, or another cavity--like the phonon laser cavity--could be filled with a 2DEG and engineered to strongly favor Kerr processes (which we have previously shown to be extremely large in the same material system \cite{chatterjee2024abinitio}), creating anharmonicity that could lead to an acoustic qubit. The on-chip phonon laser could then be used, together with modulation, to perform single-qubit rotations on that qubit in the cavity. All these and more enable nonclassical phononic resource states, thus paving the way to circuit and linear mechanical computing architectures.

\begin{acknowledgements}

This work was supported by the Laboratory Directed Research and Development program at Sandia
National Laboratories. This work was also performed, in part, at the Center for Integrated Nanotechnologies, an Office of Science User Facility operated for the U.S. Department of Energy (DOE) Office of Science. Sandia National Laboratories is a multimission laboratory managed and operated by National Technology \& Engineering Solutions of Sandia, LLC, a wholly owned subsidiary of Honeywell International, Inc., for the U.S. DOE’s National Nuclear Security Administration under contract DE-NA-0003525. The views expressed in the article do not necessarily represent the views of the U.S. DOE or the United States Government.

In addition, this material is based on research sponsored in part by the Defense Advanced Research Projects Agency (DARPA) through a Young Faculty Award (YFA) under grant D23AP00174-00 and through DARPA contract DARPA-PA-23-03-01. The views and conclusions contained herein are those of the authors and should not be interpreted as necessarily representing the official policies or endorsements, either expressed or implied, by DARPA, the Department of the Interior, or the US Government.


\end{acknowledgements}

\appendix

\section{Imaginary Part of Lindhard Chi(1) in High-Mobility/Low-Drift-Velocity Calculation}
\label{sec: Imaginary Part of Chi(1) High-Mobility/Low-Drift-Velocity Calculation}

Here, we derive the imaginary part of $\chi^{(1)}_L$ in the high-mobility/low-drift-velocity limit. We start by integrating over the valid phase-space region $S'$ (see Fig.~\ref{fig:phasespacelowmobility}(b)) as follows:
\begin{equation} \label{eq: raw phase space integral}
\textrm{Im}[\chi^{(1)}_L(-\omega_0)] = \beta \int_{\substack{\mathrm{valid} \\ \mathrm{states}}} dk_x k_{y,\mathrm{span}}(k_x) \textrm{Im}[f^{(1)}_{k_x}(-\omega_0)],
\end{equation}
where $\beta$ is a constant function of the 2DEG and piezoelectric material parameters \cite{chatterjee2024abinitio}, proportional to the square of the dipole moment for an electron that has undergone interaction with the phonon field and inversely proportional to the 2DEG thickness:
\begin{equation}
\beta = \frac{q_e^2 v_s^2}{2 \pi^2 \hbar \epsilon_0 t_\mathrm{2DEG} \omega_0^2},
\end{equation}
and $f^{(1)}_{k_x}$ corresponds to the closeness of the electron-phonon interaction to resonance (i.e., the inverse of the complex detuning) and is thus a function of the electron transition frequency $\omega_{k_x-q,k_x} = \omega_{k_x - q} - \omega_{k_x}$:
\begin{equation}
f^{(1)}_{k_x}(-\omega_0) \approx \frac{1}{\omega_{k_x-q,k_x} + \omega_0 - i\gamma},
\end{equation}
where $\gamma$ relates inversely to the electron mobility $\mu$ as $\gamma = q_e/(m\mu)$. Note that this only takes into account the resonant term, since in the high-mobility limit, the near-resonance process will dominate. Consequently, $f^{(1)}$ becomes a function of the electron-phonon detuning $\Delta \omega = \omega_{k_x,k_x-q} - \omega_0$, yielding:
\begin{equation}
\textrm{Im}[f^{(1)}(\Delta \omega)] \approx \frac{\gamma}{(\Delta \omega)^2 + \gamma^2}.
\end{equation}
It is convenient to convert the phase-space integral in Eq.~\eqref{eq: raw phase space integral} to an integral over $\Delta \omega$. We note that $dk_x$ is linear in $d (\Delta \omega)$, since any shift in the electron transition frequency $\omega_{k_x,k_x-q}$ is proportional to the shift in $k_x$:
\begin{equation}
dk_x = \frac{m}{\hbar q} d(\Delta \omega).
\end{equation}
In the high-mobility limit, the contributions to the integral will be dominated by the near-resonance region (see Fig.~\ref{fig:phasespace}). There, the deviation of $k_{y,\mathrm{span}}$ from the value at resonance approximately varies linearly with $\Delta \omega$:
\begin{equation}
k_{y,\mathrm{span}}(\Delta \omega) \approx \frac{2 m}{\hbar k_F} (\omega' - \Delta \omega),
\end{equation}
where $\omega'$ is defined as follows:
\begin{equation}
\omega' = \frac{\hbar q (k_d - k_c)}{m} = \frac{\omega_0 (v_d - v_s)}{v_s}.
\end{equation}
Note that the $k_{y,\mathrm{span}}$ at resonance (i.e., $2 m \omega'/(\hbar k_F)$) increases with the drift velocity in the low-drift-velocity limit, becoming positive for $v_d > v_s$. This matches the intuition laid out earlier in this work. Substituting the above expressions into Eq.~\eqref{eq: raw phase space integral}, we calculate the imaginary part of $\chi^{(1)}_L$ in the following manner:
\begin{align}
\begin{split} \label{eq: imaginary chi(1) high mobility low drift}
\textrm{Im}[\chi^{(1)}_L] &\approx \beta \frac{2 m^2}{\hbar^2 q k_F} \int_{-\infty}^{\infty} d(\Delta \omega) (\omega' - \Delta \omega) \frac{\gamma}{(\Delta \omega)^2 + \gamma^2} \\
&= \beta \frac{2 m}{\hbar k_F} (k_d - k_c) \int_{-\infty}^{\infty} d(\Delta \omega)\frac{\gamma}{(\Delta \omega)^2 + \gamma^2} \\
&= \beta \frac{2 m^2}{\hbar^2 k_F} (v_d - v_s) \pi \\
&= \frac{q_e^2 v_s^2 (v_d - v_s) m^2}{\pi \hbar^3 \epsilon_0 t_\mathrm{2DEG} \omega_0^2 k_F},
\end{split}
\end{align}
where in the second line, we kept only the integrand term that is even in $\Delta \omega$. Specifically, since $\textrm{Im}[f^{(1)}(\Delta \omega)]$ is even in $\Delta \omega$, we dropped the second term in $k_{y,\mathrm{span}}$.

\section{Imaginary Part of Lindhard Chi(1) in Low-Mobility Limit}
\label{sec: Imaginary Part of Chi(1) Low-Mobility Limit}

We turn to deriving the imaginary part of $\chi^{(1)}_L$ in the low-mobility limit. In this regime, all possible electron-phonon interactions will fall within the near-resonance window. Therefore, $f^{(1)}$ needs to incorporate both resonant (emissive) and counter-resonant (absorptive) terms. We label these terms as $f_1$ and $f_2$, respectively:
\begin{align}
\begin{split}
f_{1,k_x'}^{(1)}(-\omega_0) &= \frac{1}{\omega_{k_x-q,k_x} + \omega_0 - i\gamma} \\
&= \frac{1}{\omega_{k_x'+k_d-q,k_x'+k_d} + \omega_0 - i\gamma},
\end{split}
\\
\begin{split}
f_{2,k_x'}^{(1)}(-\omega_0) &= \frac{1}{\omega_{k_x+q,k_x} - \omega_0 + i\gamma} \\
&= \frac{1}{\omega_{k_x'+k_d+q,k_x'+k_d} - \omega_0 + i\gamma}.
\end{split}
\end{align}
We solve for the net imaginary part of $\chi^{(1)}_L$ by integrating the resonant emissive terms and the counter-resonant absorptive terms over the regions $S$ and $S'$, respectively (see Fig.~\ref{fig:phasespacelowmobility}):
\begin{widetext}
\begin{align}
\begin{split} \label{eq: low-mobility chi(1) integral}
&\textrm{Im}[\chi^{(1)}_L(-\omega_0)] \\
&= \beta \bigg(\int_S dk_x' k_{y,\mathrm{span}}(k_x') \textrm{Im}[f^{(1)}_{1,k_x'}(-\omega_0)] + \int_{S'} dk_x' k_{y,\mathrm{span}}(k_x') \textrm{Im}[f^{(1)}_{2,k_x'}(-\omega_0)]\bigg) \\
&= \beta \bigg(\int_{S'} dk_x'' k_{y,\mathrm{span}}(-k_x'') \textrm{Im}[f^{(1)}_{1,-k_x''}(-\omega_0)] + \int_{S'} dk_x' k_{y,\mathrm{span}}(k_x') \textrm{Im}[f^{(1)}_{2,k_x'}(-\omega_0)]\bigg) \\
&= \beta \int_{S'} dk_x' k_{y,\mathrm{span}}(k_x') \textrm{Im}[f^{(1)}_{\mathrm{net},k_x'}(-\omega_0)] \\
&= \beta \gamma \int_{S'} dk_x' k_{y,\mathrm{span}}(k_x') \bigg(\frac{1}{(\omega_{-k_x'+k_d-q,-k_x'+k_d} + \omega_0)^2 + \gamma^2} - \frac{1}{(\omega_{k_x'+k_d+q,k_x'+k_d} - \omega_0)^2 + \gamma^2}\bigg) \\
&= \beta \gamma \int_{S'} dk_x' k_{y,\mathrm{span}}(k_x') \bigg(\frac{1}{(\omega_{k_x'-k_d+q,k_x'-k_d} + \omega_0)^2 + \gamma^2} - \frac{1}{(\omega_{k_x'+k_d+q,k_x'+k_d} - \omega_0)^2 + \gamma^2}\bigg),
\end{split}
\end{align}
\end{widetext}
where in the second step, we established the coordinate $k_x'' = -k_x'$, in the third step, we folded both of the integrals over $S'$ into the single coordinate $k_x'$ (using the fact that $k_{y,\mathrm{span}}$ is even about $k_x' = 0$) and defined the quantity $f_{\mathrm{net},k_x'}^{(1)} = f_{1,-k_x'}^{(1)} + f_{2,k_x'}^{(1)}$, in the fourth step, we expressed the imaginary part of this quantity in terms of the electronic transition frequencies, and in the fifth step, we used the evenness of the electronic energy spectrum (i.e., the fact that $\omega_{k_{x,f},k_{x,i}} = \omega_{-k_{x,f},-k_{x,i}}$) to re-write the first term (representing the emission process).

Next, we seek to approximate the above integral in the low-mobility limit. Substituting Eq.~\eqref{eq: average detuning} into Eq.~\eqref{eq: low-mobility chi(1) integral}, we approximate the imaginary part of $\chi^{(1)}_L$ in the low-mobility as follows:
\begin{align}
\begin{split} \label{eq: low mobility imaginary chi(1)}
&\textrm{Im}[\chi^{(1)}_L(-\omega_0)] \\
&= \beta \gamma \int_{S'} dk_x' k_{y,\mathrm{span}}(k_x') \bigg(\frac{1}{((\Delta \omega)_{k_x'} - (v_d - v_s)q)^2 + \gamma^2} \\
&\quad - \frac{1}{((\Delta \omega)_{k_x'} + (v_d - v_s)q)^2 + \gamma^2}\bigg) \\
&\approx \beta \gamma \int_{S'} dk_x' k_{y,\mathrm{span}}(k_x') \frac{4 q (v_d - v_s) (\Delta \omega)_{k_x'}}{\gamma^4} \\
&\approx \frac{4 \beta q (v_d - v_s)}{\gamma^3} \int_0^{k_F} dk_x' \frac{2 q k_x'}{\sqrt{k_F^2 - k_x'^2}} \frac{\hbar q k_x'}{m} \\
&= \bigg(\frac{4 \beta \omega_0 (v_d - v_s)}{v_s \gamma^3}\bigg) \bigg(\frac{\pi \hbar \omega_0^2 k_F^2}{2 m v_s^2}\bigg) \\
&= \bigg(\frac{v_d - v_s}{v_s}\bigg) \frac{q_e^2 \omega_0 k_F^2}{\pi \epsilon_0 m t_\mathrm{2DEG} \gamma^3}.
\end{split}
\end{align}
Here, in the second line, we apply the low-mobility approximation $\gamma \gg (\Delta \omega)_{k_x'},\omega_0$ (corresponding to the fact that all valid transitions approximately occur with the same probability in the low-mobility regime, as previously discussed), yielding an expression analogous to that calculated for the imaginary part of $\chi^{(1)}$ in Ref.~\cite{chatterjee2024abinitio}, but with the replacement $v_s \rightarrow v_d - v_s$. In the third line, we apply the long-acoustic-wavelength assumption that since $k_F \gg q$, and since the average wavevector for the electrons in the region $S'$ is on the order of $k_x' = k_F$, we can approximate that $k_x' \gg q$, yielding $(\Delta \omega)_{k_x'} \approx \hbar q k_x'/m$. In the fourth line, we solve the integral from the third line as in Ref.~\cite{chatterjee2024abinitio}. Finally, in the fifth line, we substitute the definition of $\beta$.

\section{Gain Comparison in Low-Drift-Velocity/High-Mobility Limit}
\label{sec: Gain Comparison in Low-Drift-Velocity/High-Mobility Limit}

Here, we seek to compare our results for the gain to that determined in previous studies. In particular, the gain was previously derived by Mosekilde in the high-mobility/low-drift-velocity limit \cite{mosekilde1974quantum}. In this section, we will use the Mosekilde results to calculate the corresponding real and imaginary parts of $\chi^{(1)}$ as well as the acoustoelectric gain. We will then seek to compare these to our result for the high-mobility/low-drift-velocity limit.

First, we verify that the Mosekilde derivation specifically requires a simultaneously high-mobility and low-drift-velocity electron gas. To this end, it is stated in the work that the results apply for the case of $ql \gg 1$, where $q$ is the acoustic wavevector and $l$ is the electron mean-free-path. Assuming that the electrons participating in electron-phonon interactions are concentrated near the Fermi surface (as is the case for the long-acoustic-wavelength limit $q \ll k_F$), we can relate the mean-free-path to the Fermi wavevector and the electronic decay rate $\gamma$ as $l = v_F/\gamma = \hbar k_F/(m \gamma)$. Substituting the definition of mobility $\mu = q_e/(m \gamma)$, this yields the mobility condition $\mu \gg q_e/(\hbar q k_F)$, which matches our definition for the high-mobility limit \cite{chatterjee2024abinitio}. Regarding the drift velocity, Mosekilde states the assumption that the drift wavevector $k_d$ is much smaller than the characteristic electron wave numbers \cite{mosekilde1974quantum}. This implies that $k_d \ll k_F$ and thus $v_d \ll v_F$, corresponding to the low-drift-velocity limit discussed above.

Having established the regime in which the Mosekilde results apply, we start our quantitative analysis by breaking down $Q(q,\omega)$ from Eq. (29) \cite{mosekilde1974quantum}. Since our system features a well-defined speed of sound $v_s$ such that $\omega = v_s q$, we can reduce $Q$ to a function of just $\omega$. Note also that $\epsilon(q,\omega) = \chi^{(1)}(\omega)$. We can thus relate the real and imaginary parts of $Q$ to the real and imaginary parts of $\chi^{(1)}$ in the following manner:
\begin{align}
\textrm{Re}[\chi^{(1)}(\omega)] &= \frac{\textrm{Re}[Q(\omega)]}{q^2} + 1 \approx \frac{v_s^2}{\omega^2} \textrm{Re}[Q(\omega)], \\
\textrm{Im}[\chi^{(1)}(\omega)] &=  \frac{\textrm{Im}[Q(\omega)]}{q^2} = \frac{v_s^2}{\omega^2} \textrm{Im}[Q(\omega)],
\end{align}
where the approximation in the first equation is due to the assumption that $\textrm{Re}[\chi^{(1)}] \gg 1$. Examining the imaginary part of $Q$ as given in Mosekilde's Eq. (32) and substituting the definitions of $B$, $y$, $a$ and $f_0(a)$ from Eqs. (34) through (37) \cite{mosekilde1974quantum}, we find the following result for $\textrm{Im}[\chi^{(1)}]$:
\begin{align}
\begin{split} \label{eq: imaginary part of chi(1) Mosekilde}
\textrm{Im}[\chi^{(1)}(\omega)] &= -\frac{v_s^2}{\omega^2} \bigg(\frac{q_e^2 m^2 k_B T}{2 \pi \epsilon_0 \hbar^4 q}\bigg) \bigg(\frac{\hbar \omega}{k_B T}\bigg) \bigg(\frac{v_d - v_s}{v_s}\bigg) \\
&\quad \times \Big(e^{\hbar^2(q^2 - 4k_F^2)/(8m k_B T)} + 1\Big)^{-1} \\
&\approx -\frac{q_e^2 v_s^2 (v_d - v_s) m^2}{2 \pi \hbar^3 \epsilon_0 \omega^2},
\end{split}
\end{align}
where in the second line, we took the limit of the exponential as $T \rightarrow 0$, in addition to substituting $q = \omega/v_s$ in the first parenthetical term. Note that this matches, up to a negative sign, our result from Eq.~\eqref{eq: imag chi(1) high mobility low drift} for the high-mobility/low-drift-velocity limit (with the exception of the replacement $t_\mathrm{2DEG} \rightarrow 2/k_F$ to convert from a 2DEG to a bulk electronic distribution). It is also useful to note that since $q \ll 2k_F$ in the relevant regime, $f_0(a) \approx 1$ as long as the temperature $T$ is well below the Fermi temperature $T_F = E_F/k_B$. Intuitively, this results from the fact that if $T \ll T_F$, the thermal broadening in phase space is much narrower than the Fermi circle radius. As a result, as the temperature is increased from absolute zero in this regime, any suppression of interaction probabilities that were prevalent at zero-temperature is almost exactly cancelled out by an increase in interaction probabilities that were invalid at zero-temperature, causing the overall carrier-phonon interaction probability to remain constant.

Next, we examine the real part of $Q$ from Mosekilde's Eq. (33) and substitute the definitions of $B$ and $G$ from Eqs. (34) and (38) \cite{mosekilde1974quantum} to determine $\textrm{Re}[\chi^{(1)}]$. Here, the most important task is to analytically simplify the integrand in Eq. (33) using the defintion of $G$ from Eq. (38). In the limits $T \ll T_F$ and $q \ll k_F$, we find the following:
\begin{align}
\begin{split}
&G\bigg(\frac{1}{2} \hbar q - x\bigg) - G\bigg(\frac{1}{2} \hbar q + x\bigg) \\
&= \ln{\Bigg(\frac{1 + e^{(E_F - (\frac{1}{2} \hbar q - x)^2/(2m))/(k_B T)}}{1 + e^{(E_F - (\frac{1}{2} \hbar q + x)^2/(2m))/(k_B T)}}\Bigg)} \\
&\approx 
\begin{cases}
\ln{\bigg(e^{\frac{1}{k_B T} (\frac{\hbar q x}{2 m} - \frac{-\hbar q x}{2 m})}\bigg)}, & E_F > \bigg(\frac{1}{2} \hbar q \pm x\bigg)^2/(2 m), \\
0, & E_F < \bigg(\frac{1}{2} \hbar q \pm x\bigg)^2/(2 m)
\end{cases} \\
&\approx
\begin{cases}
\frac{\hbar q x}{m k_B T}, & x \lessapprox \sqrt{2 m E_F}, \\
0, & x \gtrapprox \sqrt{2 m E_F}
\end{cases},
\end{split}
\end{align}
where in the second line, we applied the low-temperature approximation $T \ll T_F$, while in the third line, we applied the fact that $q \ll k_F$. Regarding the latter, the detailed calculation shows a separation of the result as a function of $x$ into 3 regions: $x < \sqrt{2mE_F} - \hbar q/2$, $x > \sqrt{2mE_F} + \hbar q/2$, and an intermediate region $\sqrt{2mE_F} - \hbar q/2 < x < \sqrt{2mE_F} + \hbar q/2$. Since $\sqrt{2mE_F} = \hbar k_F$, the long-acoustic-wavelength limit $q \ll k_F$ ensures that the intermediate region is negligible, and that the result for $x < \sqrt{2mE_F} - \hbar q/2$ ($x > \sqrt{2mE_F} + \hbar q/2$) applies generally for $x \lessapprox \sqrt{2 m E_F}$ ($x \gtrapprox \sqrt{2 m E_F}$). We thus solve the integral in Mosekilde's Eq. (33), substitute $B$ from Eq. (34), and convert from $\textrm{Re}[Q]$ to $\textrm{Re}[\chi^{(1)}]$ as follows:
\begin{align}
\begin{split}
\textrm{Re}[\chi^{(1)}(\omega)] &\approx \frac{v_s^2}{\omega^2} \frac{2}{\pi} \bigg(\frac{q_e^2 m^2 k_B T}{2 \pi \epsilon_0 \hbar^4 q}\bigg) \int_0^{\sqrt{2mE_F}} \frac{dx}{x} \frac{\hbar q x}{m k_B T} \\
&= \frac{q_e^2 v_s^2 m}{\pi^2 \hbar^3 \epsilon_0 \omega^2} \sqrt{2mE_F} \\
&= \frac{q_e^2 v_s^2 m k_F}{\pi^2 \hbar^2 \epsilon_0 \omega^2},
\end{split}
\end{align}
where in the final line, we substituted the relationship $\sqrt{2mE_F} = \hbar k_F$ between the Fermi energy and the Fermi wavevector. As desired, this matches our result for the high-mobility/low-drift-velocity regime (see Eq.~\eqref{eq: real chi(1) high mobility low drift}), with the exception of the replacement $t_\mathrm{2DEG} \rightarrow \pi/k_F$ to convert from 2DEG to bulk, and the result is invariant with temperature in the regime $T \ll T_F$. The 2DEG-to-bulk conversion factors differ slightly (by a factor of $\pi/2$) between the imaginary and real parts of $\chi^{(1)}$, which can be explained by the fact that the thickness at which a 2DEG transitions to a bulk is a range rather than a well-defined value.

Finally, we verify how the gain varies with the real and imaginary parts of $\chi^{(1)}$. Here, Mosekilde's Eq. (26) converts the dielectric constant to an acoustoelectric gain \cite{mosekilde1974quantum}, and we simplify this expression as follows:
\begin{align}
\begin{split}
\beta &= K^2 q \textrm{Im}[(\chi^{(1)}(\omega))^{-1}] \\
&= K^2 q \textrm{Im}\Bigg[\frac{1}{\textrm{Re}[\chi^{(1)}(\omega)] + i\textrm{Im}[\chi^{(1)}(\omega)]}\Bigg] \\
&= \frac{K^2 q}{|\chi^{(1)}(\omega)|^2} \textrm{Im}\Big(\textrm{Re}[\chi^{(1)}(\omega)] - i\textrm{Im}[\chi^{(1)}(\omega)]\Big) \\
&= -\frac{\epsilon_0^2 K^2 \omega}{v_s} \frac{\textrm{Im}[\chi^{(1)}(\omega)]}{|\epsilon(\omega)|^2}.
\end{split}
\end{align}
We note that this matches, up to a negative sign, the gain per unit length that we laid out in Eq.~\eqref{eq: acoustoelectric gain in dB} (without the factor $10/\ln{(10)}$ required to convert to dB) given $t_\mathrm{2DEG} = L_z$ (since the bulk piezoelectric material features equivalent mode depth and electron gas thickness), provided that the electromechanical-coupling-squared is defined as $K^2 = e^2/(\epsilon_0 \kappa) = C^2/\epsilon_0$, where $e$ and $\kappa$ are the piezoelectric coupling coefficient and elasticity, respectively. The negative sign is cancelled out by the fact that the imaginary part of $\chi^{(1)}$ calculated from Mosekilde's work is opposite to our result (see the discussion after Eq.~\eqref{eq: imaginary part of chi(1) Mosekilde}).

\section{Amplified Phonon Wave and Quantum Noise}
\label{sec: Amplified Phonon Wave and Quantum Noise}

In this section, we will calculate the amplified phonon wave and the quantum noise in the phonon amplifier. We first consider a unitary evolution without the Lindbladian operator defined in the previous section. We will calculate the noise characteristics including the Lindbladian decoherence operator. 

The initial quantum state for the electrons that can contribute for the amplification process is
\begin{equation}
	|\Psi_a (t=0) \rangle = \prod_{k=1}^{N} |h \rangle_k. \label{eq:atom_initial}
\end{equation}
Note that all $N$ electrons can potentially change to $|l\rangle_k$ state while emitting a phonon with an energy $\hbar \omega$. To be clear, we will divide the entire amplifier length into $M$ segments with identical lengths. $N$ is the number of ``excited'' electrons in each segment. 

A quantum operator $A$ evolves in time as
\begin{equation}
	A(t) = e^{\frac{i}{\hbar} H t} A(0) e^{- \frac{i}{\hbar} H t}.
\end{equation}
To calculate the evolution of operator, we use the ``truncated'' expansion of the evolution \cite{inoue2014quantum}:
\begin{align} \label{eq: generic expansion time-evolution state}
	e^{\xi B} A e^{- \xi B} \simeq A +\xi [B,A] + \frac{\xi^2}{2} [B,[B,A]],
\end{align}
where we truncate the third-order of $\xi$ or higher assuming that the time the phonon field spends in a particular segment is short enough.

We are particularly concerned with the following operators $b$ (annihilation operator for the acoustic wave field), $n (=b^\dagger b)$ (phonon number operator), $x_1 = (b + b^\dagger)/2$ (in-phase phonon quadrature operator), $x_2 = (b - b^\dagger)/(2i)$ (out-phase phonon quadrature operator), $x_1^2$, $x_2^2$, $n^2$. The squared operators are related to the quantum noise.  

We start with the following beam-splitter Hamiltonian, consisting of a sum over $N$ electron-phonon interactions, each with a time-varying coupling coefficient $g_j(t)$:
\begin{equation} \label{eq: interaction Hamiltonian in waveguide}
H = \sum_{j = 1}^N \hbar (g_j^*(t) |h \rangle \langle l|_j b + g_j(t) |l \rangle \langle h|_j b^\dagger).
\end{equation}
Let us first examine the phonon number operator $n$:
\begin{widetext}
\begin{align}
	n(\tau) &= e^{\frac{i}{\hbar} \int_0^{\tau} H dt} n e^{-\frac{i}{\hbar} \int_0^{\tau} H dt} \\
    &= e^{\frac{i}{\hbar} \expect{H}_t \tau} n e^{-\frac{i}{\hbar} \expect{H}_t \tau} \nonumber \\
	&= n - i \tau \sum_{j=1}^N (-g_j^* |h \rangle \langle l|_j b + g_j |l \rangle \langle h|_j b^\dagger) \nonumber \\
	& \quad + \tau^2 \sum_{j=1}^N \Big(|g_j|^2 \left\{  (|h \rangle \langle l|_j  | l \rangle \langle h |_j - |l\rangle \langle h|_j | h \rangle \langle l|_j) n + |h \rangle \langle l|_j |l \rangle \langle h|_j  \right\} + \sum_{j' \neq j} g_{j'}^* g_j \ket{h}\bra{l}_{j'} \ket{l}\bra{h}_j\Big) \nonumber \\
	&= n - i \tau \sum_{j=1}^N (-g_j^* |h \rangle \langle l|_j b + g_j |l \rangle \langle h|_j b^\dagger) + \tau^2 \sum_{j=1}^N \Big(|g_j|^2 \left\{ ( |h \rangle \langle h|_j - |l \rangle \langle l|_j )n +|h \rangle \langle h|_j     \right\} + \sum_{j' \neq j} g_{j'}^* g_j \ket{h}\bra{l}_{j'} \ket{l}\bra{h}_j\Big),
\end{align}
\end{widetext}
where $g_j = \expect{g_j(t)}_t$, i.e., the expectation value for the coupling coefficient $g_j(t)$ over the time period $\tau$. Then, we calculate for $b$:
\begin{align}
	b(\tau) &= b - i \tau \sum_{j=1}^N g_j |l \rangle \langle h|_j \\
    &\quad + \frac{\tau^2}{2} \sum_{j=1}^N |g_j|^2 \left(  |h \rangle \langle l|_j | l \rangle \langle h|_j - |l \rangle \langle h|_j |h \rangle \langle l|_j    \right) b \nonumber \\
	&= \left\{  1 + \frac{\tau^2}{2} \sum_{j=1}^N |g_j|^2 \left( |h \rangle \langle h|_j - |l \rangle \langle l|_j \right)  \right\}b \\
    &\quad - i \tau \sum_{j=1}^N g_j |l \rangle \langle h|_j.
\end{align}
Here, $\tau$ is the propagation time through one segment. We then take advantage of the knowledge about the atom initial state in \eqref{eq:atom_initial}, and take the partial expected operator of the phonon number at the end of the first segment:
\begin{equation}
	\langle n(\tau) \rangle_a = \langle \Psi_a (0) | n (\tau) | \Psi_a (0) \rangle = G_0 n + (g \tau)^2  N,
\end{equation}
where $G_0 = 1 + (g \tau)^2 N$, and $g$ is defined as the square-root of the expectation value of $|g_j|^2$ over the $N$ electrons, i.e., $g = \expect{|g_j|^2}_j^{1/2} = \expect{\Big|\expect{g_j(t)}_t\Big|^2}_j^{1/2}$. We also take the partial expected operator of the phonon field as:
\begin{align}
	\langle b(\tau) \rangle_a &= \left\{ 1 + \frac{(g \tau)^2}{2} N  \right\} b \approx \sqrt{G_0} \,b,
\end{align}
where we approximated $\sqrt{1 + (g \tau)^2 N} \approx 1 + (g\tau)^2 N/2$.

These operators go through evolution in the second segment as
\begin{align}
	n(2 \tau) &= e^{\frac{i}{\hbar} \expect{H}_t \tau} \langle n (\tau) \rangle_a e^{-\frac{i}{\hbar} \expect{H}_t \tau} \nonumber \\
	&= G_0 e^{\frac{i}{\hbar} \expect{H}_t \tau} n e^{-\frac{i}{\hbar} \expect{H}_t \tau} + (g \tau)^2 N,
\end{align}
and
\begin{align}
	b(2 \tau) &= \sqrt{G_0} b(\tau).
\end{align}
Taking the partial expectation operator using the atom's initial condition in the second segment given also as equation \eqref{eq:atom_initial}, we obtain the partial expected operators after the second segment as
\begin{equation}
	\langle n( 2 \tau) \rangle_a = G_0^2 n + (g\tau)^2 N (1 + G_0),
\end{equation}
and
\begin{equation}
	\langle b(2 \tau) \rangle_a = (\sqrt{G_0})^2 b.
\end{equation}
After $M$ segments, we obtain the following partial expected operator for the phonon number operator and the phonon field operator:
\begin{align}
	\langle n_\mathrm{out} \rangle_a &= G_0^M n + (g \tau)^2 N \sum_{k=0}^{M-1} G_0^k = G n + G - 1, \nonumber \\
	\langle b_\mathrm{out} \rangle_a &= (\sqrt{G_0})^M b = \sqrt{G} b, \label{eq:output-operators}
\end{align}
where we set $G = G_0^M$. 

Next, we calculate the variance operators. For this, let us define the quadrature operators $x_1 = (b + b^\dagger)/2$, $x_2 = (b - b^\dagger)/(2i)$. We are interested in calculating the evolution of $X_1 = x_1^2, X_2 = x_2^2$. We first calculate $X_1$ in the first segment of the amplifier:
\begin{widetext}
\begin{align}
\begin{split}
	&X_1 (\tau) \\
	&= X_1 - i \tau \sum_{j=1}^N (g_j |l \rangle \langle h|_j - g_j^* |h \rangle \langle l |_j) x_1 \\
	& \quad + \tau^2 \left\{ \sum_{j=1}^N |g_j|^2 ( |h\rangle \langle l|_j |l \rangle \langle h |_j - |l \rangle \langle h|_j |h \rangle \langle l|_j ) X_1 + \frac{1}{4} \sum_{j=1}^N |g_j|^2 ( | l \rangle \langle h|_j |h \rangle \langle l |_j + | h \rangle \langle l|_j | l \rangle \langle h|_j ) \right. \\
    & \quad \quad \left. + \frac{1}{4} \sum_{j=1}^N \sum_{j' \neq j} (g_j^* \ket{h}\bra{l}_j - g_j \ket{l}\bra{h}_j) (g_{j'} \ket{l}\bra{h}_{j'} - g_{j'}^* \ket{h}\bra{l}_{j'}) \right\} \\
	&= X_1 - i \tau \sum_{j=1}^N (g_j |l \rangle \langle h|_j - g_j^* |h \rangle \langle l |_j) x_1  + \tau^2 \sum_{j=1}^N |g_j|^2 \left\{ (|h \rangle \langle h|_j - |l \rangle \langle l|_j ) X_1 + \frac{1}{4} \left( |l \rangle \langle l|_j + |h \rangle \langle h|_j \right)  \right\} \\
    &\quad + \frac{\tau^2}{4} \sum_{j=1}^N \sum_{j' \neq j} (g_j^* \ket{h}\bra{l}_j - g_j \ket{l}\bra{h}_j) (g_{j'} \ket{l}\bra{h}_{j'} - g_{j'}^* \ket{h}\bra{l}_{j'}).
\end{split}
\end{align}
\end{widetext}
Its average over the atom state is
\begin{equation}
	\langle X_1 (\tau) \rangle_a = X_1 + (g \tau)^2 \left( N X_1 + \frac{N}{4} \right) = G_0 X_1 + \frac{N (g \tau)^2}{4}.
\end{equation}
The evolution in the next segment is calculated as
\begin{align}
	\langle X_1 (2 \tau) \rangle_a &= \langle e^{\frac{i \tau}{\hbar} \expect{H}_t} \langle X_1 (\tau) \rangle_a e^{-\frac{i \tau}{\hbar} \expect{H}_t} \rangle_a \nonumber \\
	&= G_0^2 X_1 + \frac{N(g \tau)^2}{4} (  1 + G_0 ).
\end{align}
After a sequential calculation over $M$ segments, we obtain
\begin{equation}
	\langle X_{1, \mathrm{out}} \rangle_a = G_0^M X_1 + \frac{ N (g \tau)^2}{4} \sum_{k=0}^{M-1} G_0^k = G X_1 + \frac{1}{4} (G-1).
\end{equation}
Similarly, we also obtain
\begin{equation}
	\langle X_{2,\mathrm{out}} \rangle_a = G X_2 + \frac{1}{4} (G-1). 
\end{equation}

We also calculate the operator $N_b = n^2$ in the following. The evolution in the first segment is
\begin{widetext}
\begin{align}
	N_b (\tau) &= e^{\frac{i \tau}{\hbar} \expect{H}_t} n^2 e^{-\frac{i \tau}{\hbar} \expect{H}_t} = \left( e^{\frac{i \tau}{\hbar} \expect{H}_t} n e^{-\frac{i \tau}{\hbar} \expect{H}_t} \right) \left( e^{\frac{i \tau}{\hbar} \expect{H}_t} n e^{-\frac{i \tau}{\hbar} \expect{H}_t} \right) \nonumber \\
	&= N_b - i \tau \sum_{j=1}^N \left\{ (-g_j^* |h \rangle \langle l |_j b + g_j|l \rangle \langle h|_j b^\dagger ) n + n (-g_j^* |h \rangle\langle l|_j b + g_j |l \rangle \langle h|_j b^\dagger  )  \right\} \nonumber \\
	&\quad + 2 \tau^2 \sum_{j=1}^N |g_j|^2 \left\{  ( |h \rangle \langle l|_j |l \rangle \langle h|_j - |l \rangle \langle h|_j |h \rangle \langle l|_j) N_b + |h \rangle \langle l |_j | l \rangle \langle h|_j n   \right\} \nonumber \\
    &\quad + \tau^2 \sum_{j=1}^N |g_j|^2 \left\{  ( |h \rangle \langle l|_j |l \rangle \langle h|_j + |l \rangle \langle h|_j |h \rangle \langle l|_j  )n + | h \rangle \langle l |_j  | l \rangle \langle h|_j  \right\} \nonumber \\
    &\quad + 2 \tau^2 \sum_{j=1}^N \sum_{j' \neq j} g_{j'}^* g_j \ket{h}\bra{l}_{j'} \ket{l}\bra{h}_j n \nonumber \\
	&=  N_b - i \tau \sum_{j=1}^N \left\{ (-g_j^* |h \rangle \langle l |_j b + g_j|l \rangle \langle h|_j b^\dagger ) n + n (-g_j^* |h \rangle\langle l|_j b + g_j |l \rangle \langle h|_j b^\dagger  )  \right\} \nonumber \\
	&\quad + 2\tau^2 \sum_{j=1}^N |g_j|^2 \left\{ ( |h \rangle \langle h|_j - |l \rangle \langle l|_j ) N_b +  |h \rangle \langle h|_j n \right\} + \tau^2 \sum_{j=1}^N |g_j|^2 \left\{  ( |h \rangle \langle h|_j + |l \rangle \langle l|_j  )n + |h \rangle \langle h |_j \right\} \nonumber \\
    &\quad + 2 \tau^2 \sum_{j=1}^N \sum_{j' \neq j} g_{j'}^* g_j \ket{h}\bra{l}_{j'} \ket{l}\bra{h}_j n.
\end{align}
\end{widetext}
Then, we take the partial average operator over the atom state using the atom state in equation \eqref{eq:atom_initial}:
\begin{align}
\begin{split}
	\langle N_b (\tau) \rangle_a &= N_b + (g \tau)^2 \left( 2 N N_b + 3 N n  + N    \right) \\
    &\simeq G_0^2 N_b + (g \tau)^2 (3 N n + N),
\end{split}
\end{align}
where we approximated $1 + 2 (g \tau)^2 N \approx (1 + (g \tau)N)^2$ in the limit $ g \tau N \ll 1$. This operator evolves in the second segment as
\begin{align}
\begin{split}
	\langle N_b (2 \tau) \rangle_a &= \langle e^{ \frac{i \tau}{\hbar} \expect{H}_t} \langle N_b (\tau) \rangle_a e^{- \frac{i \tau}{\hbar} \expect{H}_t} \rangle_a \nonumber \\
	&= G_0^2 \langle N_b (\tau) \rangle + (g \tau)^2 (3 N \langle n (\tau) \rangle_a + N) \nonumber \\
	&= G_0^4 N_b + (g \tau)^2 (3 N )(\langle n (\tau) \rangle_a + G_0^2 n) \\
    &\quad + (g \tau)^2 N (1 + G_0^2).
\end{split}
\end{align} 
This sequential calculation leads to
\begin{widetext}
\begin{align}
	\langle N_{b, \mathrm{out}} \rangle_a &= G_0^{2M} N_b + (g \tau)^2 (3 N) \sum_{k=0}^{M-1} G_0^{2(M-1-k)} \langle n (k \tau) \rangle_a + ( g \tau)^2 N_2 \sum_{k=0}^{M-1} G_0^{2k} \nonumber \\
	&= G^2 N_b + 4 G(  G-1) n + 2 (G-1)^2 - G( G-1)n + (G - 1) \nonumber \\
	&= G^2 N_b + 3G (G-1)n + (2G-1)(G-1).
\end{align}
\end{widetext}
In summary, we calculated the following:
\begin{align}
	\langle n_\mathrm{out} \rangle_a &= G n + G - 1 \approx G n, \nonumber \\
	\langle b_\mathrm{out} \rangle_a & = \sqrt{G} b, \nonumber \\
	\langle X_{1, \mathrm{out}} \rangle_a &= G X_1 + \frac{1}{4} (G-1) \approx G X_1, \nonumber \\
	\langle X_{2, \mathrm{out}} \rangle_a &= G X_2 + \frac{1}{4} (G-1) \approx G X_2, \nonumber \\
	\langle N_{b, \mathrm{out}} \rangle_a &= G^2 N_b + 3 G (G - 1) n + (2G-1)(G-1). \label{eq:final-operators}
\end{align}

\subsection{Coherent-state phonon input wave}

From these, we can now calculate the various quantities. Let us first assume that the input phonon field is in a coherent state $|\alpha \rangle$. Then, the average phonon number at the end of the amplifier is, using equations \eqref{eq:final-operators}:
\begin{equation}
	\langle \alpha |n_\mathrm{out} | \alpha \rangle = G n_0 + (G - 1), \label{eq:photon-number}
\end{equation}
and the average field is
\begin{equation}
	\langle \alpha | b_\mathrm{out} | \alpha \rangle = \sqrt{G} \alpha. \label{eq:output-field}
\end{equation}
Here, we set $n_0 = \langle \alpha | n | \alpha \rangle$ and we used that $\langle \alpha | b | \alpha \rangle = \alpha$. 

Equation \eqref{eq:photon-number} indicates that the output phonons are of two types having different characteristics: (1) stimulated emission phonons whose mean phonon number is proportional to the input mean phonon number with a gain $G$, and (2) spontaneously emitted phonons whose mean phonon number is unaffected by the input mean phonon number but is (nearly) proportional to the gain $G$. On the other hand the field in equation \eqref{eq:output-field} indicates that the mean output phonon field amplitude is simply proportional to the mean input amplitude with a gain $\sqrt{G}$. This shows that the phonon waves that is stimulated emitted adds on top of the original input phonon wave coherently in phase. 

Next, we calculate quantum noise in the output. For this, we calculate the following:
\begin{align}
	\delta x_{1,\mathrm{out}}^2 &= \langle \alpha | X_{1, \mathrm{out}} | \alpha \rangle - \langle \alpha | x_{1,\mathrm{out}} |\alpha\rangle^2 \nonumber \\
	&= G \langle \alpha | x_1^2 | \alpha \rangle + \frac{1}{4} (G-1) - G \langle \alpha | x_1 | \alpha\rangle^2  \nonumber \\
	&= \frac{1}{2} (G-1) + \frac{1}{4},
\end{align}
where we used
\begin{align}
	\langle \alpha | x_1 ^2 |\alpha \rangle &= \frac{1}{4} \langle \alpha | b^2 + (b^\dagger)^2 + b b^\dagger + b^\dagger b | \alpha \rangle \nonumber \\
	&= \frac{1}{4} \langle \alpha | b^2 + (b^\dagger)^2 + 2 b^\dagger b + 1 | \alpha \rangle, \nonumber \\
	&= \frac{1}{4} \left( \alpha^2 + (\alpha^*)^2 + 2 |\alpha|^2 + 1 \right), \nonumber \\
	\langle \alpha | x_1 | \alpha \rangle^2 &= \frac{1}{4} \left( \alpha + \alpha^* \right)^2 = \frac{1}{4} (\alpha^2 + (\alpha^*)^2 + 2 |\alpha|^2),
\end{align}
therefore, $\langle \alpha | x_1^2 | \alpha \rangle - \langle \alpha | x_1 |\alpha \rangle^2 = 1/4$. Similarly, we obtain
\begin{equation}
	\langle \alpha | X_{2, \mathrm{out}} | \alpha \rangle - \langle \alpha | x_{1, \mathrm{out}}| \alpha \rangle^2 = \frac{1}{2} (G-1) + \frac{1}{4}.
\end{equation}
These results indicate that the quadrature noise increases symmetrically regardless whether it is in in-phase or out-phase quadrature. Also, the noise has two contributions: (1) the first term $(G-1)/2$ is added noise from the spontaneous emission of the amplifier, and (2) the second term, the original shot noise $(1/4)$.

\subsection{Fock-state phonon input wave}

Next, we examine the output from a Fock-state phonon wave input $|n_b\rangle$ where $n_b$ stands for the number of phonons of the Fock state. Using the output operator given in equation \eqref{eq:output-operators}, we calculate
\begin{equation}
	\langle n_b |n_\mathrm{out} | n_b \rangle = G n_b + (G - 1),
\end{equation}
which indicates that the output phonons have two contributions: the amplified number of phonons $(G n_b)$, and the number of spontaneously emitted phonons $(G - 1)$. Because the spontaneously emitted phonons are not exactly in any integer Fock state, the quantity $(G-1)$ corresponds only to the averaged number of spontaneously emitted phonons. 

Then, we can calculate the noise (variance) in the output phonon numbers by calculating the following:
\begin{equation}
	\delta n_\mathrm{out}^2 = \langle n_b | \langle N_{b, \mathrm{out}} \rangle_a | n_b \rangle - \langle n_b | \langle n_\mathrm{out} \rangle_a |n_b \rangle^2.
\end{equation}
We use the equations \eqref{eq:final-operators}, and obtain the following
\begin{widetext}
\begin{align}
\begin{split}
	\delta n_\mathrm{out}^2 &= G^2 \langle n_b |n^2 |n_b \rangle + 3 G (G-1) \langle n_b | n | n_b \rangle + G(G-1) - (G \langle n_b | n | n_b \rangle + G -1)^2 \nonumber \\
	&= G^2 n_b^2 + 3 G ( G - 1) n_b + (2G-1)(G-1) - (G^2 n_b^2 + 2 G(G-1)n_b + (G-1)^2) \nonumber \\
	&= G( G-1)  n_b + G(G - 1).
\end{split}
\end{align}
\end{widetext}
Therefore, the phonon number fluctuation has two contributions: (1) the first term $G(G-1)n_b$ is the (spontaneous-emission)-signal beat noise, and (2) the second term $G(G-1)$ is the noise of the spontaneous emitted phonons.

\section{Detailed Derivation of Electron-Phonon Interaction}
\label{sec: Detailed Derivation of Electron-Phonon Interaction}

In this section, we derive the detailed time-evolution of the phonon population due to interaction with the carriers. We are specifically interested in the population dynamics at and near the phonon population at clamping. Conceptually, clamping should occur when the interaction timescale is on par with the reset time, and we will thus no longer assume that the former is much longer than the latter. For convenience of notation, we re-write the Hamiltonian from Eq.~\eqref{eq: interaction Hamiltonian in waveguide} as follows:
\begin{equation}
H = \sum_{j} \hbar (g_j(t) b^{\dag} c_j + g_j^*(t) b c_j^{\dag}),
\end{equation}
where $c_j = \ket{l}\bra{h}_j$. We are specifically interested in the time-evolution from $t = 0$ to $\tau$, where $\tau$ is the reset time for the carrier population. To that end, the composite wavefunction's evolution is equivalently governed by the average Hamiltonian $\expect{H}_\tau$ (averaged over the period 0 to $\tau$), as demonstrated by the fact that we can express the propagator in the following 2 forms:
\begin{equation}
U(\tau) = e^{-i \int_0^{\tau} dt H(t)} = e^{-i \tau \expect{H}_\tau}.
\end{equation}
The average Hamiltonian is calculated simply by taking the time-averaged value of the coupling coefficients:
\begin{equation}
\expect{H}_\tau = \sum_{j} \hbar (g_j b^{\dag} c_j + g_j^* b c_j^{\dag}),
\end{equation}
where $g_j = \expect{g_j}_\tau$. 

We now use this averaged Hamiltonian to derive the time-evolution of the populations for the individual 2-level carrier systems. It is important to note that for all systems $j$, the standard fermionic relationships $c_j^{\dag}c_j + c_j c_j^{\dag} = 1$, $c_j^2 = 0$, and $(c_j^{\dag})^2 = 0$ apply. The time-derivative of the carrier population for a given system $j$ is calculated as follows:
\begin{align}
\begin{split}
\frac{d}{dt} (c_j^{\dag}c_j) &= -i \Big(g_j b^{\dag} [c_j^{\dag}c_j,c_j] + g_j^* b [c_j^{\dag}c_j,c_j^{\dag}]\Big) \\
&= -i \Big(-g_j b^{\dag} c_j c_j^{\dag} c_j + g_j^* b c_j^{\dag} c_j c_j^{\dag}\Big),
\end{split}
\end{align}
where we omitted the terms containing $c_j^2$ or $(c_j^{\dag})^2$. As expected, the first derivative of the carrier population depends on the phase dynamics of the phonon and carrier states. We abstract these out by taking the second derivative:
\begin{align}
\begin{split}
&\frac{d^2}{dt^2} (c_j^{\dag}c_j) \\
&= -\sum_{j'} \Big(-g_j g_{j'} b^{\dag} [c_j c_j^{\dag} c_j, c_{j'}] - g_j g_{j'}^* [b^{\dag} c_j c_j^{\dag} c_j, b c_{j'}^{\dag}] \\
&\quad\quad + g_j^* g_{j'} [b c_j^{\dag} c_j c_j^{\dag}, b^{\dag} c_{j'}] + g_j^* g_{j'}^* b [c_j^{\dag} c_j c_j^{\dag}, c_{j'}^{\dag}]\Big). 
\end{split}
\end{align}
The first and fourth commutators go to 0 since they contain double-lowering and double-raising operators, respectively:
\begin{align}
[c_j c_j^{\dag} c_j, c_{j'}] &= \delta_{j,j'} \Big(c_j c_j^{\dag} c_j^2 - c_j^2 c_j^{\dag} c_j\Big) = 0, \\
[c_j^{\dag} c_j c_j^{\dag}, c_{j'}^{\dag}] &= \delta_{j,j'} \Big(c_j^{\dag} c_j (c_j^{\dag})^2 - (c_j^{\dag})^2 c_j c_j^{\dag}\Big) = 0.
\end{align}
On the other hand, the second and third commutators simplify to alternating sequences of raising and lowering operators:
\begin{align}
\begin{split}
&[b^{\dag} c_j c_j^{\dag} c_j, b c_{j'}^{\dag}] \\
&= 
\begin{cases}
(b^{\dag} b - b b^{\dag}) c_j c_j^{\dag} c_j c_{j'}^{\dag}, & j \neq j', \\
b^{\dag} b (c_j c_j^{\dag})^2 - b b^{\dag} (c_j^{\dag} c_j)^2, & j = j'
\end{cases}
\\
&=
\begin{cases}
-c_j c_j^{\dag} c_j c_{j'}^{\dag}, & j \neq j', \\
-(c_j^{\dag} c_j)^2 + b^{\dag} b \Big((c_j c_j^{\dag})^2 - (c_j^{\dag} c_j)^2\Big), & j = j'
\end{cases},
\end{split}
\\
\begin{split}
&[b c_j^{\dag} c_j c_j^{\dag}, b^{\dag} c_{j'}] \\
&= -[b^{\dag} c_j c_j^{\dag} c_j, b c_{j'}^{\dag}]^{\dag} \\
&=
\begin{cases}
c_j^{\dag} c_j c_j^{\dag} c_{j'}, & j \neq j', \\
(c_j^{\dag} c_j)^2 - b^{\dag} b \Big((c_j c_j^{\dag})^2 - (c_j^{\dag} c_j)^2\Big), & j = j'
\end{cases}.
\end{split}
\end{align}
Substituting these commutators back into the time-evolution of the carrier population, we find that the second-derivative contains a term scaling in the phonon population $b^{\dag}b$ that conserves the 2-level-system $j$, while another term couples disparate 2-level-systems $j$ and $j'$:
\begin{align}
\begin{split}
\frac{d^2}{dt^2} (c_j^{\dag}c_j) &= |g_j|^2 b^{\dag}b \Big(2 (c_j c_j^{\dag})^2 - 2 (c_j^{\dag} c_j)^2\Big) \\
&\quad - \sum_{j'} \Big(g_j g_{j'}^* c_{j'}^{\dag} c_j c_j^{\dag} c_j + g_j^* g_{j'} c_j^{\dag} c_j c_j^{\dag} c_{j'}\Big).
\end{split}
\end{align}
The first term represents the change in the carrier population due to direct carrier-phonon interaction, while the second term represents the change due to phonon-induced interaction with other carriers. Quantitatively, therefore, the amplitude of the first term scales roughly as the product of the phonon number and the effective carrier number (i.e., the number of carriers in the near-resonance region), while the amplitude of the second term scales roughly as the square of the effective carrier number. Applying the assumption that the phonon number at clamping is much greater than the effective carrier number, we omit the second term from the above expression, yielding:
\begin{align}
\begin{split}
\frac{d^2}{dt^2} (c_j^{\dag}c_j) &\approx |g_j|^2 b^{\dag}b \Big(2 (c_j c_j^{\dag})^2 - 2 (c_j^{\dag} c_j)^2\Big) \\
&= 2 |g_j|^2 b^{\dag}b \Big(1 - c_j^{\dag}c_j\Big) \\
&\approx -4 |g_j|^2 n \bigg(c_j^{\dag}c_j - \frac{1}{2}\bigg),
\end{split}
\end{align}
where in the second line, we apply the relationship $c_j c_j^{\dag} = 1 - c_j^{\dag} c_j$, and in the third line, we make the approximation that the phonon population remains roughly static at a number $n$, due to the assumption that the effective number of 2-level systems is far less than $n$ (which ensures that the oscillation range of the phonon number is negligible compared to $n$). This shows that the excited-state population of each 2-level carrier-state pair undergoes a Rabi oscillation with a median value of $1/2$ and an angular oscillation frequency of $2|g_j|\sqrt{n}$. For 2-level-systems initialized in the excited state or ground state, respectively (i.e., $c_j^{\dag}c_j(0) = 1$ or $c_j^{\dag}c_j(0) = 0$, respectively), we find the following excited-state population at the reset time $\tau$:
\begin{align}
\begin{split}
c_j^{\dag}c_j(\tau) &\approx \pm\frac{1}{2} \cos{\Big(2|g_j|\sqrt{n}\tau\Big)} + \frac{1}{2} \\
&=
\begin{cases}
1 - \sin^2{\Big(|g_j|\sqrt{n}\tau\Big)}, & j \textrm{ excited}, \\
\sin^2{\Big(|g_j|\sqrt{n}\tau\Big)}, & j \textrm{ ground},
\end{cases}
\end{split}
\end{align}
where the first (second) case represents the set of 2-level systems initially in the excited state (ground state). As such, given an initially excited 2-level system, the excited-state population shifts by $-\sin^2{\Big(|g_j|\sqrt{n}\tau\Big)}$, while for a 2-level system initially in the ground state, the excited-state population shifts by $+\sin^2{\Big(|g_j|\sqrt{n}\tau\Big)}$. Since the sum of the phonon and carrier populations is conserved during the period $t = 0$ to $\tau$, the change in the phonon population from $t = 0$ to $\tau$ is opposite of the change in the total carrier excited-state population:
\begin{equation}
\Delta (b^{\dag}b)(\tau) \approx \bigg(\sum_{j \textrm{ excited}} - \sum_{j \textrm{ ground}}\bigg) \sin^2{\Big(|g_j|\sqrt{n}\tau\Big)}.
\end{equation}

\bibliography{ref}

\end{document}